\documentclass[aps,prd,showkeys]{revtex4-2}

\usepackage{amsmath}
\usepackage{amssymb}
\usepackage{natbib}
\usepackage{hyperref}
\usepackage{xcolor}
\usepackage{graphicx}
\usepackage{dcolumn}
\usepackage[loop]{animate}
\usepackage{comment}
\usepackage{ulem}

\graphicspath{{Pictures/},{Pictures/r-mode geometry/},{Pictures/r-mode examples/},{Pictures/r-mode frequencies/},{Pictures/Stitch/},{Pictures/Slides/}}

\newcommand{\bs}{\boldsymbol}
\newcommand{\p}{\partial}
\newcommand{\divv}{\text{div}}
\newcommand{\ctg}{\text{ctg}}
\newcommand{\const}{\text{const}}
\newcommand{\Ai}{\text{Ai}}
\newcommand{\Bi}{\text{Bi}}

\begin{document}

%%%%%%%%%%%%%%%%%%%%%%%%%%%%%%%%%%%%%%%%%%%%%%%%%%%%%%%%%%%%%%%%%%%%%%%%%%%%%%%%%%%%%%%%%%%%%%%%%
%%%%%%%%%%%%%%%%%%%%%%%%%%%%%%%%%%%%%%%%%%%%%%%%%%%%%%%%%%%%%%%%%%%%%%%%%%%%%%%%%%%%%%%%%%%%%%%%%
%%%%%%%%%%%%%%%%%%%%%%%%%%%%%%%%%%%%%%%%%%%%%%%%%%%%%%%%%%%%%%%%%%%%%%%%%%%%%%%%%%%%%%%%%%%%%%%%%
\title{Non-analytic behavior of the relativistic r-modes \\ 
in slowly rotating neutron stars}

\date{\today}

\author{K. Y. Kraav}
\author{M. E. Gusakov}
\author{E. M. Kantor}
\affiliation{Ioffe Institute, Polytekhnicheskaya 26, St.-Petersburg 194021, Russia}

\begin{abstract}

An inconsistency between the theoretical analysis and numerical calculations of the relativistic $r$-modes puzzles the neutron star community since the Kojima's finding of the continuous part in the $r$-mode oscillation spectrum in 1997. In this paper, after a brief review of the Newtonian $r$-mode theory and of the literature devoted to the continuous spectrum of $r$-modes, we apply our original approach to the study of relativistic oscillation equations. Working within the Cowling approximation, we derive the general equations, governing the dynamics of discrete relativistic $r$-modes for both barotropic (isentropic) and nonbarotropic stars. A detailed analysis of the obtained equations in the limit of extremely slow stellar rotation rate reveals that, because of the effect of inertial reference frame-dragging, the relativistic $r$-mode eigenfunctions and eigenfrequencies become {\it non-analytic} functions of the stellar angular velocity, $\Omega$. We also derive the explicit expressions for the $r$-mode eigenfunctions and eigenfrequencies for very small values of $\Omega$. These expressions explain the asymptotic behavior of the numerically calculated eigenfrequencies and eigenfunctions in the limit $\Omega\to 0$. All the obtained $r$-mode eigenfrequencies take discrete values in the frequency range, usually associated with the continuous part of the spectrum. No indications of the continuous spectrum, at least in the vicinity of the Newtonian $l=m=2$ $r$-mode frequency $\sigma=-4/3 \ \Omega$, are found.

\end{abstract}


\maketitle

%%%%%%%%%%%%%%%%%%%%%%%%%%%%%%%%%%%%%%%%%%%%%%%%%%%%%%%%%%%%%%%%%%%%%%%%%%%%%%%%%%%%%%%%%%%%%%%%%
%%%%%%%%%%%%%%%%%%%%%%%%%%%%%%%%%%%%%%%%%%%%%%%%%%%%%%%%%%%%%%%%%%%%%%%%%%%%%%%%%%%%%%%%%%%%%%%%%
%%%%%%%%%%%%%%%%%%%%%%%%%%%%%%%%%%%%%%%%%%%%%%%%%%%%%%%%%%%%%%%%%%%%%%%%%%%%%%%%%%%%%%%%%%%%%%%%%
\section{Introduction}

The generic property of some oscillation modes in rotating stars to be unstable with respect to the emission of gravitational waves, known as the CFS-instability \cite{chandra1970, fs1978_1, fs1978_2, friedman1978}, drives one to the conclusion that, under favorable conditions, these modes can potentially become ``visible'' to the gravitational-wave detectors. Should such a gravitational signal from a mode be observed, it would serve as a valuable source of information on the properties of superdense matter in the stellar interiors, which cannot be studied in terrestrial experiments. Among the variety of oscillation modes that neutron stars can exhibit, the $r$-modes are believed to be the most promising ones from this point of view, since they appear to be the most unstable. In fact, in the absence of dissipation, the $r$-modes become unstable at any rotation frequency of the star \cite{andersson1998, fm1998}.

Extending the traditional mode classification developed by Cowling \cite{cowling1941}, one can attribute the $r$-modes to the class of stellar oscillations, whose main restoring force is the Coriolis force. Their oscillation frequencies vanish in non-rotating stars, and, for slowly rotating stars, the motion of fluid elements induced by such an oscillation is, with high accuracy, characterized as purely toroidal (for this reason the $r$-modes are sometimes called ``quasi-toroidal''). The searches for the $r$-mode gravitational-wave signal are already being conducted \cite{LIGO2021, LIGO21_2}, and there are tentative indications of possible presence of $r$-mode signatures in the X-ray spectra of few observed sources \cite{sm2014_1,sm2014_2}. The fact that the $r$-modes have not yet been observed can be either attributed to insufficient gravitational-wave detectors sensitivity or to the mode suppression by various dissipative processes, operating in the stellar matter. Understanding of the $r$-mode physics becomes of paramount importance as the new more sensitive detectors, such as the Einstein Telescope or the Cosmic Explorer (see, e.g., \cite{he19,et20}), will come into operation.

While the Newtonian theory of $r$-modes is well-developed and has already reached maturity (see the extensive reviews by Andersson \& Kokkotas \cite{ak2001}  and Haskell \cite{haskell14} and references therein), the attempts to find the generalization of the Newtonian $r$-modes within the framework of General Relativity (GR) lead to various contradictory results. An application of the traditional techniques in the slow rotation limit, inherited from the Newtonian theory, immediately predicts the presence of the continuous part in the relativistic $r$-mode oscillation spectrum (e.g., \cite{kojima1997, kojima1998, bk1999, kh1999, kh2000, laf2001, rk2001}), whereas the straightforward $r$-mode numerical calculations do not lead to any indications of the continuous spectrum and predict the discrete spectrum, as in the Newtonian theory (e.g., \cite{yl2002, lfa2003, yoshidaetal2005, gk2008, idrisyetal2015}). A number of attempts has been made to regularize the spectrum, but none of them, in our opinion, can be considered as successful. It was shown that the continuous spectrum does not disappear with the inclusion of the gravitational radiation effect \cite{yf2001, rk2002}. Accounting for the higher order terms, neglected in the slow-rotation approximation, could potentially regularize the spectrum \cite{law2004}, but the fully relativistic calculation including such terms has never been attempted, to our best knowledge. Finally, inclusion of the viscous dissipation in the problem does regularize the spectrum \cite{ponsetal2005, gualtierietal2006}, but such regularization was studied only in the case when the rotational corrections are small compared to the viscous ones, which is exactly the opposite to the situation in real neutron stars.

This set of discordant results discussed in the literature is collectively known as ``the problem of the continuous $r$-mode spectrum''. At this point, it is not clear, whether the presence of the continuous spectrum is a physical phenomenon or it is just an artifact of the made assumptions and simplifications. However, the fact that the numerical calculations, performed beyond the slow-rotation approximation, do not show any continuous spectrum, indicates that the latter emerges as a result of the slow-rotation approximation breakdown, i.e., the traditional techniques of the Newtonian theory cannot be applied to calculating the relativistic $r$-modes in the slowly rotating neutron stars.

In this paper, we develop a new approach to the analysis of the relativistic oscillation equations that allows us to highlight some previously overlooked properties of relativistic $r$-modes, that turn out to be of key importance to the solution of the continuous spectrum problem. The outline of the paper is the following. In section II we describe the theoretical model of a neutron star adopted in our study, and also introduce some useful notations and conventional definitions. In section III, we briefly revisit the traditional approach to the calculation of the $r$-mode eigenfrequencies and eigenfunctions, developed for Newtonian stars. There we discuss the nonrelativistic linearized equations, that govern the dynamics of small stellar oscillations, and then provide some technical details, related to the idea of the derivation of the Newtonian $r$-mode equations. Section IV is devoted to the theoretical investigation of the $r$-modes in General Relativity. We begin with the discussion of the linearized equations that govern small stellar oscillations in GR. We show that the traditional approach, inherited from the Newtonian theory, inevitably leads to the continuous $r$-mode spectrum, instead of the discrete one, which is consistent with the analytical results, known in the literature. Then we develop a new approach to the study of the relativistic oscillation equations, that allows us to restore the traditional discrete $r$-mode spectrum and to obtain the equations, governing the dynamics of the discrete relativistic $r$-modes. We finish section IV with the discussion of the appropriate boundary conditions for the obtained equations. Then, in section V, we show the numerical results of the $r$-mode calculations for different stellar rotation rates. It turns out that the behavior of the relativistic $r$-modes for extremely small values of $\Omega$ significantly differs from that of the Newtonian ones. We focus on this issue in section VI, where we provide a detailed analysis of the obtained equations in the limit of vanishing rotation rate, derive explicit formulas for the $r$-mode eigenfrequencies and eigenfunctions, and discuss the slow-rotation approximation breakdown. Finally, section VII contains a summary and discussion of our results, as well as some concluding remarks.

%%%%%%%%%%%%%%%%%%%%%%%%%%%%%%%%%%%%%%%%%%%%%%%%%%%%%%%%%%%%%%%%%%%%%%%%%%%%%%%%%%%%%%%%%%%%%%%%%
%%%%%%%%%%%%%%%%%%%%%%%%%%%%%%%%%%%%%%%%%%%%%%%%%%%%%%%%%%%%%%%%%%%%%%%%%%%%%%%%%%%%%%%%%%%%%%%%%
%%%%%%%%%%%%%%%%%%%%%%%%%%%%%%%%%%%%%%%%%%%%%%%%%%%%%%%%%%%%%%%%%%%%%%%%%%%%%%%%%%%%%%%%%%%%%%%%%
\section{The model of a neutron star}

From the hydrodynamic viewpoint, a neutron star can be thought of as a liquid mixture of several particle species, bound by gravitational forces. Throughout the text these particle species will be referred to as the components of the matter, and Latin indices $(i,j,k,\dots)$ will be employed to relate different physical quantities to one or another component (for example, $f_k$ stands for the value of a quantity $f$, associated with the component $k$). Then the stellar matter is characterized by the pressure $p$, energy density $\varepsilon$, enthalpy density $w=\varepsilon+p$, temperature $T$, as well as by a set of number densities $n_k$ and chemical potentials $\mu_k$ of its different components. These parameters are not independent: first of all, they should satisfy certain thermodynamic relations, and, second, they should obey a certain equation of state (EOS), provided by the microscopic theory.

In this paper, we focus on the case of the quasineutral, non-magnetized and degenerate matter, and, for the sake of simplicity, we assume that all components of the matter are normal, i.e., are not superfluid or superconducting. The typical frequencies of the shear torsional oscillations are about an order of magnitude lower \cite{sio16,tews17} than the $r$-mode eigenfrequencies in the recycled neutron stars. This means that the crustal shear modulus is small for the problem of $r$-modes in millisecond pulsars. Thus, in what follows, we treat the crust as a liquid. Next, since the magnetic field enters the hydrodynamic equations of the normal matter only via quadratic terms [see Eqs.~(23)-(24) from \cite{dommes2020} for the corresponding terms in the stress-energy tensor], for the case of normal quasineutral non-magnetized matter its perturbations do not appear in the linearized equations of the theory that describe small stellar oscillations. The matter degeneracy
allows us to ignore the temperature dependence of any thermodynamic quantity $f$ and consider it as a function $f(n_k)$, depending only on the set of number densities $n_k$. The thermodynamic relations in degenerate matter take the form
\begin{gather}
\label{thermodynamics}
d\varepsilon=\mu_k dn_k, \qquad dp=n_k d\mu_k, \qquad w=\mu_k n_k.
\end{gather}
Here and hereafter we imply the summation over the repeated Latin indices. At the same time, the EOS of the degenerate matter can be written as $\varepsilon=\varepsilon(n_k)$. Knowing the exact form of this relation, one, with the use of thermodynamic equations, can also obtain the explicit dependencies for all the thermodynamic quantities:
\begin{gather}
\label{EOS}
p=p(n_k), \qquad \varepsilon=\varepsilon(n_k), \qquad
w=w(n_k), \qquad \mu_m=\mu_m(n_k).
\end{gather}

Depending on the properties of the matter, we can roughly divide the stellar interior into two different regions. We will assume that in the outer region any thermodynamic quantity can actually be considered as a function of only one parameter, for example, the energy density.  The EOS of the matter that possesses this property, is said to be barotropic (or isentropic) and can be written in the form $p=p(\varepsilon)$. This outer region throughout the text will be referred to as the crust of a neutron star. We believe that such an approach is justified for the following reasons. First, although we do not account for the superfluidity in this paper, numerical calculations predict (see, e.g. \cite{drddwcp16}) that neutrons should be superfluid in the outer core and crust. In this case the outer core consisting of neutrons, protons, and electrons is effectively barotropic \cite{kg14}. Similarly, the bulk of the crust can be treated as barotropic, except for the narrow regions around chemical inhomogeneities \cite{finn87}. These regions can hardly accumulate significant fraction of oscillation energy and thus should not affect the global mode. In the inner region, on the contrary, it is impossible to parametrize each thermodynamic quantity with only one chosen parameter, thus one generally has to deal with Eqs.~\eqref{EOS}. The EOS obeying this property is called nonbarotropic (nonisentropic), and the inner region with such a matter will further be referred to as the core of a neutron star.

We should also mention that, whether the EOS of the stellar matter is barotropic or not, the matter in thermodynamic and chemical equilibrium can be described by the relation of the form $p_0=p_0(\varepsilon_0)$. This form of the EOS is used in calculations of stellar equilibrium configurations. It is possible to show \citep{hartle1967} that in a slowly and uniformly rotating neutron star in equilibrium the spatial dependency of thermodynamic quantities reduces to the form $f_0=f_0(a)$ with the new spatial coordinate $a$, introduced to replace the radial coordinate $r$ in spherical $(r,\theta,\varphi)$ coordinate system:
\begin{gather}
\label{r(a,theta)}
r(a,\theta)=a+\Omega^2 \zeta(a,\theta),
\end{gather}
where $\Omega$ is the angular velocity of the star, and the function $\zeta(a,\theta)$ describes the oblateness of the star %due to 
caused by rotation. Note that this statement is not only valid for the Newtonian stars, but also for calculations in GR. A number of techniques has been developed to compute $\zeta(a,\theta)$ for both Newtonian and relativistic stellar models, see \cite{tassoul1978, hartle1967, hartle1968, cr1963}.

%%%%%%%%%%%%%%%%%%%%%%%%%%%%%%%%%%%%%%%%%%%%%%%%%%%%%%%%%%%%%%%%%%%%%%%%%%%%%%%%%%%%%%%%%%%%%%%%%
%%%%%%%%%%%%%%%%%%%%%%%%%%%%%%%%%%%%%%%%%%%%%%%%%%%%%%%%%%%%%%%%%%%%%%%%%%%%%%%%%%%%%%%%%%%%%%%%%
%%%%%%%%%%%%%%%%%%%%%%%%%%%%%%%%%%%%%%%%%%%%%%%%%%%%%%%%%%%%%%%%%%%%%%%%%%%%%%%%%%%%%%%%%%%%%%%%%
\section{r-modes in Newtonian gravity}\label{Newt sec}

%%%%%%%%%%%%%%%%%%%%%%%%%%%%%%%%%%%%%%%%%%%%%%%%%%%%%%%%%%%%%%%%%%%%%%%%%%%%
%%%%%%%%%%%%%%%%%%%%%%%%%%%%%%%%%%%%%%%%%%%%%%%%%%%%%%%%%%%%%%%%%%%%%%%%%%%%
\subsection{General linearized equations in Newtonian gravity}

The macroscopic state of a neutron star in Newtonian physics is characterized by the parameters of the stellar matter, discussed in the previous section, the velocity vector field ${\bs v}({\bs r},t)$ of the macroscopic material flows inside the star, and its gravitational potential $\phi$. The closed system of equations, describing dynamical processes in the star, consists of the Euler equation, continuity equations for different particle species, Poisson's equation for the gravitational potential, the EOS of the matter, and thermodynamic relations.

Now, let us consider a small perturbation over the equilibrium configuration of a neutron star, rotating uniformly with the angular velocity ${\bs\Omega}$, directed along the $z$-axis. Within the Eulerian treatment any perturbed quantity $f$ in a given point $({\bs r},t)$ of time and space is decomposed as a sum $f(\bs{r},t)=f_0(\bs{r})+\delta f({\bs r},t)$ of its equilibrium value $f_0$ and its Eulerian perturbation $\delta f$ that represents a small deviation of $f$ from the equilibrium, such that $|\delta f/f_0|\ll 1$. In the study of stellar oscillations, however, this treatment can become inconvenient. The reason is, for example, that a given perturbation may induce the deformation of the surface of a neutron star, which, in turn, leads to the formation of the areas near the surface, where $f_0=0$ and $\delta f\neq 0$, thus the condition $|\delta f/f_0|\ll 1$ does not hold.

The Lagrangian treatment offers an alternative approach to the description of small perturbations, free of these shortcomings.
Let ${\bs r}(t)$ be the perturbed trajectory of the chosen fluid element and $\bs{r}_0(t)$ be the trajectory of the same element in the equilibrium configuration. The vector ${\bs \xi}(\bs{r},t)={\bs r}(t)-\bs{r}_0(t)$, called the Lagrangian displacement, shows the variation of the trajectory of the fluid element, induced by a perturbation. Formally, we have the map $\Psi$ from the manifold that corresponds to the support of the equilibrium star, to the manifold that corresponds to the support of the perturbed star, and this map transforms the unperturbed trajectories into the perturbed ones. This map uniquely induces the corresponding map $\Psi_*$, called ``pushforward'' that transfers tensor fields of the equilibrium star to the perturbed star. That is, while $\Psi$ maps the point ${\bs r}$ into the point ${\bs r}+{\bs \xi}$, the pushforward $\Psi_*$ converts the tensor field $f({\bs r},t)$ in the point ${\bs r}$ into the transformed tensor field $(\Psi_*f)({\bs r}+{\bs \xi},t)$ in the point ${\bs r}+{\bs \xi}$. Since the displacement vector is assumed to be small, tensor fields before and after such transformation (up to linear terms in ${\bs\xi}$) are related as $(\Psi_*f)({\bs r},t)=f({\bs r},t)-\mathcal{L}_{\xi}f({\bs r},t)$, which is simply the definition of the Lie derivative $\mathcal{L}_\xi f$ of a tensor field $f$ along the vector field ${\bs\xi}$, being the generator of the map $\Psi$. Following the paper of Friedman and Shutz \cite{fs1978}, we introduce the Lagrangian perturbation of a tensor field $f$ as
\begin{gather}
\Delta f({\bs r},t)=f({\bs r},t)-(\Psi_*f_0)(\bs r,t)=\delta f({\bs r},t)+\mathcal{L}_{\xi}f_0({\bs r}).
\end{gather}
From the physical viewpoint the Lagrangian perturbation of a tensor field $f$ shows the change in its components with respect to the reference frame, which is embedded in the fluid and sensitive to the perturbations of the star. Particularly, the Lagrangian perturbation of the scalar field can be written as $\Delta f({\bs r},t)=f[{\bs r}(t),t]-f_0[{\bs r_0}(t)]$ and shows the difference between the perturbed field $f({\bs r},t)$, measured along the perturbed trajectory ${\bs r}(t)$ of a fluid element, and the unperturbed field $f_0({\bs r}_0)$, measured along the unperturbed trajectory ${\bs r}_0(t)$ of the same fluid element. Note that thus defined perturbations $\Delta=\delta+\mathcal{L}_\xi$ generally differ from those, introduced, e.g., by Unno et al. \cite{unno1979} as $\Delta_\text{U}=\delta+({\bs \xi \bs \nabla})$, and we have $\Delta f=\Delta_\text{U}f$, only if $f$ is a scalar field.

Using the definition
$\bs{v}({\bs r},t)=d\bs{r}(t)/dt$, it is now easy to show \citep{fs1978} that
\begin{gather}
\label{xi_Newt}
\delta{\bs v}=\frac{\p {\bs \xi}}{\p t}+({\bs v}_0\bs{\nabla}){\bs\xi}-({\bs \xi}\bs{\nabla}){\bs v}_0, \qquad \Delta{\bs v}=\frac{\p \bs{\xi}}{\p t}.
\end{gather}
Although these relations may not be very useful in the Newtonian theory, where one often solves the problem in the corotating reference frame with $\bs{v}_0=0$, they can be easily generalized to the relativistic case and allow one to properly introduce the relativistic counterpart of the Lagrangian displacement, as will be discussed in section IV. Further in this section, we write the equations in the corotating reference frame, in which case $\delta {\bs v}=\Delta \bs v=\p{\bs\xi}/\p t$.

In terms of the Lagrangian displacements the closed system of equations, governing the dynamics of small stellar oscillations ${\bs \xi}\sim \delta f\sim\Delta f\sim e^{i\sigma_r t}$, where $\sigma_r$ is the oscillation frequency in the corotating reference frame, can be written in the following form: 
\begin{gather}
\label{gen_system_Newt}
\left\{
\begin{gathered}
-\sigma_r^2\bs{\xi}+2i\sigma_r[\bs{\Omega}\times\bs{\xi}]=c^2\biggl(\frac{\delta w}{w_0^2}{\bs\nabla}p_0-\frac{1}{w_0}{\bs\nabla}\delta p\biggr)-{\bs\nabla}\delta\phi \\
\delta n_k+\divv(n_{k0}{\bs\xi})=0 \\
\nabla^2\delta\phi=\frac{4\pi G}{c^2}\biggl(\delta w+2 \delta p\biggr) \\
p_0=p_0(n_{k0}), \quad \mu_{m0}=\mu_{m0}(n_{k0}), \quad w_0=\mu_{k0}n_{k0} \\
\delta p=\biggl(\frac{\p p}{\p n_{k}}\biggr)_0 \delta n_k, \quad \delta w=\mu_{k0}\delta n_k+\delta p.
\end{gathered}
\right.
\end{gather}
Here $c$ stands for the speed of light and $G$ is the gravitational constant. The first equation is the linearized Euler equation, written in the corotating reference frame. The second and third equations are the linearized continuity equations and linearized Poisson's equation for the gravitational potential, respectively. The rest of the equations are obtained with the use of  thermodynamic relations \eqref{thermodynamics} and the EOS of the matter \eqref{EOS}. 

Generally, the solution of this system is equivalent to the simultaneous solution of the two second-order differential equations in partial derivatives. In order to simplify the problem, we ignore the perturbed gravitational field in the Euler equation (the so-called Cowling approximation \citep{cowling1941}), so that the Poisson's equation could be decoupled and the rest of the equations could be solved independently. This is equivalent to the solution of one second-order differential equation in partial derivatives and greatly simplifies the problem. At the same time, Cowling approximation provides, with reasonable accuracy, information on the qualitative behavior of many different oscillation modes, arising in neutron stars (see \cite{cox1980} and references therein). In particular, for Newtonian $r$-modes the relative error in oscillation eigenfrequencies [to be more specific, in the eigenfrequency corrections $\sigma^{(1)}$, defined below by Eq.\eqref{ordering_Newt}] due to Cowling approximation is less than 8\% \cite{provost1981, saio1982}.

The system \eqref{gen_system_Newt} can be easily reduced to the system of three equations for ${\bs\xi}$, $\delta p$ and $\delta w$. To this aim, we multiply the linearized continuity equations for different particle species either by $(\p p/\p n_k)_0$, or by $\mu_{k0}$, and perform the summation over the $k$-index. With the use of thermodynamic relations two equations, obtained this way, can be transformed to the form, which is free of the number density perturbations and contains only the functions ${\bs\xi}$, $\delta p$, and $\delta w$. Adopting the Cowling approximation, we can eventually rewrite the resulting equations as
\begin{gather}
\label{system_Newt_core}
\left\{
\begin{gathered}
-\sigma_r^2\bs{\xi}+2i\sigma_r[\bs{\Omega}\times\bs{\xi}]=c^2\biggl(\frac{\delta w}{w_0^2}{\bs\nabla}p_0-\frac{1}{w_0}{\bs\nabla}\delta p\biggr) \\
\delta p+w_0\biggl(\frac{c_s}{c}\biggr)^2 \divv{\bs\xi}+({\bs\xi\bs\nabla})p_0=0 \\
\delta w-\delta p+\divv(w_0{\bs\xi})-({\bs\xi\bs\nabla})p_0=0,
\end{gathered},
\qquad c_s=c\sqrt{\frac{1}{w_0}\biggl(\frac{\p p}{\p n_k}\biggr)_0 n_{k0}},
\right.
\end{gather}
where $c_s$ is the local speed of sound. Because of their origin and for the further convenience, the second and the third equations of this system will be referred to as the ``continuity equation 1'' and ``continuity equation 2'', correspondingly. Note that in the case of the barotropic EOS we have
\begin{gather}
\label{barotropic_EOS}
\delta w=\biggl(\frac{d w}{d p}\biggr)_0 \delta p, \qquad c_s=c\sqrt{\biggl(\frac{d p}{d\varepsilon}\biggr)_0},
\end{gather}
which makes these continuity equations equivalent to each other and allows one to transform the right-hand side of the Euler equation to the form of the pure gradient, so that the system reduces to,
\begin{gather}
\label{system_Newt_crust}
\left\{
\begin{gathered}
-\sigma_r^2\bs{\xi}+2i\sigma_r[\bs{\Omega}\times\bs{\xi}]=-c^2{\bs\nabla}\biggl(\frac{\delta p}{w_0}\biggr), \\
\delta p+w_0\biggl(\frac{c_s}{c}\biggr)^2 \divv{\bs\xi}+({\bs\xi\bs\nabla})p_0=0.
\end{gathered}
\right.
\end{gather}
As we shall see in the following sections, the difference of the right-hand sides of the Euler equations for barotropic and nonbarotropic matter significantly affects the mathematical properties of the problem. Thus, generally, the study of the global oscillatory modes splits into the study of the system \eqref{system_Newt_core} in the core of the star and the study of the system \eqref{system_Newt_crust} in the crust of the star with appropriate boundary conditions, imposed at the stellar center, at the crust-core interface and at the surface of the star.

Finally, for the sake of mathematical convenience, instead of dealing with $\xi^\theta$ and $\xi^\varphi$, we introduce the functions $Q$ and $T$, such that
\begin{gather}
\xi^{\theta}=\frac{\p Q}{\p \theta}+\frac{1}{\sin\theta}\frac{\p T}{\p\varphi}, \qquad \xi^\varphi=\frac{1}{\sin\theta}\frac{\p Q}{\p\varphi}-\frac{\p T}{\p\theta}.
\end{gather}
It is easy to verify that, once $Q$ and $T$ are expanded in spherical harmonics, these formulas simply reduce to the decomposition of the vector field ${\bs \xi}$ in vector spherical harmonics \citep{barrera1985}. The function $T$ is often referred to as the toroidal component of the motion. One can show that any purely toroidal vector field is divergenceless.

%%%%%%%%%%%%%%%%%%%%%%%%%%%%%%%%%%%%%%%%%%%%%%%%%%%%%%%%%%%%%%%%%%%%%%%%%%%%
%%%%%%%%%%%%%%%%%%%%%%%%%%%%%%%%%%%%%%%%%%%%%%%%%%%%%%%%%%%%%%%%%%%%%%%%%%%%
\subsection{The traditional approach to the Newtonian $r$-mode calculations}

As was mentioned above, rotation of a star brings to life special classes of oscillatory modes, whose main restoring force is the Coriolis force. They arise because of the rotation itself and have no counterparts in non-rotating stars. In particular, there is a class of such oscillations that obey the following two conditions. First, their frequency should vanish when $\Omega=0$. Second, they should be predominantly toroidal, that is, only toroidal component should survive in the limit $\Omega\to 0$. In the literature oscillations, satisfying these conditions, are referred to as the $r$-modes (see the extensive reviews by Andersson \& Kokkotas \cite{ak2001} and Haskell \cite{haskell14}, and references therein for more details). Now we, following Provost et al. \citep{provost1981}, briefly explain the main ideas of the traditional approach to the Newtonian $r$-mode calculations. Here we do not focus on the explicit form of the equations, since in the following sections we provide the detailed derivation of their relativistic counterparts. 

Let us consider the $r$-modes in slowly rotating neutron stars. Since the rotation is slow, we can present all the quantities of interest in the form of $\Omega$-series. Once we assume that the oscillation frequency in the leading order is proportional to $\Omega$, the analysis of Eqs.~\eqref{system_Newt_core} and \eqref{system_Newt_crust} shows that both in the nonbarotropic core and in the barotropic crust the first terms of these series are
\begin{gather}
\label{ordering_Newt}
\sigma_r/\Omega=\sigma^{(0)}_r+\Omega^2\sigma^{(1)}+ o(\Omega^2), \qquad T=T^{(0)}+\Omega^2 T^{(1)}+o(\Omega^2), \\ 
Q=\Omega^2 Q^{(1)} + o(\Omega^2), \qquad \xi^r=\Omega^2 \xi^{(1)} + o(\Omega^2), \\
\label{ordering_Newt2}
\delta p=\Omega^2 p^{(1)}+o(\Omega^2), \qquad \delta w=\Omega^2 w^{(1)}+o(\Omega^2),
\end{gather}
and that only even powers of $\Omega$ survive in all equations. 
The latter property allows one to develop the traditional perturbation theory with $\Omega^2$ being the small parameter. These series will further be referred to as the {\it traditional $r$-mode ordering.}

The mode eigenfrequency $\sigma^{(0)}_r$ can be easily found from the leading order equations. Let us have a look, for example, at the equations in the core. Since in the leading order the vector $\bs{\xi}$ is purely toroidal, its divergence vanishes, and continuity equations of the system \eqref{system_Newt_core} become identities. The $r$-, $\theta$- and $\varphi$-components of the Euler equation are:
\begin{gather}
\label{order0_Newt}
\left\{
\begin{gathered}
2 i \sigma^{(0)}_r\sin\theta\frac{\p T^{(0)}}{\p\theta}=c^2\biggl(\frac{w^{(1)}}{w_0^2}\frac{d p_0}{da}-\frac{1}{w_0}\frac{\p p^{(1)}}{\p a}\biggr), \\
\frac{\sigma^{(0)2}_r}{\sin\theta}\frac{\p T^{(0)}}{\p\varphi}-2 i \sigma^{(0)}_r\cos\theta\frac{\p T^{(0)}}{\p\theta}=\frac{c^2}{w_0 a}\frac{\p p^{(1)}}{\p\theta}, \\
-\sigma^{(0)2}_r\frac{\p T^{(0)}}{\p\theta}-2 i \sigma^{(0)}_r\ctg\theta\frac{\p T^{(0)}}{\p\varphi}=\frac{c^2}{w_0 a \sin\theta}\frac{\p p^{(1)}}{\p\varphi}.
\end{gathered}
\right.
\end{gather}
Now, if we set $\p^2 p^{(1)}/\p\theta\p\varphi$, obtained from the $\theta$-component, equal to the same derivative, obtained from the $\varphi$-component of the Euler equation, we eventually arrive at the equation for the toroidal component:
\begin{gather}
\label{Newt_spec_Eq}
\frac{1}{\sin\theta}\frac{\p}{\p\theta}\sin\theta\frac{\p T^{(0)}}{\p \theta}+\biggl[-\frac{2 i}{\sigma^{(0)}_r}\frac{\p T^{(0)}}{\p\varphi}+\frac{1}{\sin^2\theta}\frac{\p^2 T^{(0)}}{\p\varphi^2}\biggr]=0.
\end{gather}
It is easy to see that this equation admits the solution of the form
\begin{gather}
\label{toroidal_comp}
\sigma^{(0)}_r=\frac{2m}{l(l+1)}, \qquad T^{(0)}=-i T^{(0)}_{lm}(a)P_{l}^m(\cos\theta)e^{i m\varphi},
\end{gather}
where $P_l^m(\cos\theta)$ is the associated Legendre polynomial, $T^{(0)}_{lm}$ is the $r$-mode amplitude, and the factor $(-i)$ is introduced for further convenience.  Note that the analogous derivation with the same outcome can be carried out for the crust of the neutron star. This result implies that the dependency of all perturbations on the azimuthal angle reduces to $e^{im\varphi}$. Further, for brevity, we omit this dependency and write all the perturbations simply as $\delta f(a,\theta)$. Without any loss of generality we can assume that $m>0$, since the solution with $m<0$ can be obtained from the solution with $m>0$ via complex conjugation. The typical geometry of the streamlines, corresponding to the purely toroidal vector field $\bs\xi$ with angular dependency \eqref{toroidal_comp} for different combinations of $l$ and $m$, is pictured in the animated Fig.\ \ref{streamlines} in Appendix \ref{r-mode geometry}.

Thus, the leading order equations allow one to specify the angular dependency of $T^{(0)}$. The situation with the calculation of the amplitude $T^{(0)}_{lm}(a)$ is more complicated and depends on whether the EOS is barotropic or not. From Eq.~\eqref{order0_Newt} it is clear that in the core $T^{(0)}_{lm}(a)$ cannot be found in the leading order, thus, the analysis of the next order equations is needed. These equations are studied using the expansions in associated Legendre polynomials,
\begin{gather}
\label{expansions}
f^{(1)}=\sum_{L\geq m} f^{(1)}_{Lm}(a) P_L^m(x), \qquad x=\cos\theta.
\end{gather}
Since the azimuthal number $m$ is fixed, in what follows, for convenience, we omit the subscript $m$ and write $T^{(0)}_l$ instead of $T^{(0)}_{lm}$ and $f^{(1)}_{L}$ instead of $f^{(1)}_{Lm}$, although these quantities may actually depend on the value of $m$. We substitute these expansions in the first order equations and use the relations
\begin{gather}
\label{Legendre_properties}
\frac{d}{dx}(1-x^2)\frac{dP_L^m}{dx}+\biggl[L(L+1)-\frac{m^2}{1-x^2}\biggr]P_L^m=0, \qquad P_L^m=0 \text{ when $L<m$}, \\
\label{Legendre_properties_2}
x P_L^m=k^{+}_L P_{L-1}^m+k^{-}_L P_{L+1}^m, \qquad (1-x^2)\frac{d P_L^m}{dx}=(l+1)k^{+}_L P_{L-1}^m-l k^{-}_L P_{L+1}^m, \\
k^+_L=\frac{L+m}{2L+1}, \qquad k^-_L=\frac{L-m+1}{2L+1}
\end{gather}
to reduce the angular dependency of all the terms in the obtained equations to one or another Legendre polynomial. Then, setting the coefficients before different Legendre polynomials to zero, we arrive at the system of the first-order ordinary differential equations (ODEs) for $f_L$. The analysis of these equations, along with the first equation of \eqref{order0_Newt} shows that in the core the problem reduces to the study of the closed subsystem of equations for $T^{(0)}_l$, $Q^{(1)}_{l\pm 1}$, and $\xi^{(1)}_{l\pm 1}$.

In the crust of the neutron star the situation is different. The reason is that the right-hand side of the Euler equation has the form of the pure gradient, which allows one to find $T^{(0)}_l(a)$ without considering the first order equations. Indeed, $\theta$- and $\varphi$-components of the Euler equation exactly coincide with their counterparts in the core, whereas the radial component takes the form [compare with the first equation of \eqref{order0_Newt}]:
\begin{gather}
2 i \sigma^{(0)}_r\sin\theta\frac{\p T^{(0)}}{\p\theta}=-c^2\frac{\p}{\p a}\biggl(\frac{p^{(1)}}{w_0}\biggr).
\end{gather}   
Now, since $p^{(1)}\sim e^{im\varphi}$, we can express $p^{(1)}$ from the $\varphi$-component of the Euler equation and substitute it to the equation above. Using the relations \eqref{Legendre_properties}, we obtain:
\begin{gather}
\label{lm only}
(l+1)^2k^+_l\biggl[\frac{d}{da}\bigl(a T^{(0)}_l\bigr)+l T^{(0)}_l\biggr]P_{l-1}^m(x)+l^2 k^-_l\biggl[\frac{d}{da}\bigl(a T^{(0)}_l\bigr)-(l+1)T^{(0)}_l\biggr]P_{l+1}^m(x)=0.
\end{gather}
This equation is not contradictory, if and only if $l=m$, in which case we immediately find
\begin{gather}
T^{(0)}_l(a)=\const\cdot a^l. 
\end{gather}
The approach to the next order equations in the crust is the same, as the one, adopted in the core: we start with the expansions \eqref{expansions} and then study the closed subsystem of the first-order ODEs for the coefficients $f_L$. But in the case of the barotropic EOS the solution of this system can be reduced  to the solution of one simple equation for the radial displacement that can be formally written as
\begin{gather}
\xi^{(1)}_{l+1}{}'+g_1(a)\xi^{(1)}_{l+1}+[g_{21}(a)\sigma^{(1)}+g_{22}(a)]T^{(0)}_l(a)=0,
\end{gather}
where $g_1(a)$, $g_{21}(a)$ and $g_{22}(a)$ are some functions of $a$ [see equations \eqref{g1coeff}-\eqref{g22coeff} for their relativistic counterparts], and the prime denotes the derivative $d/da$. This equation is easily solved analytically, 
\begin{gather}
\label{XiSolNewt}
\xi^{(1)}_{l+1}(a)=\frac{1}{\eta(a)}\biggl[\xi_0+\int\limits_{a}^1 \biggl(g_{21}(a)\sigma^{(1)}+g_{22}(a)\biggr)\eta(a)T^{(0)}_l(a)da\biggr], \qquad \eta(a)=\exp\biggl(-\int\limits_{a}^1 g_1(a)da\biggr).
\end{gather}
Here $\xi_0$ is the integration constant and the variable $a$ is normalized such that $a=1$ corresponds to the stellar surface.

Thus, in order to find the global $r$-modes one has to solve the mentioned system of ODEs in the core and the obtained ODE in the crust, supplemented by appropriate boundary conditions. Solution of these equations, as will be discussed below in Sec.\ \ref{bc section}, constitutes the eigenvalue problem that allows one to find the eigenfrequency corrections $\sigma^{(1)}$ and corresponding eigenfunctions. These eigenfunctions are traditionally distinguished by the number of nodes of the toroidal function, i.e by the number of points inside the star, where $T^{(0)}_l(a)=0$ (see, e.g, Provost et al. \cite{provost1981}). 

In numerical calculations and in the theoretical analysis it is usually convenient to operate with dimensionless equations. Let $M$ and $R$ be, respectively, the mass and radius of a given neutron star. The transition to the dimensionless counterparts of the functions and variables discussed above can be achieved by the following formal replacements in the hydrodynamic equations:
\begin{gather}
\{a,T,Q,\xi^r\}\to R\cdot \{a,T,Q,\xi^r\}, \qquad \{c,c_s\}\to \Omega_K R\cdot\{c,c_s\}, \\
\{p_0,w_0,\delta p,\delta w\}\to\frac{M(\Omega_K R)^2}{R^3}\cdot\{p_0,w_0,\delta p,\delta w\}, \qquad \left\{\Omega,\sigma_r\right\}\to\Omega_K\cdot \left\{\Omega,\sigma_r\right\},
\end{gather}
where
\begin{gather}
\Omega_K=\sqrt{\frac{GM}{R^3}}
\end{gather}
is the quantity of the order of the stellar Keplerian angular velocity. Actually, these replacements do not change the form of the equations: all thus introduced multipliers eventually cancel each other out and one is left with {\it exactly} the same equations. Therefore, all the quantities in the resulting ODEs in the core and in the crust can be considered as dimensionless. From now on, we always work with dimensionless equations.

%%%%%%%%%%%%%%%%%%%%%%%%%%%%%%%%%%%%%%%%%%%%%%%%%%%%%%%%%%%%%%%%%%%%%%%%%%%%%%%%%%%%%%%%%%%%%%%%%
%%%%%%%%%%%%%%%%%%%%%%%%%%%%%%%%%%%%%%%%%%%%%%%%%%%%%%%%%%%%%%%%%%%%%%%%%%%%%%%%%%%%%%%%%%%%%%%%%
%%%%%%%%%%%%%%%%%%%%%%%%%%%%%%%%%%%%%%%%%%%%%%%%%%%%%%%%%%%%%%%%%%%%%%%%%%%%%%%%%%%%%%%%%%%%%%%%%
\section{r-modes in General Relativity}

%%%%%%%%%%%%%%%%%%%%%%%%%%%%%%%%%%%%%%%%%%%%%%%%%%%%%%%%%%%%%%%%%%%%%%%%%%%%
%%%%%%%%%%%%%%%%%%%%%%%%%%%%%%%%%%%%%%%%%%%%%%%%%%%%%%%%%%%%%%%%%%%%%%%%%%%%
\subsection{General linearized equations in GR}

The approach to the description of stellar dynamics in General Relativity resembles the one, adopted in the Newtonian physics, although there is a number of key differences. First of all, the macroscopic flows in the star are characterized by the 4-velocity vector field $\mathfrak{u}^\mu$, instead of the vector field ${\bs v}$. Secondly, a set of metric functions $\mathfrak{g}_{\mu\nu}$, obeying the Einstein equations, describes the gravity of the neutron star, instead of the gravitational potential $\phi$, obeying the Poisson's equation. In $x^{\mu}=(ct,a,\theta,\varphi)$ coordinates the equilibrium configuration of the slowly rotating neutron star corresponds to the equilibrium 4-velocity
\begin{gather} 
u^\varphi=\frac{\Omega}{c} u^t, \qquad u^a=u^\theta=0, \qquad u_{\mu} u^\mu=-1,
\end{gather}
and its gravitational field is defined by the line element of the form \cite{hartle1967}
\begin{gather}
\label{geometry}
ds^2=g_{\mu\nu}dx^
\mu dx^\nu=-e^{2\nu(a,\theta)}c^2 dt^2+e^{2\lambda(a,\theta)}[dr(a,\theta)]^2+r^2(a,\theta)e^{2\psi(a,\theta)}[d\theta^2+\sin^2\theta (d\varphi-\Omega \omega(a)dt)^2].
\end{gather}
In this study we present the differential $dr(a,\theta)$ [see Eq.~\eqref{r(a,theta)} for the definition of $\zeta$]  and metric functions $\nu$, $\lambda$, and $\psi$ as
\begin{gather}
dr(a,\theta)=\biggl(1+\Omega^2\frac{\p\zeta}{\p a}\biggr)da+\Omega^2\frac{\p \zeta}{\p \theta}d\theta, \\
 \nu(a,\theta)=\nu_0(a)+\Omega^2\nu_2(a,\theta), \qquad
\lambda(a,\theta)=\lambda_0(a)+\Omega^2\lambda_2(a,\theta), \qquad \psi(a,\theta)=\Omega^2\psi_2(a,\theta),
\end{gather}
which is equivalent to the decompositions, adopted by Hartle \cite{hartle1967}. Here and hereafter, for convenience, we omit the subscript ``0'' when dealing with the equilibrium 4-velocity and metric tensor, and write $u^\mu$ and $g_{\mu\nu}$ instead of, respectively, $\mathfrak{u}_0^\mu$ and $\mathfrak{g}_{0,\mu\nu}$. In what follows, we also use the $(-,+,+,+)$ metric signature convention and imply the summation over the repeated Greek indices. Note that, whereas in the Newtonian gravity the rotation affects the equilibrium configuration of the star only starting with the terms of the order $\Omega^2$ and higher, in GR the rotation of the star manifests itself already in the first order in $\Omega$: the function $\omega(a)$ describes the relativistic effect of the inertial reference frame-dragging and, as we shall see, drastically changes the properties of the relativistic $r$-mode counterparts.

As in the Newtonian theory, in GR there are two ways of describing small perturbations of a neutron star. The Eulerian treatment does not undergo any changes and can be instantly applied in the study of relativistic equations. The Lagrangian approach, however, needs modifications, since the space-time manifold does not have any natural vector space structure and, thus, the Newtonian definition of the Lagrangian displacement as of the difference $\bs r(t)-\bs r_0(t)$ cannot be used anymore. The relativistic generalization of the Lagrangian treatment was developed in the papers by Taub \cite{taub1969}, Carter \cite{carter1973}, and  Friedman and Shutz \cite{friedman1978, fs1975}. It was proposed to introduce the Lagrangian displacement 4-vector  $\xi^\mu$ as the generator of the map $\Phi$ that transforms the word lines of fluid elements in the equilibrium star into the world lines of the same fluid elements in the perturbed star (similarly to how the vector field ${\bs \xi}$ serves as a generator of the Newtonian map $\Psi$). The Lagrangian perturbation of a tensor field $f$ is then defined as $\Delta f(x^\mu)=f(x^\mu)-(\Phi_*f_0)(x^\mu)$, where $\Phi_*$ is the pushforward, associated with the map $\Phi$ (compare with the definition of the Lagrangian perturbation in the Newtonian theory). As we have already discussed, the Eulerian and Lagrangian perturbations of any physical quantity $f$ are related as,
\begin{gather}
\Delta f(x^\mu)=\delta f(x^\mu)+\mathcal{L}_\xi f_0(x^\mu),
\end{gather}
where $\mathcal{L}_\xi f$ is the Lie derivative of a tensor field $f$ along the vector field $\xi^\mu$. Particularly, it can be shown \citep{friedman1978} that the Eulerian perturbation of the 4-velocity $\delta u^\mu=\mathfrak{u}^\mu-u^\mu$ is related to $\xi^\mu$ and the metric Eulerian perturbations $\delta g_{\mu\nu}=\mathfrak{g}_{\mu\nu}-g_{\mu\nu}$ as 
\begin{gather}
\label{displacement_GR}
\delta u^\mu=\frac{1}{2}u^\mu u^\rho u^\lambda \delta g_{\rho\lambda}+\perp^\mu_\rho\mathcal{L}_{u} \xi^\rho, \qquad \perp^\mu_\rho=\delta^\mu_\rho+u^\mu u_{\rho}.
\end{gather}
From these equations it is easy to verify that the Eulerian perturbation $\delta u^\mu$ automatically satisfies the linearized form of the normalization condition $\mathfrak{u}_\mu \mathfrak{u}^\mu=-1$.

The introduced Lagrangian displacement $\xi^\mu$ is defined up to the gauge transformation $\xi^\mu\to \xi^\mu +\eta^\mu$, where $\eta^\mu$ is the trivial displacement, that is the displacement that realizes the solution of the perturbed fluid equations with all Eulerian perturbations set to zero. Particularly, this means that $\perp^\mu_\rho\mathcal{L}_{u}\eta^\rho=0$, which allows us to impose the gauge condition $u_{\mu}\xi^\mu=0$ on the Lagrangian displacement $\xi^\mu$ \citep{friedman1978, fs1975}. 

The fully relativistic calculation of stellar oscillations constitutes a complex problem, since one has to deal with linearized Einstein equations in {\it rotating} neutron star. At the same time, if we ignore the perturbations of the geometry and set $\delta g_{\mu\nu}=0$ (the simplest form of the relativistic Cowling approximation), the problem greatly simplifies and reduces to studying the linearized hydrodynamic equations, as it was in the Newtonian theory. A comparison of the full relativistic calculation and the calculation within the Cowling approximation for different stellar oscillations shows \citep{lsr1990, yk1997, jc2017} that the latter can be used, with reasonable accuracy, as an estimate for the normal stellar modes and corresponding eigenfrequencies, unless one is interested in the study of $w$-modes, which do not have a Newtonian counterpart. For the relativistic $r$-modes, which we are interested in, the Cowling approximation leads to the relative error in oscillation frequency $\sigma$ ranging from 6\% to 11\% \cite{jc2017}. As will be discussed below (see Sec.\ \ref{disc sec}), the error introduced by the Cowling approximation for the relativistic $r$-modes in the $\Omega\to 0$ limit is likely to be significantly smaller because of the peculiarities of the $r$-mode behaviour in this limit.

In the Cowling approximation the dynamics of small perturbations is governed by the following closed system of equations:
\begin{gather}
\left\{
\begin{gathered}
(u^\rho\nabla_\rho)\delta u^\mu+\delta u^\rho\biggl[\nabla_\rho u^\mu+\frac{1}{w_0}\biggl(u^\mu\delta^\lambda_\rho+u^\lambda\delta^\mu_\rho\biggr)\nabla_\lambda p_0\biggr]-\perp^{\mu\rho}\biggl[\frac{\delta w}{w_0^2}\nabla_\rho p_0-\frac{1}{w_0}\nabla_\rho \delta p\biggr]=0, \\
\frac{1}{\sqrt{-\det\mathfrak{g}}}\frac{\p }{\p x^\mu}\biggl[\sqrt{-\det\mathfrak{g}}(\delta n_k u^\mu+n_{k0}\delta u^\mu)\biggr]=0, \\
p_0=p_0(n_{k0}), \quad \mu_{m0}=\mu_{m0}(n_{k0}), \quad w_0=\mu_{k0}n_{k0}, \\
\delta p=\biggl(\frac{\p p}{\p n_{k}}\biggr)_0 \delta n_k, \quad \delta w=\mu_{k0}\delta n_k+\delta p, \\
\delta u^\mu=\perp^\mu_\rho\mathcal{L}_{u} \xi^\rho, \qquad u_{\mu}\xi^\mu=0.
\end{gathered}
\right.
\end{gather}
Here $\det\mathfrak{g}\equiv\det\mathfrak{g}_{\mu\nu}=\det g_{\mu\nu}$, the covariant derivative $\nabla_\rho$ is defined for the geometry \eqref{geometry}, and the metric tensor $g_{\mu\nu}$ is used to relate covariant and contravariant tensorial components. The first equation is the linearised Euler equation, written in the laboratory reference frame, and the second equation presents the set of continuity equations for different particle species $k$. The rest of the equations are obtained from thermodynamic relations \eqref{thermodynamics}, the EOS of the matter \eqref{EOS}, the gauge condition for the Lagrangian displacement, and the formula \eqref{displacement_GR}, written in the Cowling approximation. Note that the components of the linearized Euler equation are linearly dependent: if we denote the left-hand side of this equation as $F^\mu$, then, using the normalization $u_{\mu}u^\mu=-1$ and the unperturbed Euler equation,
\begin{gather}
\label{GR equilibrium}
(u^\rho \nabla_\rho)u^\mu=-\frac{1}{w_0}\perp^{\mu\rho}\nabla_\rho p_0,
\end{gather}
it is easy to show that $u_{\mu}F^\mu=0$. Therefore, in our study it is sufficient to consider only $a$-, $\theta$- and $\varphi$-components of the Euler equation.

We can get rid of the number density perturbations by multiplying the continuity equations by either $(\p p/\p n_k)_0$ or $\mu_{k0}$ and then performing the summation over the $k$-index. As a result, in the core the equations
can be written as
\begin{gather}
\label{system_GR_core}
\left\{
\begin{gathered}
(u^\rho\nabla_\rho)\delta u^\mu+\delta u^\rho\biggl[\nabla_\rho u^\mu+\frac{1}{w_0}\biggl(u^\mu\delta^\lambda_\rho+u^\lambda\delta^\mu_\rho\biggr)\nabla_\lambda p_0\biggr]=\perp^{\mu\rho}\biggl[\frac{\delta w}{w_0^2}\nabla_\rho p_0-\frac{1}{w_0}\nabla_\rho \delta p\biggr], \\
\frac{1}{\sqrt{-\det\mathfrak{g}}}\frac{\p}{\p x^\mu}\biggl[\sqrt{-\det\mathfrak{g}} \ \delta p \  u^\mu\biggr]+w_0\biggl(\frac{c_s}{c}\biggr)^2\frac{1}{\sqrt{-\det\mathfrak{g}}}\frac{\p}{\p x^\mu}\biggl[\sqrt{-\det\mathfrak{g}} \ \delta u^\mu\biggr]+(\delta u^\mu\nabla_\mu)p_0=0, \\
\frac{1}{\sqrt{-\det\mathfrak{g}}}\frac{\p}{\p x^\mu}\biggl[\sqrt{-\det\mathfrak{g}}(\delta w-\delta p)u^\mu\biggr]+\frac{1}{\sqrt{-\det\mathfrak{g}}}\frac{\p}{\p x^\mu}\biggl[\sqrt{-\det\mathfrak{g}} \ w_0 \delta u^\mu\biggr]-(\delta u^\mu\nabla_\mu)p_0=0, \\
\delta u^\mu=\perp^\mu_\rho\mathcal{L}_{u} \xi^\rho, \qquad u_{\mu}\xi^\mu=0.
\end{gathered}
\right.
\end{gather}
Note that in the derivation of these equations we have explicitly used the equality $u^a=u^\theta=0$, and the fact that the equilibrium quantities do not depend on $t$ and $\varphi$. In the following, for simplicity, we shall refer to the second and third equations of this system as the ``continuity equation 1'' and ``continuity equation 2''. Equations in the crust of the neutron star are obtained in the same manner, but, similarly to the Newtonian case, the barotropy \eqref{barotropic_EOS} of the EOS makes two continuity equations equivalent to each other and modifies the right-hand side of the Euler equation. As a result, we have
\begin{gather}
\label{system_GR_crust}
\left\{
\begin{gathered}
(u^\rho\nabla_\rho)\delta u^\mu+\delta u^\rho\biggl[\nabla_\rho u^\mu+\frac{1}{w_0}\biggl(u^\mu\delta^\lambda_\rho+u^\lambda\delta^\mu_\rho\biggr)\nabla_\lambda p_0\biggr]=-\perp^{\mu\rho}\nabla_\rho\biggl(\frac{\delta p}{w_0}\biggr) \\
\frac{1}{\sqrt{-\det\mathfrak{g}}}\frac{\p}{\p x^\mu}\biggl[\sqrt{-\det\mathfrak{g}} \ \delta p \ u^\mu\biggr]+w_0\biggl(\frac{c_s}{c}\biggr)^2\frac{1}{\sqrt{-\det\mathfrak{g}}}\frac{\p}{\p x^\mu}\biggl[\sqrt{-\det\mathfrak{g}} \ \delta u^\mu\biggr]+(\delta u^\mu\nabla_\mu)p_0=0 \\
\delta u^\mu=\perp^\mu_\rho\mathcal{L}_{u} \xi^\rho, \qquad u_{\mu}\xi^\mu=0 .
\end{gathered}
\right.
\end{gather}
Thus, the study of global stellar oscillations splits into the study of the system \eqref{system_GR_core} in the core and system \eqref{system_GR_crust} in the crust. As we shall see, the barotropy of the EOS brings about even more serious differences between the $r$-mode properties in the core and in the crust of the neutron star, than it was for non-relativistic modes. 

Relativistic counterparts of Newtonian functions $Q$ and $T$ can be written as
\begin{gather}
\label{QT_GR}
\xi^{\theta}_\text{GR}=\frac{1}{r}\biggl[\frac{\p Q_\text{GR}}{\p \theta}+\frac{1}{\sin\theta}\frac{\p T_\text{GR}}{\p\varphi}\biggr], \qquad \xi^\varphi_\text{GR}=\frac{1}{r\sin\theta}\biggl[\frac{1}{\sin\theta}\frac{\p Q_\text{GR}}{\p\varphi}-\frac{\p T_\text{GR}}{\p\theta}\biggr].
\end{gather}
It is easy to verify that the functions $Q_{\rm GR}$ and $T_{\rm GR}$ reduce, respectively, to the functions $Q$ and $T$ introduced in the Newtonian limit. Once we expand $Q_{\rm GR}$ and $T_{\rm GR}$ in the associated Legendre polynomials (or spherical harmonics), it becomes evident that this decomposition is equivalent to that used by Regge \& Wheeler \cite{rw1957} and Thorne \& Campolattaro \cite{tc1967}. Further, we employ this decomposition in the study of relativistic equations, omitting the ``GR'' subscript. As we shall see, the expansions of $Q$, $T$ and other quantities in the associated Legendre polynomials will allow us to separate variables in the equations and reduce the problem to the study of the first order ODEs, similar to the Newtonian ones. The function $T$ will be referred to as the toroidal function throughout the text. It can be shown that the 4-divergence $\nabla_\mu A^\mu$ of a purely toroidal 4-vector $A^\mu$ vanishes in the limit $\Omega\to 0$.

%%%%%%%%%%%%%%%%%%%%%%%%%%%%%%%%%%%%%%%%%%%%%%%%%%%%%%%%%%%%%%%%%%%%%%%%%%%%
%%%%%%%%%%%%%%%%%%%%%%%%%%%%%%%%%%%%%%%%%%%%%%%%%%%%%%%%%%%%%%%%%%%%%%%%%%%%
\subsection{The problem of the continuous spectrum}

Let us consider the case of a slowly rotating neutron star and try to find the solution of relativistic perturbative equations in the form of $r$-modes, i.e., our aim will be to find predominantly toroidal oscillations, for which the $t$- and $\varphi$-dependencies of any perturbation are $\xi^\mu\sim \delta f\sim \Delta f\sim e^{i\sigma t+i m\varphi}$, and for which eigenfrequencies $\sigma=\sigma_r-m\Omega$, defined in the laboratory reference frame, vanish in the $\Omega=0$ case. The linearized equations contain only even powers of $\Omega$ and it seems that the most natural way to do this is to study relativistic equations, assuming that the Newtonian ordering \eqref{ordering_Newt} holds for GR (with the frequencies $\sigma_r$ and $\sigma_r^{(0)}$ replaced by $\sigma$ and $\sigma^{(0)}$, correspondingly). If we employ this ordering, for example, in the core, then in the leading order the continuity equations of the system \eqref{system_GR_core} become identities, whereas the $\theta$- and $\varphi$- components of the Euler equations can be written as [cf. similar Eq.~\eqref{order0_Newt} for Newtonian stars]:
\begin{gather}
\left\{
\begin{gathered}
\frac{i m(m+\sigma^{(0)})^2}{\sin\theta}T^{(0)}-2i(m+\sigma^{(0)})\cos\theta[1-\omega(a)]\frac{\p T^{(0)}}{\p\theta}=\frac{c^2 e^{2\nu_0}}{a w_0}\frac{\p p^{(1)}}{\p\theta}, \\
-(m+\sigma^{(0)})^2\frac{\p T^{(0)}}{\p\theta}+2 m (m+\sigma^{(0)})[1-\omega(a)]\ctg\theta T^{(0)}=\frac{i m c^2 e^{2\nu_0}}{\sin\theta}\frac{p^{(1)}}{aw_0}.
\end{gathered}
\right.
\end{gather}
Expressing $p^{(1)}$ from the second equation and inserting it in the first one, we arrive at the following equation for the toroidal function $T^{(0)}$
\begin{gather}
\frac{1}{\sin\theta}\frac{\p}{\p\theta}\sin\theta\frac{\p T^{(0)}}{\p\theta}+\biggl[\frac{2m[1-\omega(a)]}{m+\sigma^{(0)}}-\frac{m^2}{\sin^2\theta}\biggr]T^{(0)}=0.
\end{gather}
If we look for the solution of this equation as $T^{(0)}=-i T^{(0)}_l(a) P_l^m(\cos\theta)$, we immediately obtain: 
\begin{gather}
\biggl[\sigma^{(0)}+m-\frac{2m[1-\omega(a)]}{l(l+1)}\biggr]T^{(0)}_l(a)=0.
\end{gather}
It follows then that either the solution is trivial, or there exists a resonance point $a^*$, in which 
\begin{gather}
\sigma^{(0)}=\frac{2m[1-\omega(a^*)]}{l(l+1)}-m, \qquad T^{(0)}_l(a)\sim \delta(a-a^*).
\end{gather}  

This result implies that,  since $\omega(a)$ is a positive monotonically decreasing function of $a$, for any combination of $l$ and $m$ we have a continuous spectrum of eigenfrequencies, corresponding to resonance points $0\leq a^*\leq 1$ and taking values within the range
\begin{gather}
\label{cont_spectrum}
\frac{2m[1-\omega(0)]}{l(l+1)}-m\leq\sigma^{(0)}\leq\frac{2m[1-\omega(1)]}{l(l+1)}-m.
\end{gather}
Thus, application of the traditional approach to the relativistic $r$-mode calculations inevitably leads to the continuous eigenfrequency spectrum. For the first time, this peculiarity of the relativistic $r$-mode spectrum was revealed in the works by Kojima \citep{kojima1997, kojima1998}, where relativistic linearized equations were studied both within and beyond the Cowling approximation. The continuous spectrum was shown to emerge as a consequence of the fact that the leading order $r$-mode equations, derived by Kojima, constitute a singular eigenvalue problem instead of a regular one \citep{kojima1997, kojima1998, bk1999}. Actually, an analysis of the equations beyond the Cowling approximation indicates that a nontrivial solution of these equations can be achieved only if the $r$-mode eigenfrequency takes a value in the range
\begin{gather}
\label{cont_spectrum2}
\frac{2m[1-\omega(0)]}{l(l+1)}-m\leq\sigma^{(0)}\leq\frac{2m[1-\omega(\infty)]}{l(l+1)}-m.
\end{gather}

As far as we know, at this point there is no consensus in the literature on whether the continuous spectrum is a physical phenomenon or an artifact of the adopted assumptions and simplifications. Kojima argued \citep{kojima1997, kojima1998} that the continuity of the spectrum reflects the fact that the frequency is measured by an observer, whose angular velocity, because of the inertial reference frame-dragging, is position-dependent even for uniformly rotating stars. Kojima also expected that, although all eigenfrequencies within the continuous spectrum range are equally possible in the leading $\Omega^2$-order, some favored eigenfrequencies can be selected by higher order corrections. Further investigations of higher order equations  \citep{kh1999, kh2000} revealed, however, that no such selection takes place: solving the next-order equations, one finds continuous corrections to the leading order frequency spectrum. Moreover, in the Cowling approximation for the case of barotropic EOS these continuous corrections remained continuous even in the Newtonian limit, i.e., it seemed that the obtained pathological solution did not have proper Newtonian counterparts.

Significant contribution to understanding the $r$-modes in barotropic stars has been made in the papers by Lockitch, Andersson and Friedman \citep{laf2001, lfa2003}, where the analysis of the effect of rotation on the zero-frequency modes of non-rotating stars was discussed. It was shown that the traditional ordering cannot be applied to study the $r$-modes in barotropic stars, since it immediately leads to the overdetermined system, a fact that was overlooked in the previous Kojima's works. Moreover, it was shown that the $r$-modes in barotropic stars simply do not exist, and that the $l=m$ Newtonian $r$-modes correspond to the relativistic inertial modes with discrete spectrum [which removes the pathological solution with the non-vanishing (even in the Newtonian limit) continuous eigenfrequency correction, discussed above]. 

For nonbarotropic stars, however, the problem of the continuous spectrum survives. In the same paper by Lockitch, Andersson \& Friedman \citep{laf2001} it was noticed that for the eigenfrequencies within the range
\begin{gather}
\label{cont_spectrum2}
\frac{2m[1-\omega(1)]}{l(l+1)}-m\leq\sigma^{(0)}\leq\frac{2m[1-\omega(\infty)]}{l(l+1)}-m 
\end{gather}
the problem, actually, is regular and the corresponding solutions, further referred to as the discrete $r$-modes, are characterized by the discrete oscillation spectrum. Such $r$-mode solutions with discrete eigenfrequencies were immediately found by the authors for the uniform density stellar model, and these solutions were claimed at first as relativistic replacement to the Newtonian $r$-modes. It is, however, interesting to notice that these $r$-modes cannot be obtained in the Cowling approximation, since their eigenfrequencies take values beyond the band \eqref{cont_spectrum} (which becomes possible, if one accounts for the metric perturbations). Thus, while the authors of that study do not exclude the possibility of coexistence of discrete and non-discrete $r$-modes, they still arrive at a conclusion that $r$-modes very likely cannot be described in the Cowling approximation. But further application of their ideas to the more realistic case of slowly rotating relativistic polytropic stars, carried out by Yoshida \cite{y2001}, showed that the existence of discrete $r$-modes depends on the polytropic index and the compactness of the stellar model, and it is very likely that these modes do not exist 
under conditions, typical for neutron star interiors, and, therefore, are not important for neutron star physics. Yoshida then came to
conclusion that one should look for the $r$-mode solutions within the continuous spectrum band that could also contain some hidden discrete modes.

Such hidden modes, further referred to as the {\it isolated} modes, indeed, were found in the low-frequency approximation by Ruoff \& Kokkotas \cite{rk2001}, but it turned out that they have divergent velocity perturbations and therefore cannot be considered as physical. The authors of that study concluded then, that the physical $r$-modes should be sought beyond the range \eqref{cont_spectrum}, which, obviously, contradicts the point of view, expressed by Yoshida \cite{y2001}. At the same time Lockitch, Andersson, and Watts \cite{law2004}
argued that accounting for the higher order terms in the relativistic $r$-mode oscillation equations might regularize the problem and that the modes that are divergent in the leading order, could actually become finite, once such terms are included. However, determining the exact form of these terms is a rather complicated problem which, as far as we know, has not been solved yet.

In the discussion of the possible origin of the continuous spectrum some authors \cite{y2001, bk1999, laf2001} suggested that the so far studied equations, derived by Kojima, simply do not govern the dynamics of the ``genuine'' $r$-modes, since they do not 
allow for the gravitational radiation and/or for the dissipative mechanisms operating in the stellar matter. The idea is that these effects would produce an imaginary part in oscillation eigenfrequencies, which could potentially regularize the problem. It seems that the inclusion of the gravitational radiation does not regularize the spectrum \cite{yf2001, rk2002}, and the existence of $r$-modes in more realistic stellar models remains questionable. At the same time, the regularization based on the inclusion of the shear viscosity in the theory, actually works \cite{ponsetal2005, gualtierietal2006}: the continuous spectrum is regularized, and, as the shear viscosity is taken to be closer and closer to zero, those stellar models that previously did not admit $r$-mode solutions, regain them. However, in such regularization the rotational corrections are implicitly considered small compared to the viscous terms, which is exactly the opposite to the situation one expects to take place in neutron stars.

All the discussed above studies are performed within the slow rotation approximation and focus on the equations obtained in different orders in $\Omega$. We did not find any comments in the literature on whether the solutions to the equations, obtained within this approximation, really satisfy, with reasonable accuracy, the general relativistic oscillation Eq.~\eqref{system_GR_core} with small values of $\Omega$ or not. In our opinion, there is no guarantee that the solutions of the known equations in the slow rotation limit correspond to the ``real'' relativistic modes obtained from the general Eqs.~\eqref{system_GR_core}. Moreover, it is interesting to notice that more 
``straightforward'' calculations that do not rely on the slow rotation approximation, both in barotropic \cite{yoshidaetal2005, gk2008, donevaetal2013, jc2017} and in nonbarotropic \cite{yl2002} stars, show no indications of the continuous part of the spectrum, and there are no difficulties in obtaining the normal $r$-modes with discrete eigenfrequencies falling within the continuous-spectrum-associated frequency band. This result seems natural for the barotropic case, since, as has already been mentioned, the relativistic generalization of the Newtonian $r$-modes are the discrete inertial modes. For nonbarotropic stars, however, this result comes as a surprise, since it contradicts the theoretical predictions of the presence of the continuous spectrum, made in the slow-rotation approximation. As far as we know, there is only one work devoted to the numerical calculations of the $r$-modes in nonbarotropic stars, carried out by Yoshida \& Lee \citep{yl2002}. Interestingly, in this work discrete $r$-modes are found in the Cowling approximation, in which the problem of the continuous spectrum seems to be especially pronounced.

Summarizing, the calculation of the relativistic $r$-modes of nonbarotropic stars produces a lot of controversial results. Theoretical studies in the slow-rotation approximation fail to find the relativistic generalization of the Newtonian $r$-modes. They predict the coexistence of the continuous eigenfrequency band \eqref{cont_spectrum}, the isolated eigenfrequencies within this band, and discrete modes with eigenfrequencies \eqref{cont_spectrum2} placed beyond the continuous spectrum range. Neither isolated modes nor discrete modes can be surely considered as the true relativistic $r$-modes: the status of the former is disputable, since they have divergent velocity perturbations in the leading order of the theory \cite{rk2001,law2004}, whereas the latter may simply not exist under conditions, typical for neutron star interiors \cite{y2001}, and, moreover, cannot be found in the Cowling approximation. At the same time, numerical studies beyond the slow rotation approximation \cite{yoshidaetal2005, gk2008, donevaetal2013, jc2017,yl2002} reproduce the discrete $r$-modes, analogous to the Newtonian ones. Despite the idea \cite{laf2001} that it is impossible to find relativistic $r$-modes in the Cowling approximation, Yoshida \& Lee \cite{yl2002}, actually, do find these modes for rapidly rotating stars. The spectrum regularization via viscous dissipation \cite{ponsetal2005,gualtierietal2006} is an interesting result, but it does not provide much information on what goes wrong, if one considers the oscillations in the non-dissipative case. There is one thing, though, that seems to be clear: it is very likely that the traditional approach, based on the expansion in $\Omega$-series, is not suitable for this problem. In the following sections, we develop an alternative approach to studying the $r$-modes within the Cowling approximation. This approach reveals some previously overlooked interesting properties of relativistic $r$-modes, and allows one to obtain general $r$-mode equations, resembling the Newtonian ones and valid for arbitrary EOS. This approach also helps us to clarify the illusive nature of the continuous spectrum and the reasons for the slow rotation approximation breakdown.

%%%%%%%%%%%%%%%%%%%%%%%%%%%%%%%%%%%%%%%%%%%%%%%%%%%%%%%%%%%%%%%%%%%%%%%%%%%%
%%%%%%%%%%%%%%%%%%%%%%%%%%%%%%%%%%%%%%%%%%%%%%%%%%%%%%%%%%%%%%%%%%%%%%%%%%%%
\subsection{A new approach to the $r$-mode equations: stellar core}

The approach we are about to develop is inspired by the one, proposed in Kantor et al. \cite{kgd2021}, where a mathematically similar problem was encountered in superfluid neutron stars and the continuous spectrum emerged as one tried to account for the so-called entrainment effect between protons and neutrons (e.g., \cite{ab1975, bjk1996, gh2005}).

Now, let us assume that the frame-dragging effect is weak, and write $\omega(a)=\epsilon\tilde{\omega}(a)$, where the magnitude of the effect $\epsilon=\max\limits_{0\leq a\leq 1}\{\omega(a)\}$ is assumed to be small. On one hand, this assumption simplifies the analysis of the equations, and, on the other hand, it does not eliminate the problem of the continuous spectrum, since the formula \eqref{cont_spectrum} was obtained for the arbitrary (e.g., small) values of $\omega(a)$. As it turns out (see Sec.\ \ref{num sec} and Sec.\ \ref{spectrum sec} for more details), this assumption is expected to provide a rather accurate estimate of the spectrum at $\Omega\to 0$ even in the realistic case of (not necessarily) weak frame-dragging effect. In the following, we will also ignore background geometry corrections $\propto \Omega^2$ and set the functions $\nu_2(a)$, $\lambda_2(a)$, $\psi_2(a)$ and $\zeta(a,\theta)$ to zero. Strictly speaking, this approximation is not justified, since these corrections affect the spectrum quantitatively and should be accounted for in the most accurate calculation. But our aim in this paper is not to present such a calculation, but to provide an insight into the origin of the problem of the continuous spectrum that would give one a hint on how should such calculations be performed. Moreover, these corrections do not affect the mathematical properties of the problem.

When we deal with perturbations of the form $\xi^\mu\sim\delta f\sim \Delta f\sim e^{i\sigma t+im\varphi}$, a further reduction of the system of equations in the core \eqref{system_GR_core} is possible. The first step is to exclude the perturbation of the enthalpy density $\delta w$, using the $a$-component of the Euler equation. The second step is to exclude the pressure perturbation $\delta p$ from the $\varphi$-component of the Euler equation. As a result, we obtain the $\theta$-component of the Euler equation and two continuity equations that form a desired closed system of equations for $\xi^a$, $T$, and $Q$. We do not provide an explicit form of these equations, since at this point they are very cumbersome and non-informative.

Now, let us try again to find the $r$-mode solutions, using the three obtained equations. An important property of these equations is that, once we assume the oscillation frequency to be proportional to $\Omega$, only terms with even powers of $\Omega$ survive, so that the small parameter in the problem, associated with slow stellar rotation, is $\Omega^2$, not $\Omega$. This rather subtle observation will play a significant role in the discussion of the relativistic $r$-mode ordering in the following sections. At this point it means that we have two small parameters in the problem --- the rotation-associated parameter $\Omega^2$ and the frame-dragging effect of magnitude $\epsilon$, 
such that in the limit of the vanishing rotation rate $\Omega^2\to 0$ and $\Omega^2/\epsilon\to 0$. Physically, the latter condition
means that we account for the frame-dragging effect even for arbitrarily slow rotation rates. Note that this is fundamentally different
from the post-Newtonian studies, when one looks for the relativistic corrections to the Newtonian modes with $\epsilon=0$, so that effectively $\epsilon/\Omega^2\to 0$. The central idea behind our approach is not to discriminate between small corrections due to $\epsilon$ or $\Omega^2$, but to consider them simultaneously.  Mathematically, this is achieved by the following decomposition
\begin{gather}
\label{new_approach}
\sigma=\Omega\bigl[\sigma^{(0)}+\sigma^{(1)}\bigr], \qquad T=T^{(0)}+T^{(1)}, \qquad Q=Q^{(1)}, \qquad \xi^a=\xi^{(1)}.
\end{gather}
Here the terms $f^{(0)}$ correspond to the solution of the equations in the limit $\Omega^2\to 0$, $\epsilon\to 0$ and $\Omega^2/\epsilon\to 0$. These terms and corresponding equations will be further attributed to the leading order of the theory, although the term ``order'' is ill-defined in our case, since we simultaneously consider two different small parameters. The terms $f^{(1)}$ are assumed to be small (due to $\Omega$, $\epsilon$, or both), compared to the terms $f^{(0)}$. These terms and equations, where one does not set $\Omega^2\to 0$, $\epsilon\to 0$ and $\Omega^2/\epsilon\to 0$ will be further attributed to the next order of the theory. The procedure to obtain the next order equations will be discussed below. Note that these decompositions do not imply any explicit ordering, and the only thing we rely on is that $\epsilon$ and $\Omega^2$ are small.

In the leading order of the theory the continuity equations become identities, and the $\theta$-component of the Euler equation reduces to the equation for $T^{(0)}$ of exactly the same form
\begin{gather}
\label{restored r-mode eq}
\frac{1}{\sin\theta}\frac{\p}{\p\theta}\sin\theta\frac{\p T^{(0)}}{\p\theta}+\biggl[\frac{2m}{m+\sigma^{(0)}}-\frac{m^2}{\sin^2\theta}\biggr]T^{(0)}=0,
\end{gather}
as it was for the Newtonian stars [cf.\ Eq.\eqref{Newt_spec_Eq}]. Thus, in the leading order the traditional $r$-mode spectrum is restored, and we have
\begin{gather}
\label{restored r-mode}
\sigma^{(0)}=\frac{2m}{l(l+1)}-m, \qquad \sigma_r^{(0)}=\frac{2m}{l(l+1)}, \qquad T^{(0)}=-i T^{(0)}_l(a)P_l^m(\cos\theta).
\end{gather}

Now, let us proceed with the derivation of the next-order equations. In order to do this, we substitute our decompositions \eqref{new_approach} into our general system of equations (two continuity equations and $\theta$-component of the Euler equation with excluded $\delta p$ and $\delta w$), and then simplify them, using the leading order equations. Then we throw away small terms, using the following selection rule: if, in any particular equation, there is a term $f$, then we can safely ignore the terms $\epsilon f$ and $\Omega^2 f$ in the very same equation. Further analysis of thus obtained first order equations becomes more feasible, if we, as in the Newtonian theory, expand all the quantities of interest in the associated Legendre polynomials:
\begin{gather}
\label{legendre_decomp}
f^{(1)}=\sum_{L\geq m} f^{(1)}_{L}(a) P_L^m(x), \qquad x=\cos\theta.
\end{gather}
Then, after using the properties \eqref{Legendre_properties}--\eqref{Legendre_properties_2}, we find out that the variables in all the equations separate, and we obtain a system of the first-order ODEs for $T^{(0)}_l(a)$, $T^{(1)}_L(a)$, $Q^{(1)}_L(a)$, and $\xi^{(1)}_L(a)$. To be more specific, after the substitution of Legendre expansions, each of the next-order equations can be formally written as:
\begin{gather}
\label{formal_eq}
\mathcal{A}_{l-2}(a) P_{l-2}^m(x) + \mathcal{A}_{l-1}(a) P_{l-1}^m(x)+\mathcal{A}_l(a) P_{l}^m(x)+\mathcal{A}_{l+1}(a) P_{l+1}^m(x)+\mathcal{A}_{l+2}(a) P_{l+2}^m(x)+\sum_{L\geq m} \mathcal{B}_L(a) P_{L}^m(x)=0,
\end{gather}
where $\mathcal{A}_L(a)$ and $\mathcal{B}_L(a)$ are some functions of $a$. The terms $\mathcal{A}_L(a)$ originate from those terms in the system before the separation of variables, that contain $T^{(0)}(a,x)$ and its derivatives. The terms $\mathcal{B}_L(a)$, in turn, are produced by those terms, that contain decompositions \eqref{legendre_decomp} for $f^{(1)}(a,x)$ and their derivatives. There are two types of the first-order ODEs that follow from such equations: 
\begin{gather}
\text{1st type:} \qquad \mathcal{A}_L(a)+\mathcal{B}_L(a)=0 \quad\text{for}\quad L\in\{l,l\pm 1,l\pm 2\} \\
\text{2nd type:} \qquad\qquad\qquad \mathcal{B}_L(a)=0 \quad\text{for}\quad L\not\in\{l,l\pm 1,l\pm 2\}.
\end{gather}
Actually, the ``permitted'' values of $L$ for the first-type equations and the ``forbidden'' values of $L$ for the second-type equations may differ from the shown ones, since for some $L$ we may have $\mathcal{A}_L(a)=0$
\footnote{Consider, for example, the equation $\mathcal{A}_l(a)P_l^m(x)+\mathcal{A}_{l+2}(a)P_{l+2}^m(x)+\sum\limits_{L}\mathcal{B}_L(a)P_L^m(x)=0$. In this case the 1st type equations are $\mathcal{A}_L(a)+\mathcal{B}_L(a)=0 \text{ for } L\in\{l,l+2\}$, whereas the 2nd type equations are $\mathcal{B}_L(a)=0 \text{ for } L\not\in\{l,l+2\}$}.

A number of useful conclusions can be drawn from the analysis of the second-type equations, obtained for the $\theta$-component of the Euler equation 
\begin{multline}
\frac{i m (L-l) (L+l+1)}{l(l+1)} T^{(1)}_L=(L+1) k^-_{L-1} \biggl[(L-1) Q^{(1)}_{L-1}-(1-ag) \xi^{(1)}_{L-1}\biggr]+ \\
+L k^+_{L+1} \biggl[(L+2) Q^{(1)}_{L+1}+(1-a g)\xi^{(1)}_{L+1}\biggr], \qquad L\not\in\{l,l\pm 2\}
\end{multline}
and for the two continuity equations
\begin{gather}
a w_0'\xi^{(1)}_L+w_0\biggl[a\xi^{(1)}_L{}'+(2+a g+a\lambda_0')\xi^{(1)}_L-L(L+1)Q^{(1)}_L\biggr]=0, \qquad L\not\in\{l\pm 1\} \\
a g \xi^{(1)}_L-\biggl(\frac{c_s}{c}\biggr)^2\biggl[a\xi^{(1)}_L{}'+(2+a \lambda_0')\xi^{(1)}_L-L(L+1)Q^{(1)}_L\biggr]=0, \qquad L\not\in\{l\pm 1\},
\end{gather}
where we have introduced the gravitational acceleration $g=\nu_0'=-p_0'/w_0$ [the last equality follows from the equilibrium 
Eq.~\eqref{GR equilibrium} for $\Omega=0$]. If we express $\xi^{(1)}_L{}'$ from one of the continuity equations and substitute into the other, we immediately obtain 
%that 
$\xi^{(1)}_L=0$ for $L\not\in\{l\pm 1\}$. Then, from any of the continuity equations and the $\theta$-component of the Euler equation it is easy to see that
\begin{gather}
\xi^{(1)}_L=Q^{(1)}_L=0 \quad\text{for}\quad L\not\in\{l\pm 1\}, \qquad T^{(1)}_L=0 \quad\text{for}\quad L\not\in\{l,l\pm 2\},
\end{gather}
which is very similar to the situation one usually encounters in Newtonian studies of nonbarotropic stellar models.

Now we are ready to study the first-type equations. It turns out that the function $T^{(1)}_l$ remains undetermined in this order of the theory (similarly to how $T^{(0)}_l$ cannot be obtained from the leading order equations), and the first-type equations  form a closed system of the first order ODEs for all remaining functions $\{T^{(0)}_l,\xi^{(1)}_{l\pm 1}, Q^{(1)}_{l\pm 1}, T^{(1)}_{l\pm 2}\}$. There are three first-type equations, obtained from the $\theta$-component of the Euler equation. Two of them can be used to express $T^{(1)}_{l\pm 2}$ through $T^{(0)}_l$, $Q^{(1)}_{l\pm 1}$ and $\xi^{(1)}_{l\pm 1}$ as
\begin{gather}
\label{Texcl1}
T^{(1)}_{l-2}=\frac{i (l+1) e^{-2 \nu _0} k^+_{l-1} \left\{2 c^2 (l-2) l^2 e^{2 \nu _0} \left[(1-a g) \xi^{(1)}_{l-1}+l Q^{(1)}_{l-1}\right]+m\Omega ^2 k^+_l a^2 \left(l^3-6 l^2-8 l+8\right)T^{(0)}_l \right\}}{4 m c^2 l (2 l-1)}, \\
\label{Texcl2}
T^{(1)}_{l+2}=-\frac{i l e^{-2 \nu _0} k^-_{l+1} \left\{2 c^2 (l+1)^2 (l+3) e^{2 \nu _0} \left[(a g-1) \xi^{(1)}_{l+1}+(l+1) Q^{(1)}_{l+1}\right]-m\Omega ^2 k^-_l a^2 \left(l^3+9 l^2+7 l-9\right)T^{(0)}_l \right\}}{4 m c^2 (l+1)(2 l+3)},
\end{gather}
while the third has the form of the algebraic relation between $T^{(0)}_l$, $\xi^{(1)}_{l\pm 1}$ and $Q^{(1)}_{l\pm 1}$
\begin{multline}
\label{eq1}
c^2 l^2 (l+1)^2 e^{2 \nu _0}  [l (l+1) \sigma ^{(1)}+2 m \epsilon \omega(a)]T^{(0)}_l+ \\
+a^2 m  \Omega ^2 \left[4 m^2-l^2 (l+1)^2+l^2\left(l^3+9 l^2+7 l-9\right)  k^-_l k^+_{l+1}-(l+1)^2 \left(l^3-6 l^2-8 l+8\right) k^-_{l-1} k^+_l\right]T^{(0)}_l= \\
=2 c^2 l^2 (l+1)^2 e^{2 \nu _0} \biggl\{(l+1) k^-_{l-1} \biggl[(a g-1) \xi^{(1)}_{l-1}+(l-1) Q^{(1)}_{l-1}\biggr]+l k^+_{l+1} \biggl[(1-a g) \xi^{(1)}_{l+1}+(l+2) Q^{(1)}_{l+1}\biggr]\biggr\}.
\end{multline}
From the first continuity equation we obtain two equations of the first type:
\begin{multline}
\label{eq2}
2 a m \Omega ^2 k^+_l \biggl[4 a T^{(0)}_l{}'+a\biggl(4A-g \left(l^3+2 l+8\right)-4\lambda _0'\biggr)T^{(0)}_l+2 (F+2 l-3)T^{(0)}_l\biggr]= \\
=c^2 g l^2 e^{2 \nu _0} \left(2 (l-1) l Q^{(1)}_{l-1}-\xi^{(1)}_{l-1} (2 a A+F-1)-2 a \xi^{(1)}_{l-1}{}'\right),
\end{multline}
\begin{multline}
\label{eq3}
2 a m \Omega ^2 k^-_l \biggl[4 a T^{(0)}_l{}'+a\biggl(4 A+g (l^3+3l^2+5l-5)-4 \lambda _0'\biggr)T^{(0)}_l+2(F-2 l-5)T^{(0)}_l\biggr]= \\
=c^2 g (l+1)^2 e^{2 \nu _0} \biggl[2 (l+1) (l+2) Q^{(1)}_{l+1}-\xi^{(1)}_{l+1} (2 a A+F-1)-2 a \xi^{(1)}_{l+1}{}'\biggr],
\end{multline}
where we have introduced 
\begin{gather}
A(a)=\frac{w_0'}{w_0}+g\biggl[1+\biggl(\frac{c}{c_s}\biggr)^2\biggr], \qquad F(a)=2 a \biggl[\lambda _0'-g\biggl(\frac{ c }{c_s}\biggr)^2\biggr]+5.
\end{gather}
First-type equations, obtained from the second continuity equation, have similar form and can be used to express $Q^{(1)}_{l\pm 1}$ through $T^{(0)}_l$ and $\xi^{(1)}_{l\pm 1}$ as
\begin{gather}
\label{Qexcl1}
Q^{(1)}_{l-1}=\frac{c^2 g l^2 \left[2 a \xi^{(1)}_{l-1}{}'+(F-1) \xi^{(1)}_{l-1}\right]-2 a m \Omega ^2 k^+_l e^{-2 \nu _0} \left[a g l \left(l^2-2\right) -2(F-5-2a\lambda_0')\right]T^{(0)}_l}{2 c^2 g l^3(l-1)}, \\
\label{Qexcl2}
Q^{(1)}_{l+1}=\frac{c^2 g(l+1)^2 \left[2 a \xi^{(1)}_{l+1}{}'+(F-1) \xi^{(1)}_{l+1}\right]+2 a m \Omega ^2 k^-_l e^{-2 \nu _0}  \left[a g \left(l^3+3 l^2+l-1\right)+2(F-5-2a\lambda_0')\right]T^{(0)}_l}{2 c^2 g (l+1)^3 (l+2)}.
\end{gather}
If we substitute \eqref{Qexcl1}--\eqref{Qexcl2} into Eqs.~\eqref{eq1}--\eqref{eq3}, we arrive at the closed system of three equations for $T^{(0)}_l$, $\xi^{(1)}_{l-1}$, and $\xi^{(1)}_{l+1}$. This system can be transformed to relatively compact form, 
\begin{gather}
\label{r-modes GR}
\left\{
\begin{gathered}
T^{(0)}_l\biggl\{c^2 e^{2 \nu _0} g l (l+1)  \bigl[l (l+1) \sigma^{(1)}+2 m \epsilon \tilde{\omega}(a)\bigr]+a\Omega^2\bigl[a g \gamma_2 +4 m \gamma_1 \left(2 a\lambda_0'-F+5\right)\bigr]\biggr\} = \hspace{2cm} \\
\hspace{4cm} =c^2 g e^{2 \nu _0} \biggl[l^2 k^+_{l+1} \biggl(2 a \xi^{(1)}_{l+1}{}'+\xi^{(1)}_{l+1}\bigl[-2 a g (l+1)+F+2 l+1\bigr]\biggr)+ \\
\hspace{7cm}+(l+1)^2 k^-_{l-1} \biggl(2 a \xi^{(1)}_{l-1}{}'+\xi^{(1)}_{l-1} \bigl[2 a g l+F-2 l-1\bigr]\biggr)\biggr], \\
T^{(0)}_l{}'+T^{(0)}_l\biggl[A+g(l-1)-\frac{l}{a}\biggr]+\frac{A c^2 g (l+1)^2 e^{2 \nu _0} }{4 a m \Omega ^2 k^-_l}\xi^{(1)}_{l+1}=0, \hspace{5cm} \\
T^{(0)}_l{}'+T^{(0)}_l\biggl[A-g(l+2)+\frac{l+1}{a}\biggr]+\frac{A c^2 g l^2 e^{2 \nu _0}}{4 a m \Omega ^2 k^+_l}\xi^{(1)}_{l-1}=0 , \hspace{5cm}
\end{gathered}
\right.
\end{gather}
where $\gamma_1$ and $\gamma_2$ are numerical coefficients that can be conveniently written as
\begin{gather}
\gamma_1=\frac{l^4 k^-_l k^+_{l+1}+(l+1)^4 k^-_{l-1} k^+_l}{l^2 (l+1)^2}, \qquad
\gamma_2=\frac{2 m^2 \left[3 m^3+29 m^2 \sigma^{(0)}+2m \left(16 \sigma^{(0)2}-1\right)+6 \sigma^{(0)} \left(\sigma^{(0)2}-4\right)\right]}{(2l-1)(2l+3)\left(m+\sigma^{(0)}\right){}^2}.
\end{gather}
Being written in the Newtonian limit, the system \eqref{r-modes GR} reproduces the Newtonian $r$-mode equations in the core, whose derivation is very similar to the derivation of the relativistic equations and was briefly discussed before in Sec.\ \ref{Newt sec}.

The $l=m$ case, most intensively studied in the literature, deserves a special consideration, since the $r$-mode with $l=m=2$ is 
most CFS-unstable. In this case, the functions $Q^{(1)}_{l-1}$, $\xi^{(1)}_{l-1}$, and $T^{(1)}_{l-2}$ should be set to zero as coefficients in front of the Legendre polynomials $P_{l-1}^m(x)$ and $P_{l-2}^m(x)$ that vanish for $l=m$. One should also ignore Eqs.~\eqref{Texcl1}, \eqref{eq2},   \eqref{Qexcl1}, and the last equation of the system \eqref{r-modes GR}. The reason is that these equations are obtained via setting the coefficients before $P_{l-1}^m(x)$ and $P_{l-2}^m(x)$ to zero, whereas for $l=m$ the terms with $P_{l-1}^m(x)$ and $P_{l-2}^m(x)$ simply do not appear. Thus, for $l=m$ we are left with the first two equations of the system \eqref{r-modes GR}, which allow us to find $T^{(0)}_l$ and $\xi^{(1)}_{l+1}$, and with Eqs.~\eqref{Texcl2} and \eqref{Qexcl2}, which allow us to find $T^{(1)}_{l+2}$ and $Q^{(1)}_{l+1}$, respectively.

%%%%%%%%%%%%%%%%%%%%%%%%%%%%%%%%%%%%%%%%%%%%%%%%%%%%%%%%%%%%%%%%%%%%%%%%%%%%
%%%%%%%%%%%%%%%%%%%%%%%%%%%%%%%%%%%%%%%%%%%%%%%%%%%%%%%%%%%%%%%%%%%%%%%%%%%%
\subsection{A new approach to the $r$-mode equations: stellar crust}

Like in the Newtonian theory, the properties of the relativistic $r$-mode equations in the crust differ from the properties of the equations in the core, and $r$-modes in the crust should be studied separately. In the crust there is only one continuity equation (continuity equation 1 and continuity equation 2 become equivalent to each other because of the barotropy of the EOS), and we are dealing with the system \eqref{system_GR_crust}. Using the $\varphi$-component of the Euler equation, we can express $\delta p$ through $Q$, $T$ and $\xi^a$, and substitute $\delta p$ in all the remaining equations. As a result, we obtain three equations --- the $a$-component of the Euler equation, the $\theta$-component of the Euler equation, and the continuity equation, --- that form a closed system for the functions $Q$, $T$, and $\xi^a$. We do not explicitly write out these equations, since they are very cumbersome and non-informative.

It turns out that the function $T^{(0)}$ is completely defined by the leading order equations in our approach. In this order, since $\xi^\mu$ is purely toroidal, the continuity equation is automatically satisfied. From the $\theta$-component we again obtain Eq.~\eqref{restored r-mode eq} and, consequently, Eq.~\eqref{restored r-mode}, i.e., the traditional $r$-mode spectrum is restored. Then, from the $a$-component of the Euler equation we get [cf.\ the Newtonian equation \eqref{lm only}]
\begin{gather}
(l+1)^2 k^+_l \biggl(a T^{(0)}_l{}'-\biggl[a g (l+2)-l-1\biggr]T^{(0)}_l \biggr)P_{l-1}^m(x)+l^2 k^-_l \biggl(a T^{(0)}_l{}'+ \biggl[a g (l-1)-l\biggr]T^{(0)}_l\biggr)P_{l+1}^m(x)=0.
\end{gather}
This equation is self-consistent only if $l=m$. In the latter case, it is easy to verify that the solution to this equation is
\begin{gather}
\label{TsolGR}
T^{(0)}_l(a)=\const\cdot a^l e^{-(l-1)\nu_0}.
\end{gather}
We see that this solution differs from the Newtonian one by a relativistic factor $e^{-(l-1)\nu_0}$.

The next-order equations in the crust have a more sophisticated form than in the core. The main reason for that is that in the core we have two continuity equations that produce very similar first and second-type equations. Due to this similarity, the expansions in associated Legendre polynomials have only finite number of non-zero terms. In the crust, however, we have only one continuity equation and two components of the Euler equation, and the analysis of the second-type equations does not indicate any truncation in the expansions in Legendre polynomials. Fortunately, among the first type equations obtained from the $\theta$-component of the Euler equation and from the continuity equation we find two equations that decouple from this infinite system of equations and contain only  $Q^{(1)}_{l+1}(a)$, $\xi^{(1)}_{l+1}(a)$, and $T^{(0)}_l(a)$:
\begin{gather}
\label{Qcrust1}
\frac{a^2 \Omega^2 l (l^2-2 l-7)e^{-2 \nu _0} }{c^2(l+1)^2} T^{(0)}_l=(2 l+3) \biggl[(l+1) \sigma^{(1)}+2\epsilon \tilde{\omega}(a)\biggr]T^{(0)}_l+2 (2 l+1) \biggl[(a g-1) \xi^{(1)}_{l+1}-(l+2) Q^{(1)}_{l+1}\biggr], \\
\label{Qcrust2}
\frac{a^2 \Omega ^2 l e^{-2 \nu _0}}{c^2(2 l+1) (l+1)^2} \left[\left(l^3+3 l^2+l-1\right)\frac{c_s^2}{c^2}-4\right]T^{(0)}_l=a g \xi^{(1)}_{l+1}+\frac{c_s^2}{c^2} \biggl[(l+1) (l+2) Q^{(1)}_{l+1}-\xi^{(1)}_{l+1} \left(a \lambda _0'+2\right)-a \xi^{(1)}_{l+1}{}'\biggr].
\end{gather}
Now, if we express $Q^{(1)}_{l+1}$ from one of these equations and substitute into the other, we arrive at the first-order ODE for $\xi^{(1)}_{l+1}$
\begin{gather}
\label{XiEqGR}
\xi^{(1)}_{l+1}{}'+g_1 \xi^{(1)}_{l+1}+\biggl(g_{21}\biggl[\sigma^{(1)}+\frac{2\epsilon\tilde{\omega}(a)}{l+1}\biggr]+\Omega^2 g_{22}\biggr)T^{(0)}_l=0
\end{gather}
with the coefficients $g_{1}(a)$, $g_{21}(a)$, and $g_{22}(a)$ defined as
\begin{gather}
\label{g1coeff}
g_1(a)=\frac{l+3}{a}-g\left(l+1+\frac{c^2}{c_s^2}\right)+\lambda _0', \qquad g_{21}(a)=-\frac{(l+1)^2(2l+3)}{2a(2l+1)}, \\
\label{g22coeff}
g_{22}(a)=\frac{a\, l \left[(l+1)(3l^2+2l-9) c_s^2-8 c^2\right]e^{-2 \nu _0}}{2 c^2 (l+1)^2 (2 l+1) c_s^2}.
\end{gather}
The solution to this equation is given by the formula [cf.\ Eq.\ \eqref{XiSolNewt}]
\begin{gather}
\label{XiSolGR}
\xi^{(1)}_{l+1}=\frac{1}{\eta(a)}\biggl[\xi_0+\int\limits_a^1\biggl(g_{21}(a)\biggl[\sigma^{(1)}+\frac{2\epsilon\tilde{\omega}(a)}{l+1}\biggr]+\Omega^2 g_{22}(a)\biggr)\eta(a)T^{(0)}_l(a)da\biggr], \qquad \eta(a)=\exp\biggl(-\int\limits_a^1 g_1(a) da\biggr),
\end{gather}
where the values of $\xi_0$ and $\sigma^{(1)}$ should be determined from the boundary conditions, as will be discussed below.

%%%%%%%%%%%%%%%%%%%%%%%%%%%%%%%%%%%%%%%%%%%%%%%%%%%%%%%%%%%%%%%%%%%%%%%%%%%%
%%%%%%%%%%%%%%%%%%%%%%%%%%%%%%%%%%%%%%%%%%%%%%%%%%%%%%%%%%%%%%%%%%%%%%%%%%%%
\subsection{Boundary conditions}\label{bc section}

Since the $l=m=2$ mode is expected to be the most unstable with respect to the emission of gravitational waves, further, for simplicity, 
we restrict ourselves only to the $l=m$ case. We have to specify three types of boundary conditions: near the center of the star, $a=0$, at the crust-core interface, $a=a_{cc}$, and at the surface of the star, $a=1$.

First of all, let us discuss the boundary condition at the surface of the star. By definition, the surface of the star corresponds to the set of points in space, where the pressure vanishes. If $x^\rho_s$ are the coordinates of the surface of the equilibrium star, then $x^\rho_s+\xi^\rho_s$, where $\xi^\rho_s=\xi^\rho(x_s)$, are the coordinates of the surface of the perturbed star. The surface of the perturbed star is then defined by the condition $p(x^\rho_s+\xi^\rho_s)=0$. It is easy to see that this condition is equivalent to the equality $\Delta p_{l+1}(1)=0$, where $\Delta p_{l+1}$ is the coefficient before the Legendre polynomial $P_{l+1}^l(x)$ in the expansion \eqref{expansions} for $\Delta p$.  Since $p_0=p_0(a)$, we have
\begin{gather}
\label{Delta p}
\Delta p=\delta p+\xi^\mu\nabla_\mu p_0=\delta p - w_0 g \xi^a.
\end{gather}
Both in the crust and in the core the leading contribution to the Eulerian pressure perturbation is given by the formula
\begin{gather}
\label{delta p}
\delta p(a,\theta)=\delta p_{l+1}(a)P_{l+1}^l(\cos\theta), \quad 
\delta p_{l+1}(a)=-\frac{4 \Omega^2 a\, l\,  w_0 e^{-2\nu_0}}{c^2(l+1)^2(2l+1)}T^{(0)}_l(a),
\end{gather}
following from the $\varphi$-component of the Euler equation. It allows us to rewrite the boundary condition at the stellar surface as
\begin{gather}
\xi^{(1)}_{l+1}(1)=\xi_0=-\frac{4\Omega^2 l e^{-2\nu_0(1)}}{c^2 (l+1)^2 (2l+1)g(1)}T^{(0)}_l(1).
\end{gather}
Further, for convenience, we always normalize the solution in the crust, so as to have $T^{(0)}_l(1)=1$. In this normalization, the toroidal function in the crust and the boundary condition at the surface take the form 
\begin{gather}
\label{bc surface}
T^{(0)}_\text{$l$,~crust}(a)=a^l e^{-(l-1)[\nu_0(a)-\nu_0(1)]}, \qquad
\xi^{(1)}_{l+1}(1)=\xi_0=-\frac{4\Omega^2 l e^{-2\nu_0(1)}}{c^2 (l+1)^2 (2l+1)g(1)}.
\end{gather} 

Secondly, let us examine the solutions to the system \eqref{r-modes GR} near the stellar center. To do this, we exclude the radial displacement $\xi^{(1)}_{l+1}$ from this system and obtain the second-order ODE for $T^{(0)}_l$. We look for the asymptotic solution of this equation near the center in the form $T^{(0)}_l\simeq K a^n$, where $K=\const$ is the mode amplitude. Taking into account that in the vicinity of the center the functions, related to the equilibrium configuration, behave as
\begin{gather}
\tilde{\omega}(a)\sim F(a)\sim 1, \quad A(a)\approx aA'(0)\equiv a \tilde{A}_c, \quad g(a)\approx a g'(0)\equiv a \tilde{g}_c, \quad \nu_0(a)\approx \nu_0(0)\equiv \nu_c, \qquad \lambda(a)\sim a^2,
\end{gather}
we find that either $n=l$ or $n=-l-1$. The second solution is divergent in the stellar center and should be discarded as unphysical. Then, recalling the relation between $\xi^{(1)}_{l+1}$ and $T^{(0)}_l$, we obtain that the following asymptotic holds near the center: 
\begin{gather}
\label{asymptotics}
T^{(0)}_l(a)\approx K a^l,  \qquad \xi^{(1)}_{l+1}(a)\approx -\frac{4 l \Omega ^2 k^-_l e^{-2 \nu _c} [\tilde{A}_c+\tilde{g}_c(l-1)]}{\tilde{A}_c \tilde{g}_c c^2 (l+1)^2 }K a^l, \qquad K=\const.
\end{gather}
Since the global $r$-mode in the core should smoothly match this asymptote, the equalities   
\begin{gather}
\label{bc center}
T^{(0)}_l(a_0)=1, \qquad \xi^{(1)}_{l+1}(a_0)=-\frac{4 l \Omega ^2 k^-_l e^{-2 \nu _c} [\tilde{A}_c+\tilde{g}_c(l-1)]}{\tilde{A}_c \tilde{g}_c c^2 (l+1)^2 },
\end{gather}
with arbitrarily small value of $a_0\ll 1$, can be used as boundary conditions for the system of equations in the core.  Once the solution in the core is found, one can renormalize it in a more convenient way, as will be discussed below.

Finally, we demand that the energy and momentum currents, defined as the $(t,k)$- and $(i,k)$-components  (with spatial indices $i$ and $k$) of the stess-energy tensor,
\begin{gather}
T^{\mu\nu}=w\mathfrak{u}^\mu\mathfrak{u}^\nu+p \mathfrak{g}^{\mu\nu},
\end{gather}
should be continuous at the crust-core interface. If $x^\rho_{cc}$ are the coordinates of the boundary between the core and the crust in the unperturbed star, then $x^\rho_{cc}+\xi^\rho_{cc}$, where $\xi^\rho_{cc}=\xi^\rho(x_{cc})$, are the coordinates of this boundary in the perturbed star. Therefore, the continuity of the energy and momentum currents at the crust-core interface implies that we must have $T^{\mu\nu}(x^\rho_{cc}+\xi^\rho_{cc})|_\text{core}=T^{\mu\nu}(x^\rho_{cc}+\xi^\rho_{cc})|_\text{crust}$ for $(\mu\nu)=(t,k)$ or $(i,k)$. It is easy to see that these conditions are equivalent to the continuity conditions for the functions $\xi^\rho$, $\Delta p$ and $\Delta w$ (our equilibrium stellar model does not contain any discontinuities in the energy density). Actually, since these perturbations are not completely independent and should satisfy the previously obtained $r$-mode equations, it is enough to consider only continuity conditions for $\xi^a$ and $\Delta p$, and all the remaining conditions will then be satisfied automatically. From the relations \eqref{Delta p} and \eqref{delta p} it, in turn, follows that these continuity conditions are equivalent to
\begin{gather}
T^{(0)}_\text{$l$,~core}(a_{cc})=T^{(0)}_\text{$l$,~crust}(a_{cc}), \qquad \xi^{(1)}_{l+1,~\text{core}}(a_{cc})=\xi^{(1)}_{l+1,~\text{crust}}(a_{cc}).
\end{gather}
Since at this point we are free to choose the normalization in the core, we can adjust the amplitude of the solution in the core, so as to have $T^{(0)}_\text{$l$,~core}(a_{cc})=T^{(0)}_\text{$l$,~crust}(a_{cc})$. In this normalization the boundary conditions at the crust-core interface reduce to
\begin{gather}
\label{bc interface}
\xi^{(1)}_{l+1,~\text{core}}(a_{cc})=\xi^{(1)}_{l+1,~\text{crust}}(a_{cc}).
\end{gather}

The whole set of boundary conditions can be satisfied not for all, but only for some values of $\sigma^{(1)}$. These
values, for which all conditions are satisfied, correspond to the sought global $r$-mode solutions. In order to find the corresponding eigenfrequencies and eigenfunctions, we employ the integration scheme, discussed in Appendix \ref{Integration scheme}.

%%%%%%%%%%%%%%%%%%%%%%%%%%%%%%%%%%%%%%%%%%%%%%%%%%%%%%%%%%%%%%%%%%%%%%%%%%%%%%%%%%%%%%%%%%%%%%%%%%%%%%%%%%%%%%%%%%%%%%%%%%%%
%%%%%%%%%%%%%%%%%%%%%%%%%%%%%%%%%%%%%%%%%%%%%%%%%%%%%%%%%%%%%%%%%%%%%%%%%%%%%%%%%%%%%%%%%%%%%%%%%%%%%%%%%%%%%%%%%%%%%%%%%%%%
%%%%%%%%%%%%%%%%%%%%%%%%%%%%%%%%%%%%%%%%%%%%%%%%%%%%%%%%%%%%%%%%%%%%%%%%%%%%%%%%%%%%%%%%%%%%%%%%%%%%%%%%%%%%%%%%%%%%%%%%%%%%
\section{Numerical results}\label{num sec}

\begin{figure}[h!!!]
\centering
\begin{minipage}{0.54\linewidth}
\includegraphics[width=1.0\linewidth]{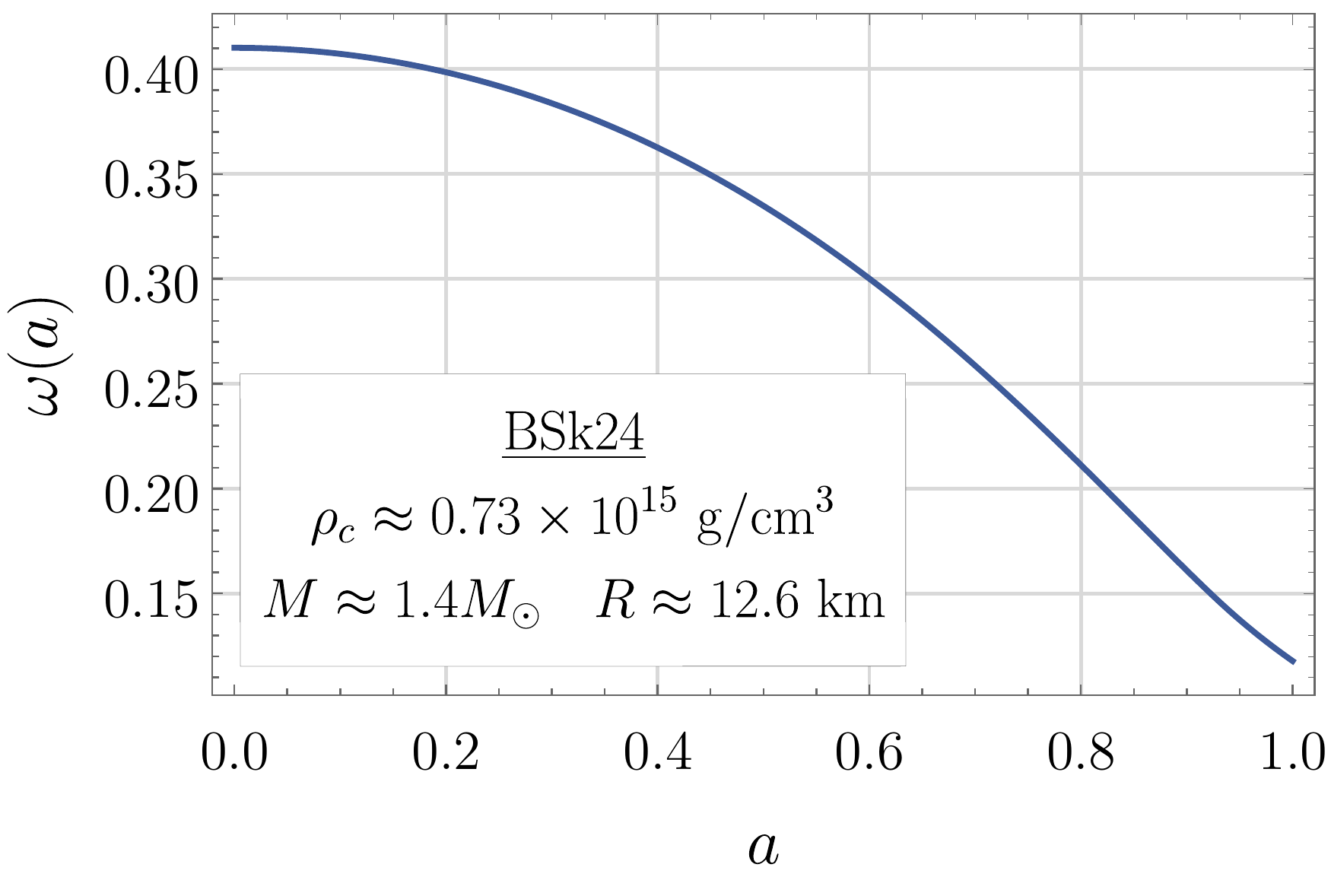}
\end{minipage}
\hspace{0.5cm}
\begin{minipage}{0.38\linewidth}
\caption{The function $\omega(a)$ that describes the effect of the inertial reference frame-dragging for the considered stellar model.}
\label{dragplot}
\end{minipage}
\end{figure}

In this section we present the results of the numerical calculation, performed for the global $l=m=2$ $r$-mode. As a microphysical input, we employ the BSk24 equation of state \cite{gorielyetal2013} that describes a neutron star with a crust and core, subdivided into the outer core with $npe$-matter (i.e., matter, consisting of neutrons, protons and electrons) and the inner core with $npe\mu$-matter (i.e., $npe$-matter with admixture of muons). Our equilibrium stellar model is obtained via solving the Hartle's equations \cite{hartle1967, hartle1968} and is characterized by the central density $\rho_c=\varepsilon(0)/c^2\approx 0.73\times 10^{15} ~\rm g/cm^3$, radius $R\approx 12.6 ~\rm km$,  mass $M\approx 1.4 M_\odot$, where $M_\odot$ is the solar mass, and the position of the crust-core interface, $a_{cc}\approx 0.92$. The function $\omega(a)=\epsilon\tilde{\omega}(a)$ that describes the effect of the inertial reference frame-dragging in this model is plotted in Fig.\ \ref{dragplot}. This is a smooth, monotonically decreasing function of $a$. The amplitude of the inertial reference frame-dragging effect for the considered stellar model equals $\epsilon\approx 0.41$. Although this value can hardly be considered as small, we still solve the equations, derived in the limit of weak frame-drag with precisely this $\epsilon$. Our assumption that the effect of the inertial reference frame-dragging is weak becomes more and more accurate as one approaches the stellar surface, since the function $\omega(a)$ and, therefore, $\tilde{\omega}(a)$ decreases with increasing radius. As we shall see, in the limit 
$\Omega\to 0$, the relativistic $r$-modes in the core become confined to a tiny region in the vicinity of the crust-core interface, where this assumption can be considered as accurate. Anyway, going beyond the small $\epsilon$ approximation does not affect the problem of the continuous spectrum and only complicates the equations.

\begin{figure}[h!!!]
\centering
\includegraphics[width=1.0\linewidth]{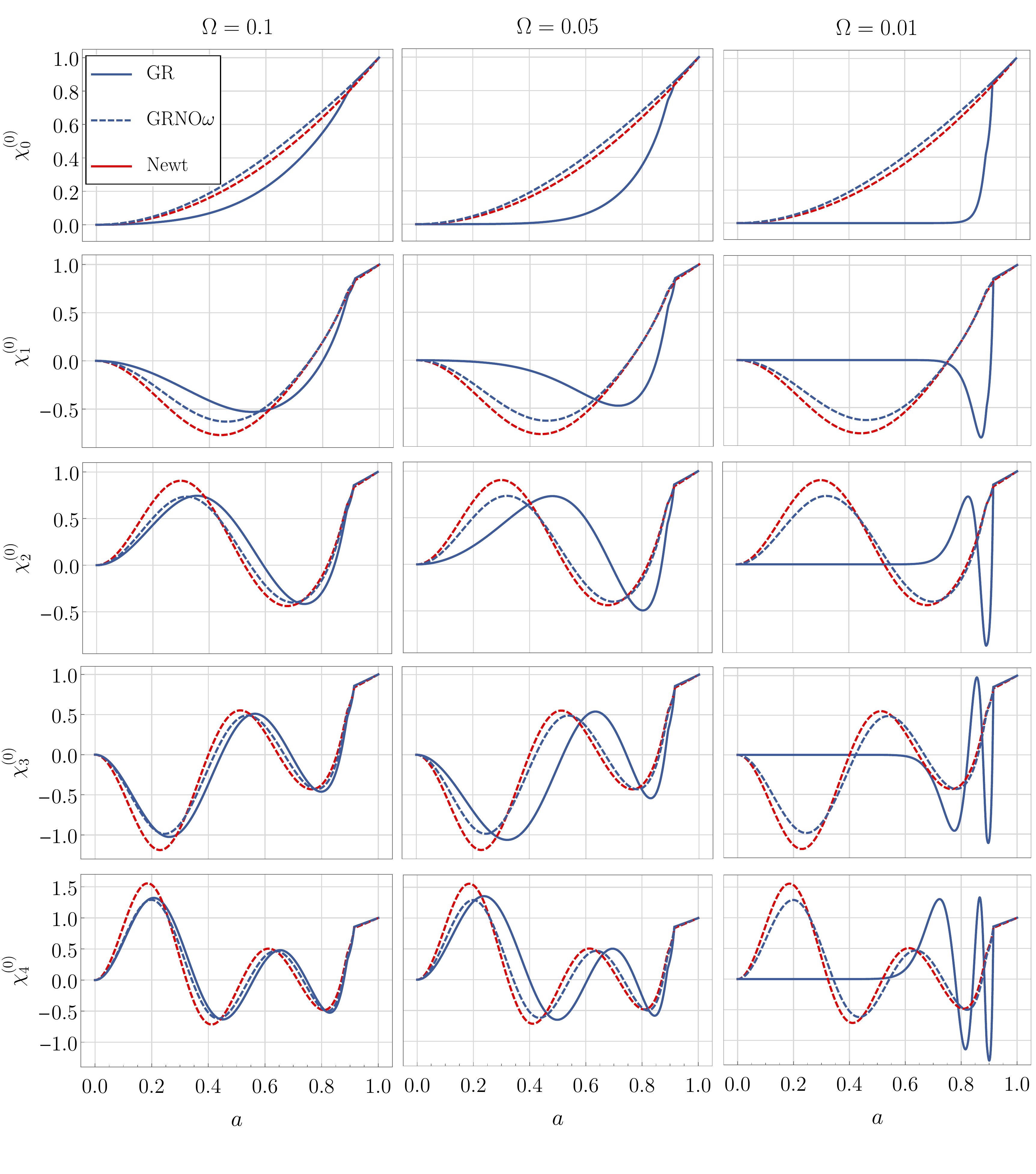}
\caption{Comparison of the toroidal $l=m=2$ eigenfunctions $\chi^{(0)}_n$ with $n$ nodes computed for different rotation rates in General Relativity (GR, solid blue lines), in General Relativity without the effect of inertial reference frame-dragging (GRNO$\omega$, dashed blue lines) and in the Newtonian theory (Newt, dashed red lines). The left, central and right columns are calculated for $\Omega=0.1$, $\Omega=0.05$, and $\Omega=0.01$, respectively. Each row corresponds to a fixed number of nodes $n$ of eigenfunctions with number $n$ increasing towards the bottom of the figure.}
\label{dragEffect}
\end{figure}

Throughout this section we use the notation $\chi^{(0)}_n$ for the toroidal eigenfunction $T^{(0)}_l$ with $n$ nodes (i.e. $n$ points inside the star, where $T^{(0)}_l$ vanishes), normalized so that $T^{(0)}_l(1)=1$. Also, by $\sigma^{(1)}_n$ we denote the corresponding eigenfrequency corrections, defined according to decomposition \eqref{new_approach}. We study the effect of GR on the $r$-mode dynamics by comparing the $l=m=2$ modes in the three different cases: relativistic $r$-modes (GR), relativistic $r$-modes in the absence of the inertial reference frame-dragging (GRNO$\omega$, corresponds to setting $\epsilon=0$ in the obtained relativistic equations), and the Newtonian $r$-modes (Newt). We present the results of our calculations for the first five eigenfunctions (number of nodes $n$ ranging from 0 to 4) for different rotation rates in Fig.\ \ref{dragEffect}. All the corresponding eigenfrequencies $\sigma^{(1)}_n$ take discrete values and are listed in Table \ref{FreqTab}.

\begin{table}[h]
\caption{Newtonian and relativistic $r$-mode eigenfrequency corrections $\sigma^{(1)}_n/\Omega^2$ for different rotation rates.}
\label{FreqTab}
\begin{ruledtabular}
\begin{tabular}{|d|d|d|d|d|d|}
\multicolumn{1}{|c|}{Number of nodes $n$} & \multicolumn{1}{c|}{Newt (any $\Omega$)} & \multicolumn{1}{c|}{GRNO$\omega$ (any $\Omega$)} & \multicolumn{1}{c|}{GR ($\Omega=0.1$)} & \multicolumn{1}{c|}{GR ($\Omega=0.05$)} & \multicolumn{1}{c|}{GR ($\Omega=0.01$)} \\
\hline\hline
0				&	-0.096 	&	-0.124	&	-14.4	&	-0.529 \times 10^2		&	-1.08 \times 10^3			\\
\hline
1				&	-7.94	&	-13.8	&	-32.4	&	-0.830 \times 10^2		&	-1.30 \times 10^3			\\
\hline
2				& 	-16.2	&	-29.6	&	-49.2	&	-1.09 \times 10^2		&	-1.46 \times 10^3			\\
\hline		
3				&	-27.1	&	-50.8	&	-70.9	&	-1.33 \times 10^2		&	-1.63 \times 10^3			\\
\hline
4				&	-40.97	&	-77.9	&	-98.4	&	-1.61 \times 10^2		&	-1.79 \times 10^3	
\end{tabular}		
\end{ruledtabular}
\end{table}

We see that, generally, the relativistic toroidal eigenfunctions and eigenfrequencies are sensitive to the value of the angular velocity $\Omega$, but, once the reference frame-dragging is switched off, they show no dependency on the rotation rate and start to behave similarly to the Newtonian ones. Such a behavior can be easily understood from the $r$-mode equations in the core [the system \eqref{r-modes GR} for $l=m$]:
\begin{gather}
\label{r-modes l=m}
\left\{
\begin{gathered}
T^{(0)}_l\biggl\{c^2 e^{2 \nu _0} g l (l+1)  \bigl[l (l+1) \sigma^{(1)}+\underline{2 l \epsilon \tilde{\omega}(a)}\bigr]+a\Omega^2\bigl[a g \gamma_2 +4 l \gamma_1 \left(2 a\lambda_0'-F+5\right)\bigr]\biggr\} = \hspace{2cm} \\
\hspace{4cm} =c^2 g e^{2 \nu _0}l^2 k^+_{l+1} \biggl(2 a \xi^{(1)}_{l+1}{}'+\xi^{(1)}_{l+1}\bigl[-2 a g (l+1)+F+2 l+1\bigr]\biggr)\\
T^{(0)}_l{}'+T^{(0)}_l\biggl[A+g(l-1)-\frac{l}{a}\biggr]+\frac{A c^2 g (l+1)^2 e^{2 \nu _0} }{4 a l \Omega ^2 k^-_l}\xi^{(1)}_{l+1}=0. \hspace{5cm}
\end{gathered}
\right.
\end{gather}
If we set $\epsilon=0$ (i.e., ignore the underlined term in the first equation), we see that the traditional $r$-mode ordering can be applied. Indeed, replacing $\sigma^{(1)}\to\Omega^2 \sigma^{(1)}$ and $\xi^{(1)}_{l+1}\to \Omega^2 \xi^{(1)}_{l+1}$, we immediately find that the angular velocity $\Omega$ disappears from these equations, as well as from the boundary conditions. Therefore, the problem becomes completely independent of the stellar rotation rate, which explains the observed behaviour of the GRNO$\omega$ modes. This also explains, why the ratios $\sigma^{(1)}/\Omega^2$ for the GRNO$\omega$ case, provided in the Table \ref{FreqTab}, do not depend on $\Omega$.

\begin{figure}[h!!!]
\centering
\includegraphics[height=0.21\textheight]{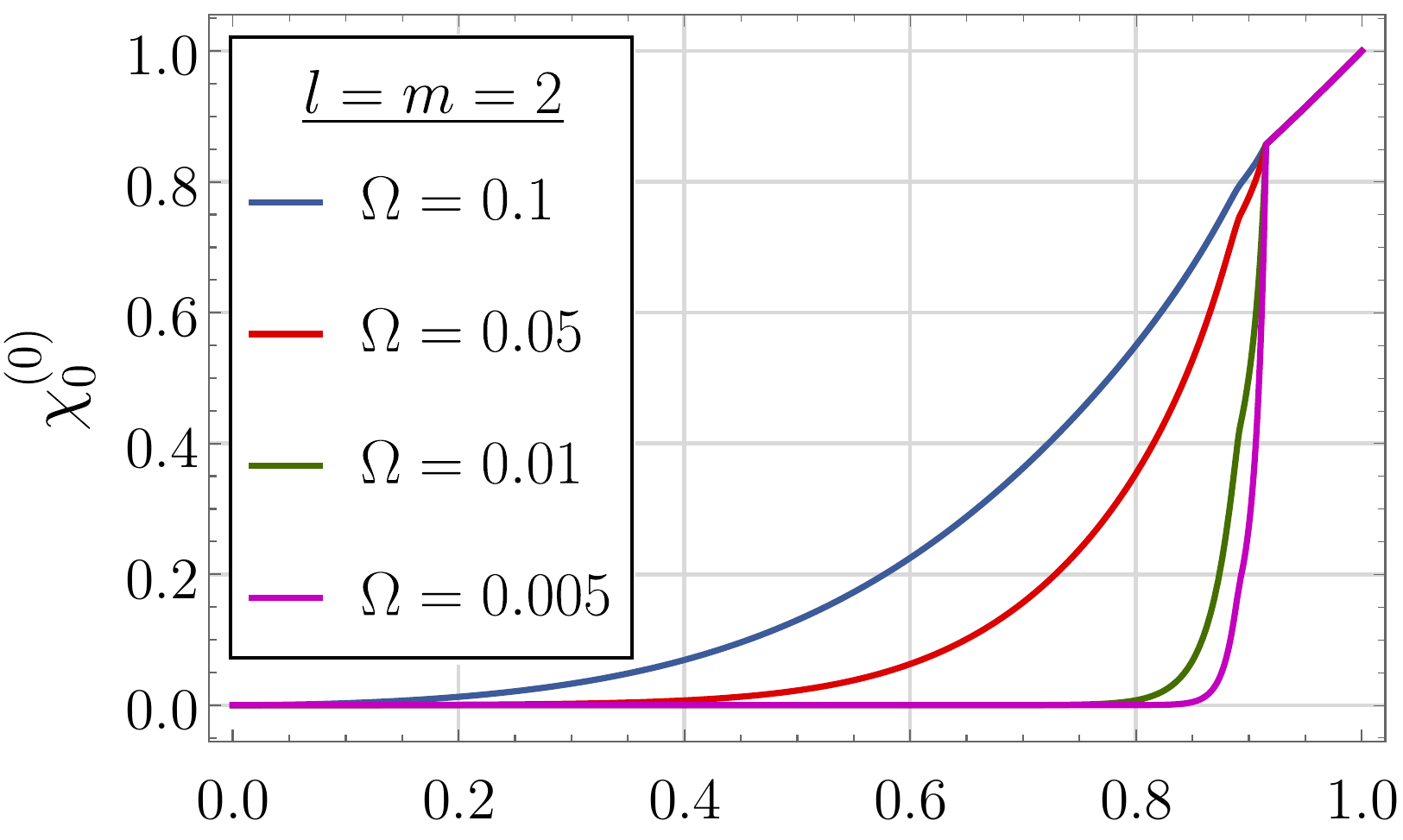}
\hspace{0.5cm}
\includegraphics[height=0.21\textheight]{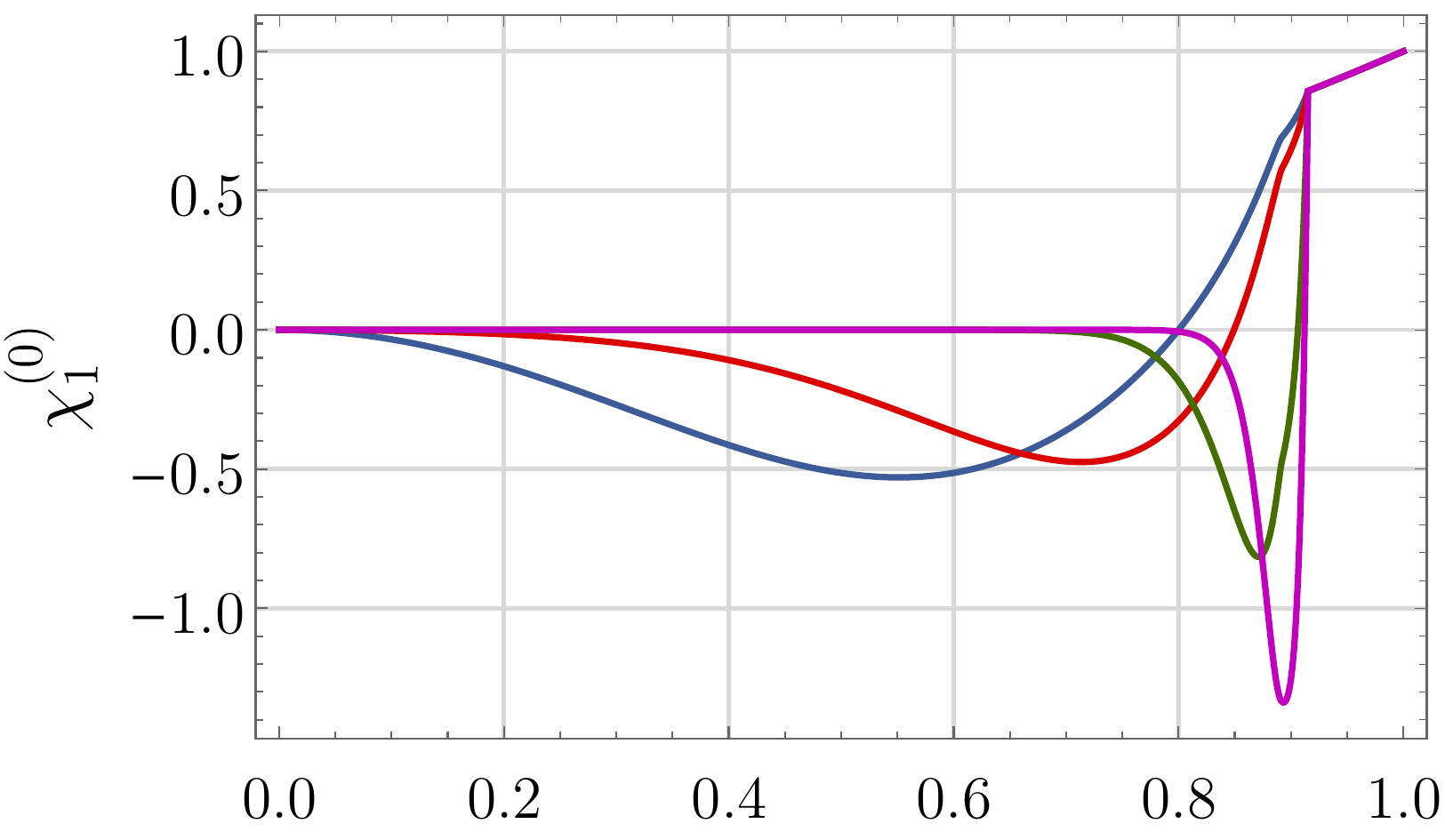}
\vfill
\includegraphics[height=0.24\textheight]{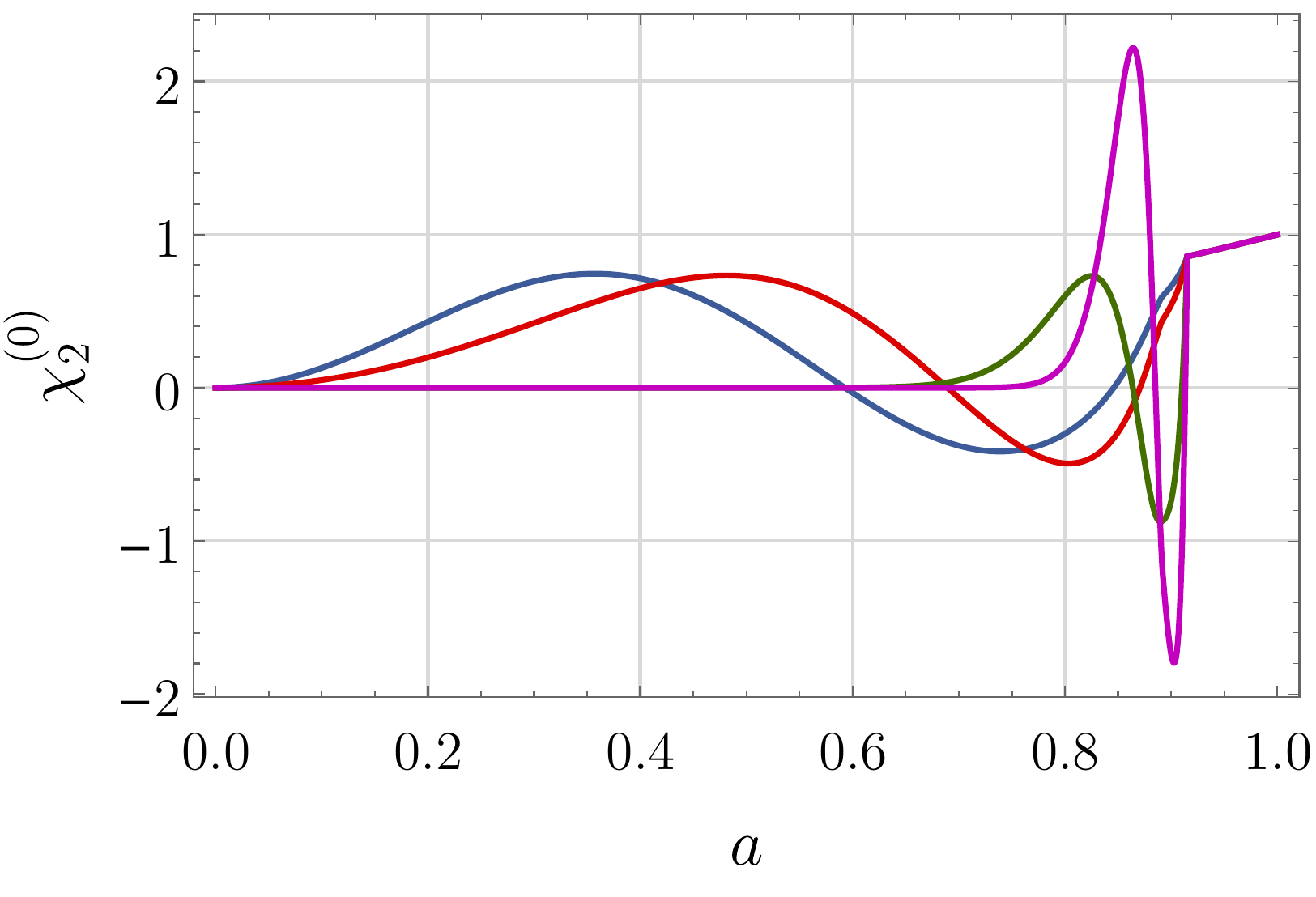}
\hspace{0.77cm}
\includegraphics[height=0.24\textheight]{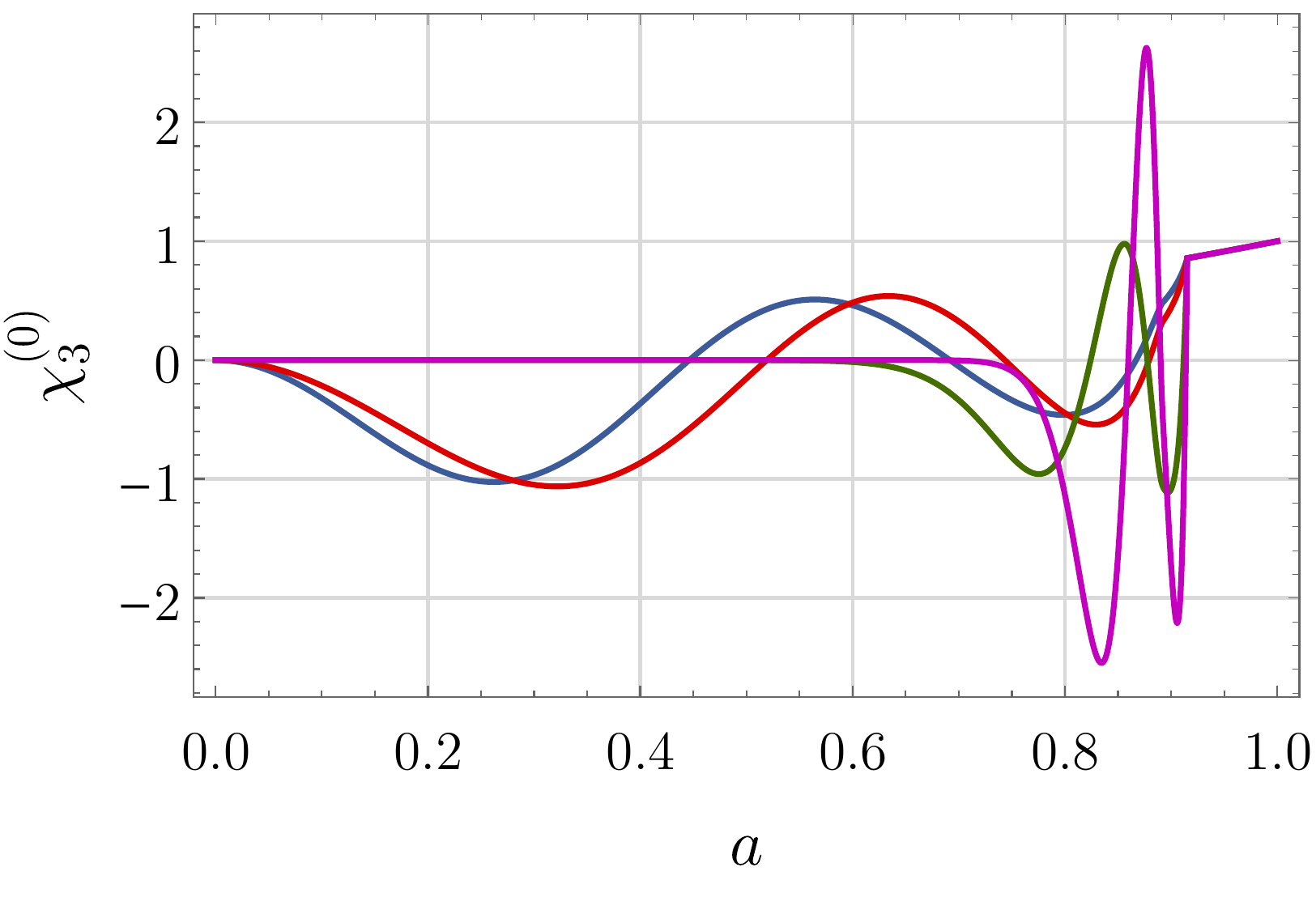}
\caption{Relativistic $r$-mode toroidal eigenfunctions $\chi^{(0)}_n$ with $n$ nodes, computed for different angular velocities $\Omega$.}
\label{mode suppression}
\end{figure}

Accounting for the effect of the inertial reference frame-dragging drastically changes the picture, and toroidal eigenfunctions and eigenfrequencies become extremely sensitive to the values of $\Omega$. This is additionally illustrated in Fig.\ \ref{mode suppression},
where we show the relativistic $r$-modes with $\epsilon\neq 0$, computed for a rather broad set of rotation rates. We see that for extremely slowly rotating stars ($\Omega\lesssim 0.01$) the modes in the core are strongly suppressed and do not vanish only in the vicinity of the crust-core interface, $a_{cc}\approx 0.92$, where the application of the weak frame-dragging effect approximation is justified. This suppression can be explained by thorough analysis of the $r$-mode equations in the $\Omega\to 0$ limit, which itself deserves a special consideration and will be addressed in the following section.

%%%%%%%%%%%%%%%%%%%%%%%%%%%%%%%%%%%%%%%%%%%%%%%%%%%%%%%%%%%%%%%%%%%%%%%%%%%%%%%%%%%%%%%%%%%%%%%%%%%%%%%%%%%%%%%%%%%%%%%%%%%%%
%%%%%%%%%%%%%%%%%%%%%%%%%%%%%%%%%%%%%%%%%%%%%%%%%%%%%%%%%%%%%%%%%%%%%%%%%%%%%%%%%%%%%%%%%%%%%%%%%%%%%%%%%%%%%%%%%%%%%%%%%%%%%
%%%%%%%%%%%%%%%%%%%%%%%%%%%%%%%%%%%%%%%%%%%%%%%%%%%%%%%%%%%%%%%%%%%%%%%%%%%%%%%%%%%%%%%%%%%%%%%%%%%%%%%%%%%%%%%%%%%%%%%%%%%%%
\section{New insight into the slow rotation approximation}

%%%%%%%%%%%%%%%%%%%%%%%%%%%%%%%%%%%%%%%%%%%%%%%%%%%%%%%%%%%%%%%%%%%%%%%%%%%%
%%%%%%%%%%%%%%%%%%%%%%%%%%%%%%%%%%%%%%%%%%%%%%%%%%%%%%%%%%%%%%%%%%%%%%%%%%%%
\subsection{The relativistic r-mode non-analyticity and ordering}\label{ordering sec}

The system \eqref{r-modes GR} governs the dynamics of relativistic $r$-modes in the core of a neutron star in the limit of weak reference frame-dragging effect and slow rotation. In its derivation, we do not use any ordering and assume only that $\epsilon$ and $\Omega^2$ take small values. It is easy to see, why the traditional ordering \eqref{ordering_Newt}--\eqref{ordering_Newt2} cannot describe relativistic $r$-modes. If we assume $T^{(0)}_l\sim 1$ and $\sigma^{(1)}\sim\xi^{(1)}_{l\pm 1}\sim\Omega^2$ and then set $\Omega^2\to 0$, from the first equation we immediately find that $\epsilon\tilde{\omega}(a) T^{(0)}_l(a)=0$. Therefore, the traditional ordering is valid, if and only if there is no reference frame-dragging, which is corroborated by our numerical results. Generally, frame-drag does not vanish
and we expect that the series \eqref{ordering_Newt}--\eqref{ordering_Newt2} cannot be applied to the study of relativistic $r$-modes.

Now, it is interesting to find out, how does the correct ordering of relativistic $r$-modes look like in the limit of slow rotation rate. Since we are dealing with two small parameters, $\epsilon$ and $\Omega^2$, we have to distinguish between the $\Omega$-ordering (series in $\Omega$, relativistic counterpart of the Newtonian ordering) and $\epsilon$-ordering (series in $\epsilon$, does not arise in the Newtonian theory, where $\epsilon=0$), so that each quantity is characterized by its $\Omega$-order and $\epsilon$-order. We start with the searches for the correct relativistic $\Omega$-ordering, and for this purpose it is sufficient to consider only the equations for the $l=m$ case (the consideration of the $l\neq m$ case is completely analogous). The system \eqref{r-modes l=m}, describing the $l=m$ modes, can be schematically represented as
\begin{gather}
\label{formal sys 0}
\left\{
\begin{gathered}
\biggl[C_{1}(a)\frac{d}{d a}+C_{2}(a)\biggr]\xi^{(1)}_{l+1}+\biggl[\Omega^2 C_{3}(a)+\sigma^{(1)}+C_4\epsilon\tilde{\omega}(a)\biggr]T^{(0)}_l=0, \\
\biggl[\frac{d}{da}+G_1(a)\biggr]T^{(0)}_l+\frac{G_2(a)}{\Omega^2}\xi^{(1)}_{l+1}=0,
\end{gathered}
\right.
\end{gather}
where $C_1(a)$, $C_2(a)$, $C_3(a)$, $C_4$, $G_1(a)$ and $G_2(a)$ are some $\Omega$- and $\epsilon$-independent functions and coefficients, whose exact form at this point is not important and can be established via straightforward comparison of this system and the system \eqref{r-modes l=m}.  Explicit expressions for these coefficients are provided in Appendix \ref{coefficients}. In the limit of vanishing rotation rate we expect that the leading contribution to the correction $\sigma^{(1)}$ is associated with the reference frame-dragging (contributions due to $\Omega$ become extremely small), therefore, from the point of view of the $\Omega$-ordering, we must have $\sigma^{(1)}\sim 1$ (nevertheless, $\sigma^{(1)}$ is small due to $\epsilon$). In order to establish the $\Omega$-ordering, let us assume that in the limit of extremely slow rotation, $\Omega\to 0$ (which implies $\Omega^2\to 0$), we have
\begin{gather}
\label{finding O-ordering}
\xi^{(1)}_{l+1}=\Omega^{k_1}X, \qquad \frac{d}{da}=\Omega^{d_1}\frac{D}{Da}, \qquad \sigma^{(1)}\sim T^{(0)}_l\sim X \sim \frac{D}{Da}\sim 1.
\end{gather}
As further analysis reveals, the full toroidal function $T^{(0)}_l(a)$ and full eigenfrequency correcton $\sigma^{(1)}$ actually contain the terms of linear $\Omega$-order, which, in the $\Omega\to 0$ limit, are discarded as small (see Appendix \ref{ordering appx} for details). Below, unless stated otherwise, under $T^{(0)}_l(a)$ and $\sigma^{(1)}$ we imply only the leading contributions without the mentioned small terms. The second equality above is formal and simply requires that the derivative $d/da$ should be considered as a ``quantity'' of order $\Omega^{d_1}$ (for example, the derivatives of $T^{(0)}_l$ or $\xi^{(1)}_{l+1}$ should be considered as quantities of the order $\Omega^{d_1}$ or $\Omega^{d_1+k_1}$, respectively). For those functions which are analytic functions of $\Omega$, we have $d_1=0$ 
\footnote{Note that the opposite is not true, and there are non-analytic functions of $\Omega$ with $d_1=0$. Consider, for example, any function of the form $f(a,\Omega)=f_1(\Omega)f_2(a)$, where $f_1(\Omega)$ is some non-analytic function of $\Omega$ that does not depend on $a$, and $f_2(a)$ is some $\Omega$-independent function.}
, whereas the case $d_1\neq 0$ signals that the functions under consideration are non-analytic functions of $\Omega$. Below this statement will be given transparent and more mathematically clear explanation. Assuming the discussed $\Omega$-ordering, we thus obtain
\begin{gather}
\label{formal sys}
\left\{
\begin{gathered}
\biggl[\underline{C_{1}(a)\Omega^{d_1}\frac{D}{D a}}+C_{2}(a)\biggr]\Omega^{k_1}X+\biggl[\Omega^2 C_{3}(a)+\underline{\sigma^{(1)}+C_4\epsilon\tilde{\omega}(a)}\biggr]T^{(0)}_l=0 \\
\biggl[\underline{\Omega^{d_1+2}\frac{D}{Da}}+G_1(a)\Omega^2\biggr]T^{(0)}_l+\underline{G_2(a)\Omega^{k_1}X}=0.
\end{gathered}
\right.
\end{gather}
At $\Omega\to 0$ some terms in this system may be small compared to the others, and only the leading terms should be retained to obtain $T^{(0)}_l$, $X$, and $\sigma^{(1)}$. In this limit the system should allow us to determine the leading eigenfrequency correction associated with $\epsilon$, therefore, the terms with $\sigma^{(1)}$ and $\tilde{\omega}(a)$ in the first equation should be of the same $\Omega$-order as at least one of the other terms (while the remaining terms in this equation should be small). This is achieved if either $k_1=0$ and $d_1\geq 0$, or $d_1+k_1=0$ and $d_1\leq 0$. It is also necessary that at least two of the three terms in the second equation were large and of the same $\Omega$-order --- otherwise we obtain the trivial solution. This, in turn, is achieved, if either $d_1=0$ and $k_1\geq 2$, or $k_1=2$ and $d_1\geq 0$, or $d_1+2=k_1$ and $k_1\leq 2$. It is easy to see that only for $d_1=-1$ and $k_1=1$ the conditions, imposed for the first equation, do not contradict the conditions, imposed for the second equation. Thus, in the limit $\Omega\to 0$ we have obtained
\begin{gather}
\xi^{(1)}_{l+1}\sim \Omega, \qquad  \frac{d}{da}\sim \frac{1}{\Omega}.
\end{gather}

According to this ordering, in the system \eqref{formal sys} we have to retain only the underlined terms and ignore everything else. Therefore, in the $\Omega\to 0$ limit the functions $T^{(0)}_l$ and $\xi^{(1)}_{l+1}$ obey the simple system of the form
\begin{gather}
\label{formal sys2}
\left\{
\begin{gathered}
C_{1}(a)\xi^{(1)}_{l+1}{}'+[\sigma^{(1)}+C_4\epsilon\tilde{\omega}(a)]T^{(0)}_l=0 \\
\Omega T^{(0)}_l{}'+\frac{G_2(a)}{\Omega}\xi^{(1)}_{l+1}=0.
\end{gathered}
\right.
\end{gather}
Using these equations it is easy to clarify, what is implied by the notation $d/da\sim 1/\Omega$. For this purpose we propose to investigate a simple toy model and consider the obtained system, assuming that coefficients $C_1(a)$ and $G_2(a)$ are constant. Then the system reduces to the single second-order equation,
\begin{gather}
\Omega^2 T^{(0)}_l{}''-q T^{(0)}_l=0,
\end{gather}
where $q$ is some constant. The solution to this equation
\begin{gather}
T^{(0)}_l=\const\cdot\exp\biggl[\pm\frac{\sqrt{q}}{\Omega}a\biggr]
\end{gather}
gives extra $1/\Omega$ factor when taking the derivative, and we have
\begin{gather}
T^{(0)}_l{}'=\pm\const\cdot\frac{\sqrt{q}}{\Omega}\exp\biggl[\pm\frac{\sqrt{q}}{\Omega}a\biggr]\sim \frac{T^{(0)}_l}{\Omega}.
\end{gather} 
Thus, the solution of our toy model is given by non-analytic functions of $\Omega$ and, therefore, cannot be expanded in Taylor $\Omega$-series near the point $\Omega=0$. Non-analyticity manifests itself when it comes to the calculation of derivatives of the sought eigenfunctions, leading to the appearance of the extra $1/\Omega$ factor. Although this interpretation of the $d/da\sim 1/\Omega$ notation comes from the consideration of the simplified equations, it also holds for the real equations with variable-dependent coefficients, as will be shown in the following section.

The system \eqref{formal sys2} can be used to determine the $\epsilon$-ordering similarly to how the system \eqref{r-modes l=m} was used to determine the $\Omega$-ordering. Since in the limit $\Omega\to 0$ the equations should allow us to find the contribution to the correction $\sigma^{(1)}$, associated with frame-dragging, the terms with $\sigma^{(1)}$ and $\tilde{\omega}(a)$ should be of the same $\epsilon$-order. We therefore expect that $\sigma^{(1)}\sim \epsilon$ and look for the $\epsilon$-ordering in the form
\begin{gather}
\sigma^{(1)}=\epsilon\sigma^{(10)},\qquad \xi^{(1)}_{l+1}\sim \epsilon^{k_2}, \qquad \frac{d}{da}\sim \epsilon^{d_2},
\end{gather}
where (yet unknown) quantity $\sigma^{(10)}$ defines the leading contribution to the eigenfrequency $\sigma^{(1)}$ in the $\Omega\to 0$ limit. All the terms in the first equation should be of the same $\epsilon$-order, therefore $k_2+d_2=1$. Similarly, the second equation implies that $k_2=d_2$. As a result, we have $k_2=d_2=1/2$ and
\begin{gather}
\xi^{(1)}_{l+1}\sim \sqrt{\epsilon}, \qquad \frac{d}{da}\sim \sqrt{\epsilon}.
\end{gather}
The last equality here should be interpreted in the same manner, as the expression $d/da\sim 1/\Omega$ was interpreted before. Combining $\Omega$- and $\epsilon$-ordering, we finally have
\begin{gather}
T^{(0)}_l\sim 1, \qquad \xi^{(1)}_{l+1}\sim \sqrt{\epsilon}\Omega, \qquad \frac{d}{da}\sim \frac{\sqrt{\epsilon}}{\Omega}.
\end{gather}
As we have already mentioned, the quantities $T^{(0)}_l(a)$ and $\sigma^{(1)}$ actually contain linear corrections in $\Omega$, which
were ignored (treated as small) in the performed analysis. If we account for these corrections and introduce a new small parameter $\kappa=\Omega/\sqrt{\epsilon}$, then the $r$-mode eigenfunctions and $\sigma^{(1)}$, up to linear terms in $\Omega$, take the form (see Appendix \ref{ordering appx})
\begin{gather}
\label{short ordering}
\sigma^{(1)}=\epsilon\sigma^{(10)}+\epsilon\kappa\sigma^{(11)}, \qquad
T^{(0)}_l=T^{(00)}_l+\kappa T^{(01)}_l, \qquad
\xi^{(1)}_{l+1}=\epsilon\kappa\xi^{(11)}_{l+1},
\end{gather}
where only the terms $T^{(00)}_l$, $\epsilon\kappa\xi^{(11)}_{l+1}$, and $\epsilon \sigma^{(10)}$ satisfy the derived above simplified system of equations. In these notations any term $f^{(ik)}$ describes the contribution of the order $\epsilon^i\kappa^k$ to the function $f$.

Summarizing, we find that the $r$-mode eigenfunctions are non-analytic functions of $\Omega$ and $\epsilon$, and that the quantities $f$ and $f'$ are of different $\Omega$- and $\epsilon$-orders. This is one of the reasons of the breakdown of the traditional approach that implicitly relies on the analyticity of eigenfunctions in the stellar angular velocity when using Taylor series \eqref{ordering_Newt}--\eqref{ordering_Newt2}. Strictly speaking, the terms ``series'', ``order'' and ``ordering'' become ill-defined, since we deal with non-analytic functions. We can now think only in terms of relative ordering. For example, the decomposition $f=f_1+\Omega f_2$ should be interpreted in the following way: both $f_1$ and $f_2$ are non-analytic functions of $\Omega$ and $\epsilon$, but the second term is $\Omega$ times smaller than the first one. The expression $T^{(0)}_l\sim 1$ means that, although $T^{(0)}_l$ depends on the small $\Omega$ and $\epsilon$, it still takes typical values of the order of unity. Keeping in mind the non-analyticity of the eigenfunctions, one can also expect that the correction $\sigma^{(10)}$, generally, may also be a non-analytic function of $\epsilon$ and $\Omega$. As we shall show, this expectation appears to be correct for all eigenfrequencies except for those of the fundamental (nodeless) harmonic.

As was mentioned before, the analysis of the $l\neq m$ case is similar to that performed for $l=m$, and it eventually leads to the same conclusions about the $r$-mode non-analyticity and to the same $r$-mode ordering. Being generalized to the $l\neq m$ case, the $r$-mode equations in the $\Omega \to 0$ limit are very similar to those derived above and are presented in Appendix \ref{general eqs}. 

Now, one can also easily establish the relativistic $r$-mode ordering in the crust. Since the toroidal function \eqref{TsolGR} is known, we have to analyze only the formula for the radial displacement \eqref{XiSolGR}:
\begin{gather}
\xi^{(1)}_{l+1}=\frac{1}{\eta(a)}\biggl[\xi_0+\int\limits_a^1\biggl(g_{21}(a)\biggl[\sigma^{(1)}+\frac{2\epsilon\tilde{\omega}(a)}{l+1}\biggr]+\Omega^2 g_{22}(a)\biggr)\eta(a)T^{(0)}_l(a)da\biggr],
\end{gather}
where $\xi_0\sim\Omega^2$ is the integration constant, found from the boundary condition at the stellar surface \eqref{bc surface}. All functions here are analytic functions of $\epsilon$ and $\Omega$. The leading contribution in the $\Omega\to 0$ limit is obviously given by
\begin{gather}
\label{leadXiCrust}
\xi^{(1)}_{l+1}=\frac{1}{\eta(a)}\int\limits_a^1 g_{21}(a)\biggl[\sigma^{(1)}+\frac{2\epsilon\tilde{\omega}(a)}{l+1}\biggr]\eta(a)T^{(0)}_l(a)da,
\end{gather}
where one should account only for the terms up to linear order in $\Omega$ in the frequency correction $\sigma^{(1)}$. We should further distinguish between two possibilities. If one considers a purely barotropic star, the vanishing of the solution at the center implies that in the $\Omega\to 0$ limit $\sigma^{(1)}\sim \epsilon$ and therefore $\xi^{(1)}_{l+1}\sim \epsilon$ and $Q^{(1)}\sim \epsilon$ [see Eq.~\eqref{Qcrust1} or Eq.~\eqref{Qcrust2}]. Rotational corrections to these quantities are then of order $\Omega^2$. When, however, a star possesses a core with a nonbarotropic EOS, the eigenfrequency, as it will be demonstrated below, contains a contribution linear in $\Omega$. In this case we have
\begin{gather}
\label{orderingCrust}
\sigma^{(1)}=\epsilon\sigma^{(10)}+\epsilon\kappa\sigma^{(11)}+\dots,
\qquad \xi^{(1)}_{l+1}=\epsilon \xi^{(10)}_{l+1}+\epsilon\kappa \xi^{(11)}_{l+1}+\dots,
\qquad Q^{(1)}_{l+1}=\epsilon Q^{(10)}_{l+1}+\epsilon\kappa Q^{(11)}_{l+1}+\dots,
\end{gather}
with the functions $\xi^{(10)}_{l+1}$ and $\xi^{(11)}_{l+1}$ defined as
\begin{gather}
\label{xi1011}
\xi^{(10)}_{l+1}(a)=\frac{1}{\eta(a)}\int\limits_a^1 g_{21}(a)\biggl[\sigma^{(10)}+\frac{2\tilde{\omega}(a)}{l+1}\biggr]\eta(a)T^{(0)}_l(a)da, \qquad 
\xi^{(11)}_{l+1}=\frac{\sigma^{(11)}}{\eta(a)}\int\limits_a^1 g_{21}(a)\eta(a)T^{(0)}_l(a)da.
\end{gather}
Here we use the same notations as in the core: the term $f^{(ik)}$ defines the contribution of the order $\epsilon^i \kappa^k$ to the quantity $f$. The functions $Q^{(10)}_{l+1}$ and $Q^{(11)}_{l+1}$ can be found from any of the Eqs.~\eqref{Qcrust1}--\eqref{Qcrust2} in the $\Omega\to 0$ limit.

%%%%%%%%%%%%%%%%%%%%%%%%%%%%%%%%%%%%%%%%%%%%%%%%%%%%%%%%%%%%%%%%%%%%%%%%%%%%
%%%%%%%%%%%%%%%%%%%%%%%%%%%%%%%%%%%%%%%%%%%%%%%%%%%%%%%%%%%%%%%%%%%%%%%%%%%%
\subsection{Finding analytically the discrete $r$-mode spectrum and eigenfunctions in the $\Omega\to 0$ limit}\label{spectrum sec}

Let us analyze more meticulously the $r$-mode equations in the core for the $l=m$ case in the limit of vanishing rotation rate. If we substitute $\xi^{(1)}_{l+1}(a)$ from the second equation of the system \eqref{formal sys2} into the first one, use that $d/da \sim 1/\kappa$ for relativistic ordering, retain only the leading-order terms and employ the explicit formulas for the coefficients $C_1(a)$, $C_4$ and $G_2(a)$ from Appendix \ref{coefficients}, we arrive at the following second order ODE for $T^{(0)}_l$:
\begin{gather}
\label{2nd order eq}
\kappa^2 T^{(0)}_l{}''-q_\sigma(a)T^{(0)}_l=0 ,\qquad q_\sigma(a)=-\frac{A(a) c^2 g(a) (2 l+3) (l+1)^3 e^{2 \nu _0(a)}[(l+1)\sigma ^{(10)}+2 \tilde{\omega}(a)]}{8 a^2 l}.
\end{gather}
Note that the asymptotic solution of this equation near the center does not match the asymptotic solution of the general system \eqref{r-modes GR}. The reason is that the correct asymptotic behavior is produced by those terms that are small due to $\Omega$ or $\epsilon$ everywhere but in the vicinity of the center. These terms do not enter the simplified system \eqref{formal sys2} [see Eqs.\ \eqref{simple_system} for the explicit form], since in the derivation of the system we implicitly did not consider the radial coordinate $a$ to be small. Therefore, Eq.~\eqref{2nd order eq} governs $r$-mode dynamics in some region $a_c\leq a\leq a_{cc}$, whereas $r$-modes in the region $0\leq a\leq a_c$ near the center are governed by the general system \eqref{r-modes GR}. The exact value of $a_c$ can be estimated from the analysis of general $r$-mode equations, and tends to zero as $\Omega\to 0$.

Mathematically, Eq.~\eqref{2nd order eq} resembles the Schr\"{o}dinger equation, and its analysis for small values of $\kappa$ can be performed using the WKB-method (see, e.g., Landau \& Lifshitz \cite{landau1965}, where the WKB-method was applied to find approximate solutions to the Schr\"{o}dinger equation, treating the Planck constant $\hbar$ as a small parameter). An equation of the same form also appears in the paper by Kantor et al.\ \cite{kgd2021}, where the $r$-modes of superfluid neutron stars were studied in the limit of vanishing rotation rate and weak entrainment. Turning points $a_t$ of both equations are defined by the condition $q_\sigma(a_t)=0$, and the analysis of these equations splits into that near and far from the turning points. 

\begin{figure}[h!!!]
\begin{center}
\includegraphics[width=0.75\linewidth]{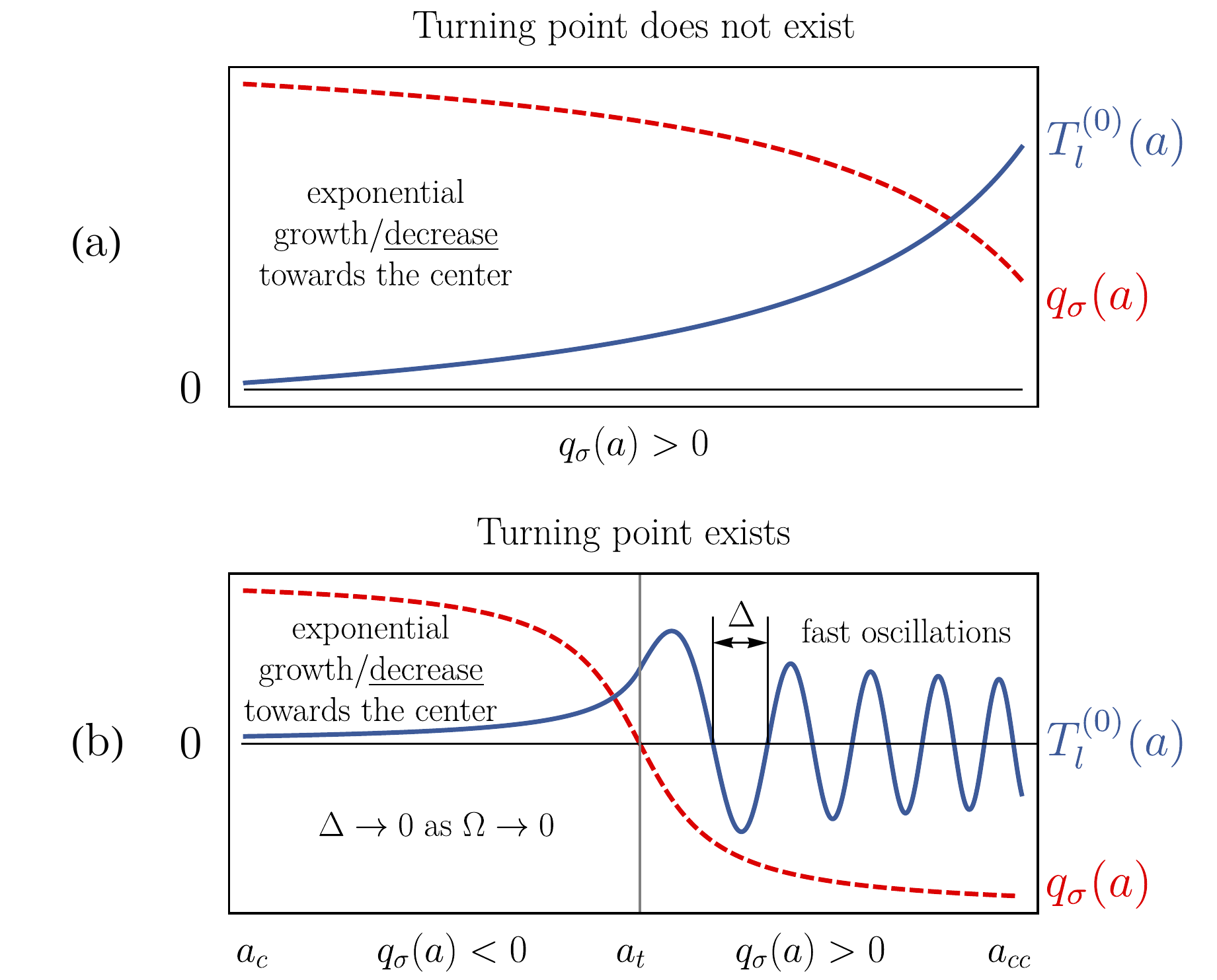}
\caption{Typical behavior of the toroidal function in the $\kappa\to 0$ limit. Panels (a) and (b) correspond to the cases, when the turning point does not exist or exists, respectively. Red dashed lines show schematically the coefficient $q_\sigma(a)$, whereas blue solid lines show the toroidal function $T^{(0)}_l(a)$ itself. Here we picture only the scenario, in which the toroidal function exponentially decreases towards the center of the star. An analogous illustration can be drawn when the toroidal function, instead, exponentially grows towards the center.}
\label{behaviours}
\end{center}
\end{figure}

Since the functions $A(a)<0$ and $g(a)>0$ do not change sign inside the star, and function $\tilde{\omega}(a)$ is monotonically decreasing function of $a$ [i.e., $\omega'(a)<0$], for each value of $\sigma^{(10)}$ there is no more than one turning point $a_t$, defined as the unique solution to the equation
\begin{gather}
\label{turning point}
(l+1)\sigma^{(10)}+2\tilde{\omega}(a_t)=0, \qquad a_t\leq a_{cc}.
\end{gather}
Typical behavior of the solution to Eq.~\eqref{2nd order eq} depends on whether the turning point exists or not (i.e., on whether it is possible to find the solution of Eq.~\eqref{turning point} satisfying $a_t\leq a_{cc}$), as shown in Fig.\ \ref{behaviours}. If the turning point does not exist [panel (a)], for $a_c\leq a\leq a_{cc}$ the solution to Eq.~\eqref{2nd order eq} exhibits either exponential growth or exponential decay towards the center of the star. Otherwise, if the turning point exists [panel (b)], such a behavior is expected 
only for $a_c<a\leq a_t$, whereas for $a_t<a<a_{cc}$ the solution is an oscillating function.

Our aim in this section is to find the explicit expressions for the $r$-mode eigenfunctions and oscillation frequencies in the $\Omega\to 0$ limit. As we have already mentioned, different $r$-mode solutions can be classified according to the number of nodes that the toroidal function $T^{(0)}_l(a)$ has inside the star. Further we will be interested in the first few solutions with small number of nodes. When the turning point exists, the wavelength of oscillations in the $a_t\leq a\leq a_{cc}$ region becomes smaller and smaller as $\kappa\to 0$. Consider, for example, the case $q_\sigma(a)=\const$, for which the distance $\Delta$ between the neighboring nodes of $T^{(0)}_l$ equals 
$\pi\kappa/\sqrt{|q_\sigma|}$. The number of nodes inside any finite interval of the region $a_t\leq a \leq a_{cc}$ tends to infinity as $\kappa$ tends to zero, therefore in the limit $\kappa\to 0$ the interval itself should considerably shrink so as not to contain an infinite number of nodes. Thus, in this limit for the turning point inside the core and for the corresponding eigenfrequencies we must have
\footnote{Since the function $q_\sigma(a)$ is positive for $a<a_{t}$, changes sign only at $a=a_t$, and $a_t\to a_{cc}$ in the $\Omega\to 0$ limit, its minimal value in the core in this limit is reached at the crust-core interface. Therefore, the condition $q_\sigma(a_{cc})\to 0$ for obtaining the leading contribution to the correction $\sigma^{(1)}$ can be equivalently presented as
\begin{gather*}
\label{minQ}
\min\limits_{a_c<a<a_{cc}}\{q_\sigma(a)\}\to 0 \qquad \text{for} \quad a_t<a_{cc}.
\end{gather*}
An analogous condition determines the spectrum of the superfluid $r$-modes, studied in Kantor et al.\ \cite{kgd2021}.
}
\begin{gather}
a_t\to a_{cc}, \qquad \sigma^{(10)}\to-\frac{2\tilde{\omega}(a_{cc})}{l+1} \qquad \text{for} \quad a_t<a_{cc}.
\end{gather}
We see that for slow rotation rates the $r$-modes in the core are localized only in the vicinity of the crust-core interface. As we have already mentioned, the weak frame-dragging effect approximation is applicable in this region and therefore should provide rather accurate predictions for the $r$-mode eigenfrequencies and oscillation spectrum.

Thus, {\it if} the turning point exists, it must be located close to the crust-core interface for small values of $\kappa$, and we have to consider the solution to Eq.~\eqref{2nd order eq} in the two different regions: region I far from the turning point with $a_c\leq a <a_{t}$, and region II in the vicinity of the turning point with $a\leq a_{cc}$, where $q_\sigma(a)$ can be Taylor-expanded as $q_\sigma(a)\approx\alpha^2(a_t-a)$. It is well-known that there exists a transition region, where both the solutions $T^{(0)}_{l,\rm I}$ and $T^{(0)}_{l,\rm II}$, obtained for the regions I and II, respectively, are valid. This region corresponds to those values of $a$ that, on one hand, are relatively far from the turning point, so that $q_\sigma(a)$ significantly differs from zero, but, on the other hand, are not too far, so that the Taylor expansion employed in the region II is still accurate enough. We demand that in the transition region these solutions should be with high accuracy equivalent to each other. If the turning point does not exist, we have to consider only the region I, which in this case spans the whole stellar core, except for a tiny region $0\leq a<a_{c}$ in the vicinity of the center.

In the region I the approximate solution to Eq.~\eqref{2nd order eq} can be written as (see Landau \& Lifshitz \cite{landau1965}):
\begin{gather}
\label{chi I}
T^{(0)}_{l,\rm I}(a)=\frac{A_{\rm I}}{q_\sigma^{1/4}}\exp\biggl(\frac{1}{\kappa}\int\limits_{a_c}^a\sqrt{q_\sigma}da\biggr)+\frac{B_{\rm I}}{q_\sigma^{1/4}}\exp\biggl(-\frac{1}{\kappa}\int\limits_{a_c}^a\sqrt{q_\sigma}da\biggr).
\end{gather}
In the region II we employ the mentioned above Taylor expansion of the function $q_\sigma(a)$, which allows us to write Eq.~\eqref{2nd order eq} in the form of the Airy equation
\begin{gather}
\label{AiryEq}
\frac{d^2}{dz^2}T^{(0)}_l-z T^{(0)}_l=0, \qquad z=(a_t-a)\biggl(\frac{\alpha}{\kappa}\biggr)^{2/3}.
\end{gather}
An arbitrary solution to this equation can be represented as a linear combination
\begin{gather}
\label{chi II}
T^{(0)}_{l,\rm II}(z)=A_{\rm II}~\Ai(z)+B_{\rm II}~\Bi(z),
\end{gather}
where $\Ai(z)$ and $\Bi(z)$ are the first and second-type Airy functions, respectively. Relatively far from the turning point $z=0$ these functions show the following asymptotic behavior:
\begin{gather}
z>0 ~~ (a<a_t): \qquad\qquad\qquad\qquad\qquad\quad z<0 ~~ (a>a_t): \\
\label{AiAsympt}
\Ai(z)\simeq\frac{1}{2\sqrt{\pi}z^{1/4}}\exp\biggl(-\frac{2}{3}z^{3/2}\biggr), \qquad \Ai(z)\simeq\frac{1}{\sqrt{\pi}|z|^{1/4}}\sin\biggl(\frac{2}{3}|z|^{3/2}+\frac{\pi}{4}\biggr),\\
\Bi(z)\simeq\frac{1}{\sqrt{\pi}z^{1/4}}\exp\biggl(\frac{2}{3}z^{3/2}\biggr),  \qquad\quad \Bi(z)\simeq\frac{1}{\sqrt{\pi}|z|^{1/4}}\cos\biggl(\frac{2}{3}|z|^{3/2}+\frac{\pi}{4}\biggr).
\end{gather}
Thus, the functions $\Ai(z)$ and $\Bi(z)$ oscillate in the region $a>a_t$, while for $a<a_t$ the function $\Ai(z)$ exponentially decreases, and function $\Bi(z)$ exponentially grows towards the center of the star. Actually, Airy functions reach their asymptotic behavior quite
fast, and for approximately $|z|\sim 2$ they, with a reasonable accuracy (with the relative error of 4-5\%), can be replaced by the corresponding asymptotic functions.

Now, as we know the form of the toroidal function in all regions, we, using the second equation of the system \eqref{formal sys2} and the expression for $G_2(a)$ from Appendix \ref{coefficients}, can obtain the explicit formulas for the radial displacement:
\begin{gather}
\xi^{(1)}_{l+1,{\rm I}}(a)=-\frac{4 a\, l\, q_\sigma^{1/4}\epsilon\kappa\, e^{-2 \nu _0}}{A c^2 g (l+1)^2(2l+1)}\biggl[A_{\rm I}\exp\biggl(\frac{1}{\kappa}\int\limits_{a_c}^a\sqrt{q_\sigma}da\biggr)-B_{\rm I}\exp\biggl(-\frac{1}{\kappa}\int\limits_{a_c}^a\sqrt{q_\sigma}da\biggr)\biggr], \\
\xi^{(1)}_{l+1,{\rm II}}(a)=-\frac{4 a\, l\, \epsilon\kappa^2 \,e^{-2 \nu _0}}{A c^2 g (l+1)^2(2l+1)}\frac{d}{da}\biggl[A_{\rm II}~\Ai(z)+B_{\rm II}~\Bi(z)\biggr].
\end{gather}

We have to check, whether we can satisfy all the boundary conditions with these solutions or not. First of all, we assume that
at the point $a=a_c$ the solution of Eq.~\eqref{2nd order eq} matches the asymptotic solution \eqref{asymptotics}. It is easy to show that these stitching conditions imply the following relation
\begin{gather}
T^{(0)}_{l,\rm I}{}'(a_c)= a_c[\tilde{A}_c+\tilde{g}_c(l-1)]T^{(0)}_{l,\rm I}(a_c),
\end{gather}
which is equivalent to
\begin{gather}
A_{\rm I}-B_{\rm I}=\frac{\kappa}{\sqrt{q_\sigma(a_c)}}a_c[\tilde{A}_c+\tilde{g}_c(l-1)](A_{\rm I}+B_{\rm I}).
\end{gather}
We see that the constants $A_{\rm I}$ and $B_{\rm I}$ are of the same order and, therefore, we can ignore the increasing towards the center exponent in the region I and write the solution approximately as
\begin{gather}
\label{regionIsol}
T^{(0)}_{l,\rm I}(a)=\frac{A_{\rm I}}{q_\sigma^{1/4}}\exp\biggl(\frac{1}{\kappa}\int\limits_{a_c}^a\sqrt{q_\sigma}da\biggr), \qquad
\xi^{(1)}_{l+1,\rm I}(a)=-\frac{4 a\, l\, \sqrt{q_\sigma}\epsilon\kappa\,e^{-2 \nu _0}}{A c^2 g (l+1)^2(2l+1)}\biggl[\frac{A_{\rm I}}{q_\sigma^{1/4}}\exp\biggl(\frac{1}{\kappa}\int\limits_{a_c}^a\sqrt{q_\sigma}da\biggr)\biggr].
\end{gather}

Secondly, if the turning point exists, the region II arises, and we require that in the transition region the obtained solution $T^{(0)}_{l,\rm I}$ and the solution $T^{(0)}_{l,\rm II}$ (and their derivatives) should coincide with high accuracy. Comparing the asymptotes of the Airy functions with the solution $T^{(0)}_{l,\rm I}$, one can show that $A_{\rm II}\gg B_{\rm II}$ and, therefore, the solution in the region II is given by the first-type Airy function:
\begin{gather}
\label{regionIIsol}
T^{(0)}_{l,\rm II}(a)=A_{\rm II}~\Ai(z), \qquad
\xi^{(1)}_{l+1,\rm II}(a)=-\frac{4 a\, l\, \epsilon\kappa^2\, e^{-2 \nu _0}}{A c^2 g (l+1)^2(2l+1)}\frac{d}{da}\biggl[A_{\rm II}~\Ai(z)\biggr].
\end{gather}

\begin{figure}[h!!!]
\centering
\includegraphics[width=1.0\linewidth]{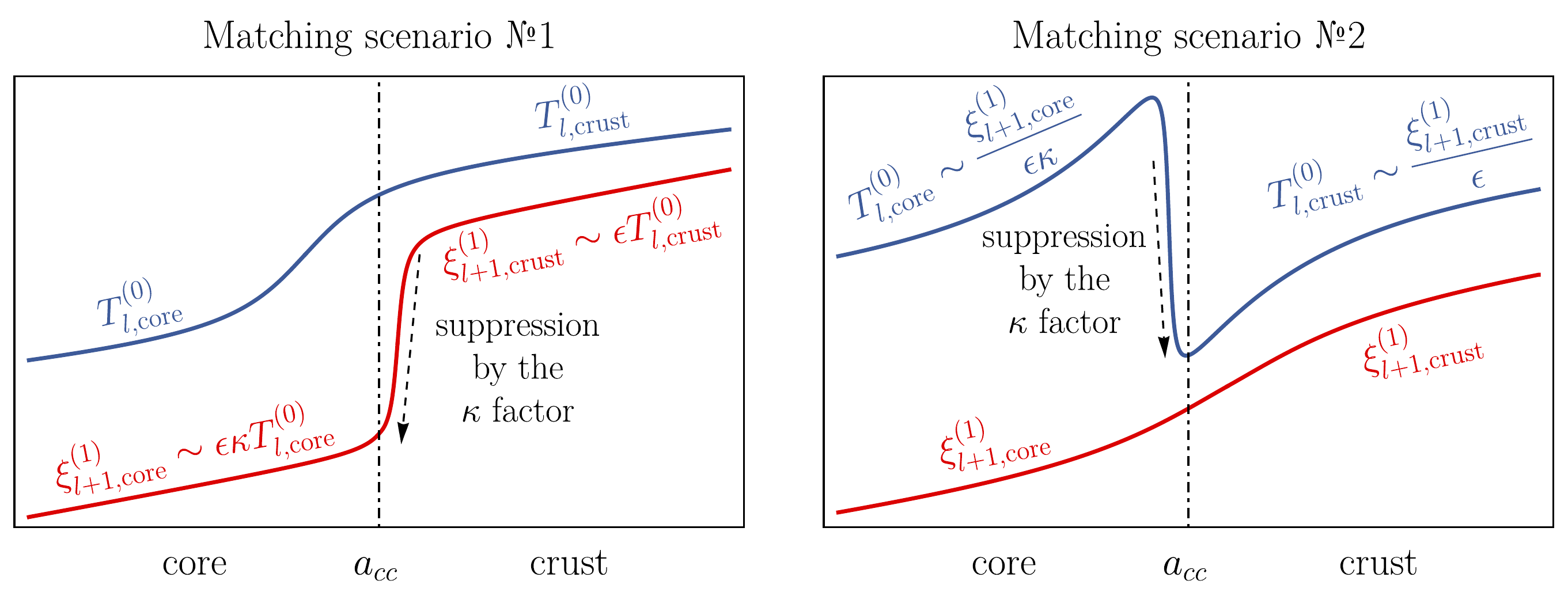}
\caption{Two different ways to satisfy the boundary conditions at the crust-core interface. The left panel presents the first way with the suppression of the $\xi^{(1)}_{l+1,\text{crust}}(a)$ at the crust-core interface $a=a_{cc}$. The right panel presents the second way with the suppression of the $T^{(0)}_{l,\text{core}}$ at the crust-core interface. The figure is schematic and does not reflect all the properties of the toroidal function and radial displacement in the vicinity of the crust-core interface (see the detailed discussion in the end of this section).}
\label{stitch}
\end{figure}

Finally, at the crust-core interface the toroidal function $T^{(0)}_l(a)$ and the radial displacement $\xi^{(1)}_{l+1}(a)$ of the core should stitch with those in the crust, which implies that
\begin{gather}
\label{ratios}
T^{(0)}_{l,\rm core}(a_{cc})=T^{(0)}_{l,\rm crust}(a_{cc}), \qquad
\frac{\xi^{(1)}_{l+1,\text{core}}(a_{cc})}{T^{(0)}_{l,\text{core}}(a_{cc})}=\frac{\xi^{(1)}_{l+1,\text{crust}}(a_{cc})}{T^{(0)}_{l,\text{crust}}(a_{cc})}.
\end{gather}
Since, generally, the ratios $\xi^{(1)}_{l+1,\text{core}}(a)/T^{(0)}_{l,\text{core}}(a)\sim \epsilon\kappa$ and $\xi^{(1)}_{l+1,\text{crust}}(a)/T^{(0)}_{l,\text{crust}}(a)\sim \epsilon$ differ one from another, there are two ways to meet these boundary conditions, as shown in Fig.\ \ref{stitch}.

The first option is to choose such specific value of $\sigma^{(1)}$, that the radial displacement $\xi^{(1)}_{l+1,\text{crust}}$ would be suppressed by a factor $\kappa$ at the crust core interface (left panel of Fig.\ \ref{stitch}). The second option is to choose such value of $\sigma^{(1)}$, for which the toroidal function $T^{(0)}_{l,\text{core}}$ would be suppressed by the factor $\kappa$ at the crust-core interface (right panel of Fig.\ \ref{stitch}). As we shall see, which one of these scenarios is actually realized, depends on whether the turning point exists or not.

For instance, consider the case, when the turning point does not exist. Then the region I spans the whole core at $a_c\leq a\leq a_{cc}$, the toroidal function cannot have nodes neither in the core, where it grows exponentially as one approaches $a_{cc}$, nor in the crust, where the solution is known [see Eq.\ \eqref{TsolGR}] and takes only positive values. Therefore, it cannot be suppressed at the crust-core interface and the scenario pictured in the right panel of Fig.\ \ref{stitch} cannot be realized. Thus, the left panel scenario takes place and the eigenfrequency $\sigma^{(1)}$ should take such value, so as to suppress $\xi^{(1)}_{l+1,\rm crust}(a)$ at the crust-core interface. From the boundary condition \eqref{ratios}, using the formulas for the radial displacement in the core \eqref{regionIsol} and in the crust \eqref{leadXiCrust}, we then find that the eigenfrequency, as expected, can be written as
\begin{gather}
\label{sigma1 main}
\sigma^{(1)}_0=\epsilon\sigma^{(10)}_0+\epsilon\kappa\sigma^{(11)}_0,
\end{gather}
where the index ``0'' refers to the fact that this solution has no nodes, and the terms $\sigma^{(10)}_0$ and $\sigma^{(11)}_0$ are given by
\begin{gather}
\label{sigma10 main}
\sigma^{(10)}_0=-\frac{2}{l+1}\biggl[\int\limits_{a_{cc}}^1 g_{21}(a)\tilde{\omega}(a)\eta(a)T^{(0)}_{l,\text{crust}}(a)da\biggr]\biggl/ \biggl[\int\limits_{a_{cc}}^1 g_{21}(a)\eta(a)T^{(0)}_{l,\text{crust}}(a)da\biggr], \\
\label{sigma11 main}
\sigma^{(11)}_0=-\frac{4 a\, l\, \eta\sqrt{q_\sigma}\,e^{-2\nu_0}}{A c^2 g (l+1)^2(2l+1)}\biggr|_{a=a_{cc}}\biggl/\biggl[\int\limits_{a_{cc}}^1 g_{21}(a)\eta(a)\bigl\{T^{(0)}_{l,\text{crust}}(a)/T^{(0)}_{l,\text{crust}}(a_{cc})\bigr\}da\biggr].
\end{gather}
Recalling the decomposition of the radial displacement in the crust \eqref{orderingCrust}, we see that the found contribution $\sigma^{(10)}$ is such that $\epsilon\xi^{(10)}_{l+1}(a_{cc})=0$, and, therefore, the leading contribution to the value of the radial displacement at the crust-core interface is $\epsilon\kappa\xi^{(11)}_{l+1}(a_{cc})$, i.e., we indeed have a suppression of the radial displacement by a factor $\kappa$. 

Now, let us consider the case, when the turning point exists. Since the scenario 1 corresponds to the solution without the turning point, this case should be described within the scenario 2, pictured in the right panel of Fig.\ \ref{stitch} (otherwise we come to contradiction). In this scenario the core toroidal function at the crust-core interface should be suppressed by a factor of $\kappa$, i.e., $A_{\rm II}~\Ai(z_{cc})\sim \kappa$. The smaller the value of $\kappa$, the better this condition can be replaced by the condition $\Ai(z_{cc})=0$, which is equivalent to
\begin{gather}
\label{atacc}
a_{t}-a_{cc}=z_n\biggl(\frac{\kappa}{\alpha}\biggr)^{2/3}, \qquad n\in \mathbb{N},
\end{gather}
where $z_n<0$ are the roots of the equation $\Ai(z)=0$, conventionally numbered so as to have $z_{n+1}<z_{n}$. Note that this result is consistent with the condition $a_t<a_{cc}$.  Note also that the condition $\Ai(z_{cc})=0$ is actually an exact boundary condition, since the equations under consideration govern only the leading order contributions to the toroidal function and radial displacement. Indeed, $T^{(0)}_l$ and $\xi^{(1)}_{l+1}$ in the studied equations correspond to the terms $T^{(00)}_l$ and $\xi^{(11)}_{l+1}$ in the relativistic $r$-mode ordering \eqref{short ordering}. At the same time, if one accounts for the rotational correction to the toroidal function and writes $T^{(0)}_l=T^{(00)}_l+\kappa T^{(01)}_l$, then it becomes clear that the suppression of the full toroidal function by the factor $\kappa$ at the crust-core interface is equivalent to the equality $T^{(00)}_l(a_{cc})=0$ at $\kappa \rightarrow 0$.

Now we can easily obtain the analytic formula for the $r$-mode spectrum. Since the turning point is close to the crust-core interface, from the definition of the coefficient $\alpha$ and from Eq.\ \eqref{turning point} we have, by expanding $\omega(a_t)$ in the Taylor series near $a=a_{cc}$
\begin{gather}
\alpha^2=\frac{q_\sigma(a_{cc})}{a_t-a_{cc}}, \qquad \sigma^{(10)}=-\frac{2 \tilde{\omega}(a_{cc})}{l+1}\biggl[1+\frac{\tilde{\omega}'(a_{cc})}{\tilde{\omega}(a_{cc})}(a_t-a_{cc})\biggr].
\end{gather}
Using the explicit form of the function $q_\sigma(a)$ and relation \eqref{atacc}, we immediately obtain
\begin{gather}
\label{sigma10 nodes1}
\alpha^2=\frac{A c^2 g(2l+3)(l+1)^3 e^{2\nu_0} \tilde{\omega}'(a)}{4 a^2 l}\biggr|_{a=a_{cc}}, 
\qquad
\sigma^{(10)}=-\frac{2 \tilde{\omega}(a_{cc})}{l+1}\biggl[1+z_n\frac{\tilde{\omega}'(a_{cc})}{\tilde{\omega}(a_{cc})}\biggl(\frac{\kappa}{\alpha}\biggr)^{2/3}\biggr].
\end{gather}
Actually, all the roots $z_n$ take the values in the region $z<-2$, where the Airy function $\Ai(z)$ can be, with reasonable accuracy, replaced by its asymptotic representation \eqref{AiAsympt}, and we can approximate $z_n$ and the $r$-mode eigenfrequencies as
\begin{gather}
\label{sigma10 nodes2}
z_n=-\biggl[\frac{3\pi}{2}\biggl(n-\frac{1}{4}\biggr)\biggr]^{2/3}, \qquad
\sigma^{(10)}_n=-\frac{2 \tilde{\omega}(a_{cc})}{l+1}\biggl\{1-\frac{\tilde{\omega}'(a_{cc})}{\tilde{\omega}(a_{cc})}\biggl[\frac{3}{2}\frac{\pi\kappa}{\alpha}\biggl(n-\frac{1}{4}\biggr)\biggr]^{2/3}\biggr\}, \qquad n\in\mathbb{N}.
\end{gather}
Note that this spectrum can alternatively  be  written in the form of the Bohr-Sommerfeld quantization rule
\begin{gather}
\label{quantization rule}
\int\limits_{a_t}^{a_{cc}}\sqrt{|q_\sigma(a)|} \,d a=\pi\kappa\biggl(n-\frac{1}{4}\biggr), \quad n\in \mathbb{N}.
\end{gather}
One can show that the latter formula, generally, is more accurate, since it does not rely on the Taylor expansion of the function $q_\sigma(a)$ near the turning point. If, for example, for a given $\Omega$, the function $q_\sigma(a)$ is not smooth enough in the region $a_t\leq a\leq a_{cc}$ (its derivative exhibits significant changes), its Taylor expansion near $a_t$ may not describe its behavior near $a_{cc}$. In this case, the formulas \eqref{sigma10 nodes1}--\eqref{sigma10 nodes2} cannot be applied to obtain the spectrum, whereas the quantization rule \eqref{quantization rule} will provide accurate results.

Let us show that all these eigenfrequencies correspond to the $r$-modes with $n$ nodes inside the core. We recall that all the nodes (if any) of the toroidal function are always located in the core. Because of the boundary condition, $T_l^{(00)}(a_{cc})=0$, one of the nodes of the derived node-possessing eigenfunctions lies exactly at the crust-core interface. We have to make sure that this node is not the ``artificial'' one, i.e., it corresponds to the real node of the full toroidal function $T^{(0)}_{l,\rm full}(a)$, defined as the solution to the system \eqref{r-modes l=m} and accounting for the next-order rotational corrections. To demonstrate this, let us consider a tiny layer near the crust-core interface, where the Airy function can be Taylor expanded as $\Ai(z)\approx\Ai'(z_n)(z-z_n)$ [recall that $\Ai(z_n)=0$]. Using this expansion and the relation \eqref{AiryEq} between $z$ and $a$, from Eq. \eqref{regionIIsol} we find that the toroidal eigenfunction and the radial displacement can be approximately related as $\xi^{(1)}_{l+1,\rm II}(a)=-\tilde{F}(a)T^{(0)}_{l,\rm II}(a)$, where $\tilde{F}(a)$ is some positive function for $a<a_{cc}$. This means that the radial displacement, obtained in the $\kappa\to 0$ limit, and toroidal function in the core approach the point $a=a_{cc}$ with different signs. On the other hand, since $a_t<a_{cc}$ and $\omega(a)>0$, from Eq.\eqref{turning point} it follows that for the node-possessing modes the combination $\sigma^{(1)}+2\epsilon\tilde{\omega}/(l+1)$ is negative in the crust. As a result, the integrand in \eqref{XiSolGR} and therefore the radial displacement in the crust take positive values  for any $a_{cc}\leq a\leq 1$. We conclude then that the radial displacement in the core should also be positive near the point $a=a_{cc}$. Therefore, near the crust-core interface, the toroidal function $T^{(0)}_{l,\rm II}(a)$ is negative. At the same time, the full toroidal function should take a positive value, since it should be matched
with $T^{(0)}_{l,\text{crust}}$, which is a positive function. This indicates that, if we account for small $\kappa$ corrections to the function $T^{(0)}_{l,\rm II}$, the node of the Airy function, located at the crust-core interface, gets shifted inside the core and produces a real node of the full toroidal function $T^{(0)}_{l,\rm full}(a)$, so that it could change sign and match the toroidal function in the crust.

Summing up, we see that in the limit $\kappa\to 0$ we have two different types of solutions, describing relativistic $r$-modes with discrete eigenfrequencies, placed within the continuous-spectrum-associated frequency band. The first type is the fundamental exponentially growing nodeless $r$-mode with no turning point and with the eigenfrequency, given by Eq.~\eqref{sigma1 main}. The second type comprises the $r$-modes with $a_t<a_{cc}$ with one or more nodes inside the core and eigenfrequencies given by the formula \eqref{sigma10 nodes1} or, approximately, \eqref{sigma10 nodes2}. For these modes the turning point is located closely to the crust-core interface, the $r$-mode eigenfunctions are exponentially suppressed on the left of the turning point and oscillate extremely rapidly
with the oscillation wavelength proportional to $\Omega^{2/3}$, in the tiny region $a_t\leq a\leq a_{cc}$ near the crust-core interface. The number $n\in\mathbb{N}$ is equal to the number of nodes of $T^{(0)}_l(a)$, concentrated in this region. Eventually, as $\kappa$ gets extremely small, all eigenfrequencies of these second-type modes reduce to $\sigma^{(10)}_n\to -2 \tilde{\omega}(a_{cc})/(l+1)$ and become almost indistinguishable.

%%%%%%%%%%%%%%%%%%%%%%%%%%%%%%%%%%%%%%%%%%%%%%%%%%%%%%%%%%%%%%%%%%%%%%%%%%%%
%%%%%%%%%%%%%%%%%%%%%%%%%%%%%%%%%%%%%%%%%%%%%%%%%%%%%%%%%%%%%%%%%%%%%%%%%%%%
\subsection{Verifying the theory and explaining numerical results}\label{verification sec}

\begin{figure}[h!!!]
\begin{minipage}{0.6\linewidth}
\includegraphics[width=0.9\linewidth]{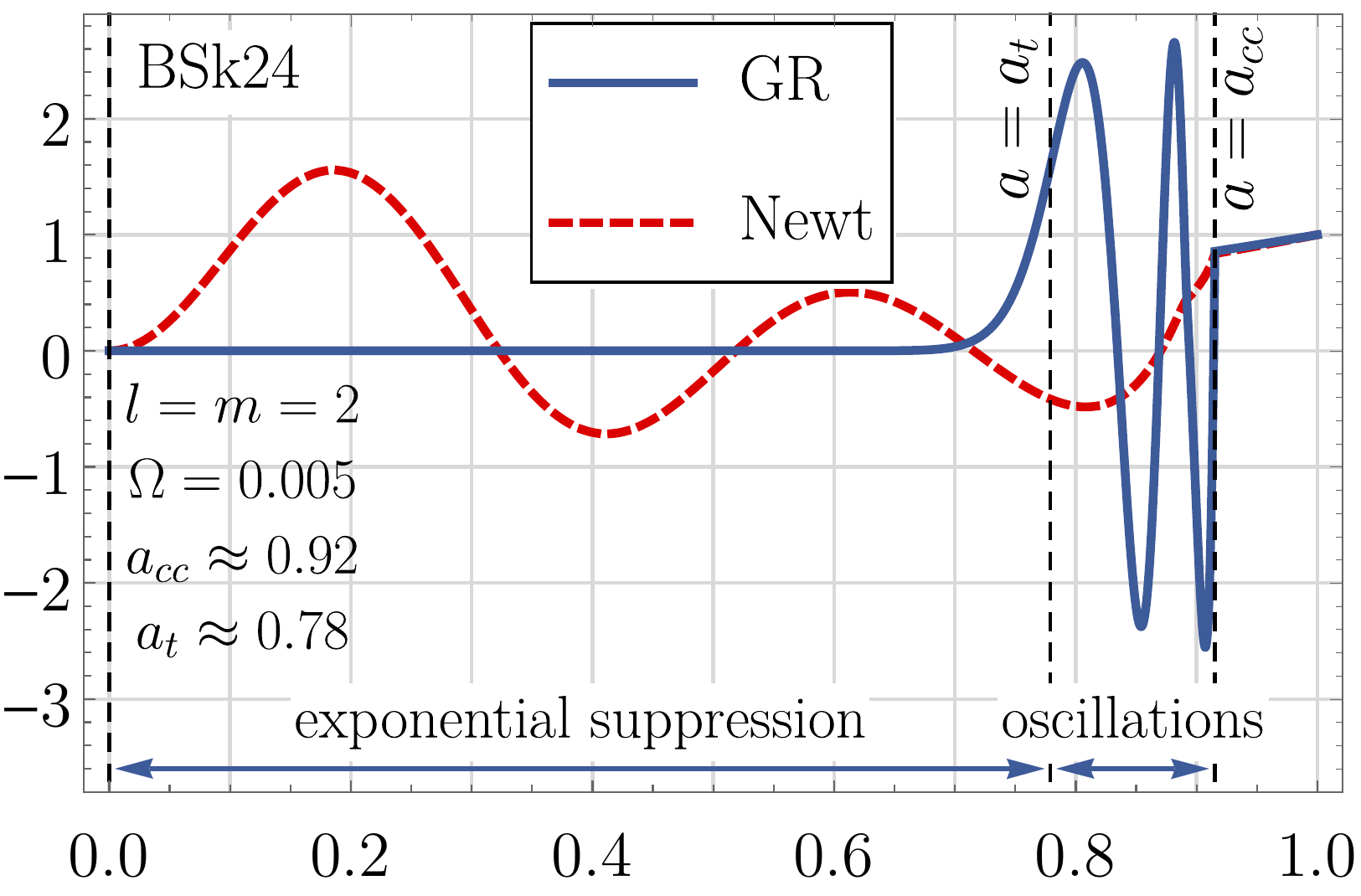}
\end{minipage}
\begin{minipage}{0.35\linewidth}
\caption{Explaining the $r$-mode suppression in the core using, as an example, the toroidal eigenfunction with 4 nodes, obtained assuming
$\Omega=0.005$. Vertical dashed lines represent the crust-core interface $a=a_{cc}$ and the turning point $a=a_t$. Red-dashed and blue solid lines show the Newtonian and relativistic $r$-modes, respectively.}
\label{explainFunc}
\end{minipage}
\end{figure}

Now we can explain the behavior of numerically obtained $r$-modes for small values of $\Omega$. Consider, for example, the relativistic $r$-mode with 4 nodes for $\Omega=0.005$, shown by blue solid line in Fig.\ \ref{explainFunc} [recall that we use the notation $\chi^{(0)}_n$ for the toroidal eigenfunction $T^{(0)}_l$ with $n$ nodes, normalized so as to have $T^{(0)}_l(1)=1$]. As expected, we see that all the nodes of the toroidal function are concentrated in the region $a_t\leq a\leq a_{cc}$. The suppression of the mode on the left of the turning point is explained by the behavior of the solution \eqref{regionIsol} in the region I.

\begin{figure}[h!!!]
\includegraphics[width=1.0\linewidth]{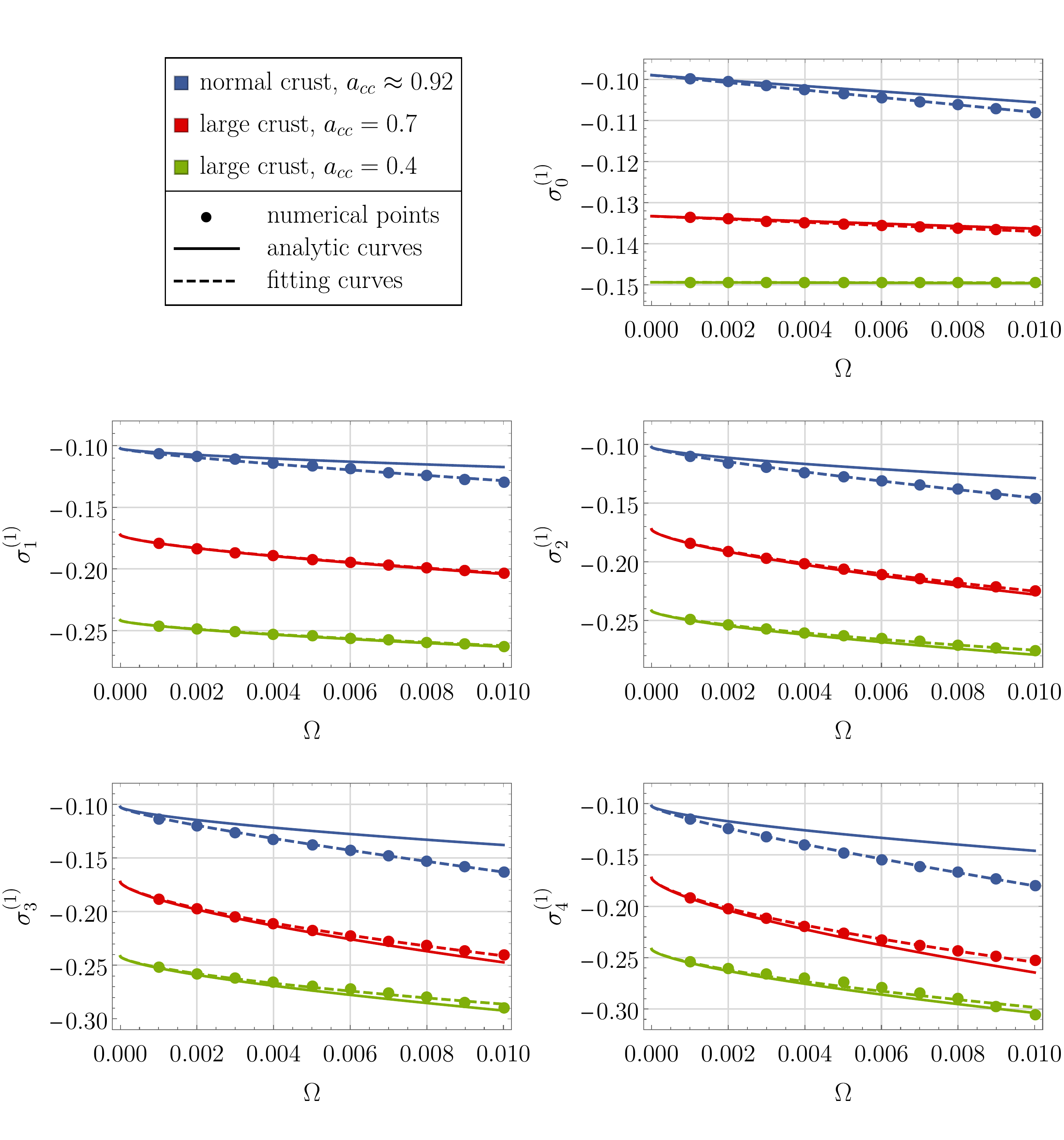}
\caption{Relativistic $r$-mode eigenfrequencies $\sigma^{(1)}_n$ for slow rotation rates, corresponding to the toroidal eigenfunctions with $n$ nodes. Blue, orange and green colors refer to the stellar models with $a_{cc}\approx 0.92$, $a_{cc}=0.7$, and $a_{cc}=0.4$, respectively. Filled circles show numerically calculated frequencies (numerical points), solid lines show the frequencies, calculated with the obtained analytical formulas (analytical curves), and dashed lines show the fit of these numerical points, accounting for the linear rotational corrections to the frequencies (fitting curves).}
\label{frequencies}
\end{figure}

We also check the accuracy of the obtained analytic expressions for the $r$-mode eigenfrequencies in the $\Omega\to 0$ limit, see Fig.\ \ref{frequencies}. Here the results for the genuine stellar model with $a_{cc}\approx 0.92$ are shown in blue, while orange and green colors correspond to modified stellar models that will be discussed a bit later. We show the calculated [via solving the system \eqref{r-modes l=m}] frequencies by filled circles and the frequencies predicted by the explicit analytic formulas \eqref{sigma1 main}-\eqref{sigma11 main} and \eqref{sigma10 nodes1} in the $\Omega\to 0$ limit by solid lines.

There are two reasons, why for the genuine stellar model the theoretical curves deviate from the numerical points with the increase of the angular velocity, as observed in the figure. The first one is that the explicit formulas \eqref{sigma10 nodes1} for the spectrum of the node-possessing $r$-modes do not account for the linear in $\Omega$ rotational corrections to the eigenfrequencies, which may become important at faster rotation rates.  One can check whether the inclusion of such terms compensates this deviation or not by trying to fit our numerical results with a simple formula
\begin{gather}
\label{fit}
\sigma^{(1)}_{n,\rm fit}(\Omega)=\sigma^{(1)}_{n,\rm analytic}(\Omega)+c_n \Omega,
\end{gather}
consistent with the expected $r$-mode ordering \eqref{short ordering}. Here the index $n$ refers to the eigenfrequency of the mode with $n$ nodes, the terms $\sigma^{(1)}_{n,\rm analytic}$ are calculated using the derived above explicit expressions \eqref{sigma1 main}-\eqref{sigma11 main} and \eqref{sigma10 nodes1} in the $\Omega\to 0$ limit, and $c_n$ are the fitting constants. Use of this fit
is justified, strictly speaking, only for the node-possessing $r$-modes ($n\neq 0$), for which $\sigma^{(1)}_{n,\rm analytic}$ is defined by Eq.~\eqref{sigma10 nodes1} and does not include linear in $\Omega$ rotational corrections. The explicit formulas \eqref{sigma1 main}--\eqref{sigma11 main} for the main harmonic eigenfrequency ($n=0$), instead, do include such correction, therefore the fit \eqref{fit} cannot be used to explain the eigenfrequency deviation of the main harmonic. Nevertheless, we can formally employ it, since the numerically obtained main harmonic eigenfrequencies still show almost linear dependency on $\Omega$. The resulting fitting curves are shown by dashed lines. The fitting curves describe the numerically obtained eigenfrequencies reasonably well, which meets our expectations concerning the relativistic $r$-mode spectrum dependency on $\Omega$ in the $\Omega\to 0$ limit. We, however, still have to explain, why the theoretically predicted slope of the line for the main harmonic does not coincide with that of the fitting formula, which brings
us to the second reason. It turns out that when we consider a two-layer stellar model, a new small parameter -- the size of the crust -- comes into play. In the original stellar model the crust forms a narrow outer layer with the relative thickness $(1-a_{cc})\approx 0.08$. As a result, the crust appears to be too small to let the radial displacement $\xi^{(1)}_{l+1}$ grow from the value $\xi_0\sim\Omega^2$ at the surface [see Eq.\eqref{bc surface}] to its expected typical value of the order $\epsilon$: the first term in the decomposition \eqref{orderingCrust} for $\xi^{(1)}_{l+1}$ remains small, so that effectively, for the considered range $10^{-3}\leq\Omega\leq 10^{-2}$ of rotation rates, we have $\xi^{(1)}_{l+1}(a)\sim \epsilon\kappa T^{(0)}_l(a)$ instead of $\xi^{(1)}_{l+1}(a)\sim\epsilon T^{(0)}_l$ (see the detailed discussion of this issue in Appendix \ref{small crust appx}). This effective violation of the ordering makes the developed theory marginally applicable for the considered values of $\Omega$ and contributes to the observed deviation of the theoretical curves from the numerical points.
 
To check the developed theory we avoid the problem associated with the small thickness of the crust by considering two artificial stellar models with significantly larger crust (see Appendix \ref{small crust appx} for details): the first with $a_{cc}=0.7$, and the second with $a_{cc}=0.4$. We show the calculated $r$-mode eigenfrequencies for these models in Fig.\ \ref{frequencies} by red and green colors, respectively. As before, we compare the frequencies predicted by the theory in the $\Omega\to 0$ limit (solid lines), the  numerically obtained frequencies (dots) and the corresponding fit \eqref{fit} (dashed lines). All the numerical eigenfrequencies lie on the fitting curves, indicating again that our predictions concerning the spectrum dependency on $\Omega$ are correct. As expected, the main harmonic frequencies in these models are described better than for the original star, especially in the model with $a_{cc}=0.4$. The fitting curves for the eigenfrequencies of all the other $r$-modes (i.e., those with the nodes, $n\neq 0$) also only slightly differ from those, calculated using the derived analytical formulas, which indicates that for these modes the linear in $\Omega$ corrections to the spectrum are relatively small (as expected) and that approximately $\sigma^{(1)}_n\approx \epsilon\sigma^{(10)}_n$ for the considered rotation rates.

%%%%%%%%%%%%%%%%%%%%%%%%%%%%%%%%%%%%%%%%%%%%%%%%%%%%%%%%%%%%%%%%%%%%%%%%%%%%%%%%%%%%%%%%%%%%%%%%%%%%%%%%%%%%%%%%%%%%%%%%%%%%%
%%%%%%%%%%%%%%%%%%%%%%%%%%%%%%%%%%%%%%%%%%%%%%%%%%%%%%%%%%%%%%%%%%%%%%%%%%%%%%%%%%%%%%%%%%%%%%%%%%%%%%%%%%%%%%%%%%%%%%%%%%%%%
%%%%%%%%%%%%%%%%%%%%%%%%%%%%%%%%%%%%%%%%%%%%%%%%%%%%%%%%%%%%%%%%%%%%%%%%%%%%%%%%%%%%%%%%%%%%%%%%%%%%%%%%%%%%%%%%%%%%%%%%%%%%%
\section{Discussion}\label{disc sec}

In this paper, in order to find the relativistic generalization of the Newtonian $r$-modes in a slowly rotating neutron star, we have developed and applied a new original approach to the study of the relativistic oscillation equations. We adopt the model of a neutron star with the barotropic (isentropic) crust and non-barotropic core. The barotropy of the EOS significantly affects the mathematical properties of the problem, so that oscillation equations in the crust and in the core are studied separately. Although the rotation of the star is assumed to be slow, we, in contrast to what is done within the traditional approach \cite{ak2001,provost1981,saio1982,kojima1997, kojima1998, bk1999, kh1999, kh2000, laf2001, rk2001}, do not rely on any preliminary postulated ordering, i.e., we do not 
assume the analyticity of the sought $r$-mode solutions in the stellar angular velocity, $\Omega$. The ordering, instead of being set from the very beginning, is determined from the $r$-mode oscillation equations, obtained within our approach. The derivation of the general equations that govern the relativistic $r$-mode dynamics, is based only on the assumptions that the angular velocity $\Omega$ of the star is small and that the effect of the inertial reference frame-dragging is weak. In the developed theory, the leading contribution to the $r$-mode eigenfrequencies exactly coincide with that of the Newtonian theory, and the numerical solutions of the obtained equations for $l=m=2$ correspond to the relativistic $r$-modes with the discrete eigenfrequency corrections. Thus, at least in the vicinity of the traditional $r$-mode frequency, we find no indications of the continuous spectrum. These results, unlike the predictions of the traditional analysis in the slow rotation limit, are consistent with numerical calculations performed beyond the slow rotation approximation for barotropic \cite{yoshidaetal2005, gk2008, donevaetal2013, jc2017} and, more importantly, for non-barotropic \cite{yl2002} stars, where the problem of the continuous spectrum has not been solved yet. Moreover, the detailed analysis of the equations in the $\Omega\to 0$ limit allows us to derive the explicit expressions for the $r$-mode eigenfunctions and corresponding discrete eigenfrequency corrections, each one uniquely characterized by the number of nodes of the toroidal function in the star. It is also interesting to notice that we have managed to find the $r$-mode solutions within the Cowling approximation, which is also possible in the Newtonian theory, whereas the discussed in the literature relativistic $r$-modes with discrete and isolated eigenfrequencies, coexisting with the continuous part of the spectrum, can be obtained only if one goes beyond this approximation. Combining all these remarks together, we conclude that the obtained $r$-mode solutions are indeed the relativistic counterparts of the Newtonian $r$-modes.

So, why does the traditional approach fail to describe the discrete relativistic $r$-modes? Let us recall that within this approach, in order to study the effects of slow rotation on stellar oscillations, one usually starts with the investigation of the perturbations of the non-rotating star, and then  finds small rotational corrections to these perturbations. In analogy to the Newtonian theory, it is assumed that the would-be $r$-modes in the non-rotating stellar model correspond to the non-oscillating solutions of the perturbation equations that form a subset of the so-called zero-frequency subspace. In a slowly rotating star, these solutions acquire small but finite oscillation frequencies and become the $r$-modes. Thus, the traditional approach relies on the perturbation theory with $\Omega$ being the small parameter, and the analysis of the oscillation equations immediately points towards the traditional $r$-mode ordering, which, in turn, immediately predicts the continuous oscillation spectrum. Note that every step and assumption of this approach is justified only if one studies the oscillations that are analytic functions of the angular velocity $\Omega$ (i.e., oscillation eigenfunctions and eigenfrequencies can be written in the form of $\Omega$-series). Now, looking at the theory of discrete relativistic $r$-modes
developed in this paper, we realize that, because of the effect of the inertial reference frame-dragging, this is actually not the case, and the obtained solutions contain non-analyticity that strongly manifests itself in the limit of extremely slow rotation rates. In the nonbarotropic core one gets, in the limit $\Omega \rightarrow 0$
[see Eqs.\ \eqref{regionIsol}-\eqref{regionIIsol}]
\begin{gather}
T^{(0)}_{l,\rm I}(a)\propto \exp\biggl(\frac{\sqrt{\epsilon}}{\Omega}\int\limits_{a_c}^a \sqrt{q_\sigma}da\biggr), \qquad T^{(0)}_{l,\rm II}(a)\propto \Ai\biggl[(a_t-a)\biggl(\frac{\alpha\sqrt{\epsilon}}{\Omega}\biggr)^{2/3}\biggr].
\end{gather}
Because of non-analyticity, these functions are undetermined at $\Omega=0$, i.e., they, in principle, cannot be considered as rotational corrections to any zero-frequency subspace perturbation of a non-rotating star. For the same reason, the discrete $r$-modes in the core cannot be studied with usual perturbative techniques, i.e., treating $\Omega$ as a small parameter and expanding all the quantities in Taylor series in $\Omega$. As for the barotropic crust, the obtained $r$-mode solutions would have been analytic functions of $\Omega$, had it not been for the non-analyticity in the core that affects the solution in the crust through the boundary conditions at the crust-core interface. As a result, we see that, because of the effect of inertial reference frame-dragging, the correct relativistic ordering in the $\Omega\to 0$ limit (with all the reservations concerning the validity of the term ``ordering'' in this situation, see the discussion in section \ref{ordering sec}),
\begin{gather}
T_{\rm GR, core}\sim 1, \qquad \sigma^{(1)}\sim\epsilon, \qquad \xi^a_{\rm GR, core}\sim \sqrt{\epsilon}\Omega, \qquad Q_{\rm GR, core}\sim \epsilon, \qquad \biggl(\frac{d}{da}\biggr)_{\rm GR, core}\sim \frac{\sqrt{\epsilon}}{\Omega}, \\
T_{\rm GR, crust}\sim 1, \qquad \sigma^{(1)}\sim\epsilon, \qquad \xi^a_{\rm GR, crust}\sim \epsilon, \qquad Q_{\rm GR, crust}\sim \epsilon, \qquad \biggl(\frac{d}{da}\biggr)_{\rm GR, crust}\sim 1
\end{gather}
drastically differs from the traditional one,
\begin{gather}
T_{\rm Newt}\sim 1, \qquad \sigma^{(1)}\sim\Omega^2, \qquad \xi^a_{\rm Newt}\sim \Omega^2, \qquad Q_{\rm Newt}\sim \Omega^2, \qquad \biggl(\frac{d}{da}\biggr)_{\rm Newt}\sim 1.
\end{gather}
Interestingly, if one turns the effect of the inertial reference frame-dragging off, one immediately obtains that the traditional ordering again holds for such relativistic $r$-modes, both in the nonbarotropic core and barotropic crust. Summarizing, we see that the frame-dragging effect is responsible for the slow rotation approximation breakdown, since it leads to the $r$-mode non-analyticity in the core and alters the ordering both in the core and in the crust.

Not only the $r$-mode eigenfunctions, but also the oscillation frequencies $\sigma_n$ of $r$-modes  with nonzero number of nodes $n$
become non-analytic because of the frame-dragging effect. In the limit $\Omega\to 0$ their spectrum is given by the following {\it analytical} formula
\begin{gather}
\sigma_n=\frac{2\Omega}{l+1}\biggl[1-\omega(a_{cc})\biggl\{1+z_n\frac{\omega'(a_{cc})}{\omega(a_{cc})}\biggl(\frac{\Omega}{\alpha\sqrt{\epsilon}}\biggr)^{2/3}\biggr\}\biggr]-l\Omega, \quad \Ai(z_n)=0, \quad n\in\mathbb{N}
\end{gather}
with $\alpha$ defined by Eq.~\eqref{sigma10 nodes1}. Here each oscillation frequency is determined by the corresponding zero $z_n$ of the Airy function. Notably, $\sigma_n$ depends only on the value of the frame-dragging function [and its derivative, $\omega'(a)$] 
at one point $a=a_{cc}$. It is worth noting that this formula has no counterparts in the Newtonian theory, where the eigenfrequency corrections for $r$-modes with different number of nodes can only be found numerically.

There are still several legitimate questions concerning the obtained $r$-mode solutions that deserve a separate discussion. As we have mentioned before, it was shown by Lockitch, Andersson \& Friedman \cite{laf2001} that the $r$-modes in relativistic barotropic stars do not exist, and that the $l=m=2$ Newtonian $r$-modes correspond to the relativistic inertial modes. Although we have managed to find the $r$-mode solutions in the barotropic crust, our results, actually, do not contradict these predictions. The reason is that by the term ``$r$-modes'' one usually implies oscillation modes, for which
all the velocity components except for the toroidal one are suppressed {\it in a slowly rotating star} (when $\Omega$ is small).
In this sense, strictly speaking, our solution in the crust cannot be referred to as the $r$-mode. It does describe the predominantly toroidal oscillation mode, but in our case all the velocity components except for the toroidal one are small because of the {\it weak effect of  inertial reference frame-dragging}.   The same remark applies to the obtained solution in the core. Whereas the radial displacement in the core, indeed, becomes small because of the slow rotation rate, the component $Q$ of the motion is small because of the small $\epsilon$. Thus, what we find in this paper cannot be termed an ``$r$-mode'' in the conventional sense, but, nevertheless, we still call it an ``$r$-mode'', since it is the predominantly toroidal solution of the oscillation equations that presents the relativistic generalization of the Newtonian $r$-modes. 

Now, what happens if one drops the small $\epsilon$ approximation? We expect that the predominantly toroidal solutions to the relativistic oscillation equations cease to exist. For the barotropic stars the obtained solutions will presumably correspond to the hybrid inertial modes discussed in \cite{laf2001} (but in the Cowling approximation). As for the nonbarotropic stars, our preliminary results indicate that the discussed in this paper $r$-modes will become non-analytic inertial hybrid modes. And still, the corresponding oscillation frequencies in the $\Omega\to 0$ limit should not drastically differ from those, given by the derived explicit formulas. The reason is that for extremely small values of $\Omega$  the $r$-mode eigenfunctions noticeably differ from zero only in the stellar crust and in the vicinity of the crust-core interface, where the effect of the inertial reference frame-dragging indeed can be considered as weak. Note that such behaviour of the eigenfunctions also implies that the Cowling approximation in the $\Omega\to 0$ limit should lead to significantly smaller eigenfrequency errors, than 6-11\%, obtained in \cite{jc2017}. Indeed, while the perturbations of the gravitational field in the crust do not affect the physics of the inner layers of the star, we also expect the perturbations of the gravitational field to be exponentially suppressed almost everywhere in the core. Anyway, finding the discrete relativistic $r$-modes beyond the Cowling approximation should not be a problem anymore, as we are now aware of the $r$-mode non-analyticity, and we have managed to find the discrete $r$-modes within the Cowling approximation, where the problem of the continuous spectrum seemed to be the most critical. 

Finally, let us remark that, although the traditional approach fails to describe the discrete $r$-modes in nonbarotropic stars, the question whether the (discussed in the literature) oscillation modes, possessing the continuous oscillation spectrum do or do not correspond to some real oscillation modes that can be found beyond the slow-rotation approximation, still has to be answered. As we have mentioned before, such numerical studies encounter no signatures of the continuous oscillation spectrum. In our opinion, this can indicate that some internal inconsistency of the theory, based on the traditional ordering, takes place. Moreover, such inconsistency has already been revealed in the study of the $r$-modes in barotropic stars: the system of equations based on the traditional ordering turned out to be overdetermined \cite{laf2001}. Revealing analogous or any other inconsistency for nonbarotropic stars is a very interesting problem, whose investigation goes beyond the scope of the present work.

%%%%%%%%%%%%%%%%%%%%%%%%%%%%%%%%%%%%%%%%%%%%%%%%%%%%%%%%%%%%%%%%%%%%%%%%%%%%%%%%%%%%%%%%%%%%%%%%%%%%%%%%%%%%%%%%%%%%%%%%%%%%%
%%%%%%%%%%%%%%%%%%%%%%%%%%%%%%%%%%%%%%%%%%%%%%%%%%%%%%%%%%%%%%%%%%%%%%%%%%%%%%%%%%%%%%%%%%%%%%%%%%%%%%%%%%%%%%%%%%%%%%%%%%%%%
%%%%%%%%%%%%%%%%%%%%%%%%%%%%%%%%%%%%%%%%%%%%%%%%%%%%%%%%%%%%%%%%%%%%%%%%%%%%%%%%%%%%%%%%%%%%%%%%%%%%%%%%%%%%%%%%%%%%%%%%%%%%%
\section*{Acknowledgements}
The results of this study were presented at the conference ``The Modern Physics of Compact Stars and Relativistic Gravity 2021''  (Yerevan, Armenia, September 27-30, 2021).

\appendix

%%%%%%%%%%%%%%%%%%%%%%%%%%%%%%%%%%%%%%%%%%%%%%%%%%%%%%%%%%%%%%%%%%%%%%%%%%%%%%%%%%%%%%%%%%%%%%%%%%%%%%%%%%%%%%%%%%%%%%%%%%%%%
%%%%%%%%%%%%%%%%%%%%%%%%%%%%%%%%%%%%%%%%%%%%%%%%%%%%%%%%%%%%%%%%%%%%%%%%%%%%%%%%%%%%%%%%%%%%%%%%%%%%%%%%%%%%%%%%%%%%%%%%%%%%%
%%%%%%%%%%%%%%%%%%%%%%%%%%%%%%%%%%%%%%%%%%%%%%%%%%%%%%%%%%%%%%%%%%%%%%%%%%%%%%%%%%%%%%%%%%%%%%%%%%%%%%%%%%%%%%%%%%%%%%%%%%%%%
\section{Typical $r$-mode streamlines}\label{r-mode geometry} 
 
\begin{figure}[h!]
\centering
\animategraphics[width=0.67\linewidth]{4}{slide-}{1}{10} 
\caption{Animated streamlines of the Lagrangian displacement vector field ${\bs \xi}$ for the $r$-modes on the stellar surface for different combinations of $l$ and $m$, as seen in the corotating reference frame. The time flow, which is individual for each pair $(l,m)$, is measured, for convenience,in units of the corresponding oscillation periods. This gives an impression that the $r$-modes in the figure oscillate with the same frequency.}
\label{streamlines}
\end{figure}

%%%%%%%%%%%%%%%%%%%%%%%%%%%%%%%%%%%%%%%%%%%%%%%%%%%%%%%%%%%%%%%%%%%%%%%%%%%%%%%%%%%%%%%%%%%%%%%%%%%%%%%%%%%%%%%%%%%%%%%%%%%%%
%%%%%%%%%%%%%%%%%%%%%%%%%%%%%%%%%%%%%%%%%%%%%%%%%%%%%%%%%%%%%%%%%%%%%%%%%%%%%%%%%%%%%%%%%%%%%%%%%%%%%%%%%%%%%%%%%%%%%%%%%%%%%
%%%%%%%%%%%%%%%%%%%%%%%%%%%%%%%%%%%%%%%%%%%%%%%%%%%%%%%%%%%%%%%%%%%%%%%%%%%%%%%%%%%%%%%%%%%%%%%%%%%%%%%%%%%%%%%%%%%%%%%%%%%%%
\section{Integration scheme}\label{Integration scheme}
To find the relativistic $r$-mode eigenfunctions and eigenfrequencies for a fixed rotation rate $\Omega$, we employ the following algorithm:

\begin{itemize}

\item[(0)] We choose the range of values $[\sigma^{(1)}_\text{left},\sigma^{(1)}_\text{right}]$ that we want to check for the presence of eigenfrequencies $\sigma^{(1)}$. We look for the solution of the equations for the values of $\sigma^{(1)}$ equidistantly located within this range.

\item[(1)] The $n$-th iteration of the algorithm starts with some value of the frequency $\sigma^{(1)}=\sigma^{(1)}_n$, for which we find the solution of Eq.~\eqref{XiEqGR} in the crust with the boundary condition \eqref{bc surface}, corresponding to the normalization $T^{(0)}_\text{$l$,~crust}(1)=1$. Let us denote the corresponding value of the radial displacement at the crust-core interface as $\xi^{(1)}_{l+1,~\text{crust}}(a_{cc})\equiv \xi_\text{crust, $n$}$.  Although we know the analytic solution of this equation, its numerical calculation turns out to be faster and more practical. Also note that the quantity $\xi_\text{crust}$ is a linear function of $\sigma^{(1)}$ [see the formula \eqref{XiSolGR}], therefore there is no need to solve every time Eq.~\eqref{XiEqGR}, but it is sufficient to find $\xi_\text{crust}$ only for two arbitrary values, say, $\sigma^{(1)}_1$ and $\sigma^{(1)}_2$. Knowing the corresponding values of $\xi_\text{crust}$ one completely finds the linear dependency that can  be used further to find $\xi_\text{crust}$ without solving Eq.~\eqref{XiEqGR}.

\item[(2)] For the same value $\sigma^{(1)}_n$ we solve the system \eqref{r-modes GR} in the core with boundary conditions \eqref{bc center}, normalize the obtained solution so as to have $T^{(0)}_\text{$l$,~core}(a_{cc})=T^{(0)}_{l,\rm crust}(a_{cc})$, and then find the corresponding value of the radial displacement at the crust-core interface $\xi^{(1)}_{l+1,~\text{core}}(a_{cc})\equiv \xi_\text{core, $n$}$.

\item[(3)] We evaluate the residual parameter $\delta_n\equiv \delta(\sigma^{(1)}_n)=(\xi_\text{core, $n$}-\xi_\text{crust, $n$})/|\xi_\text{crust, $n$}|$, and then compare the values of $\delta_n$ and $\delta_{n-1}$. If $\delta_{n-1}$ and $\delta_{n}$ are of opposite signs, then this range may contain the sought eigenfrequencies, which can be found with the use of ordinary numeric routines.
\end{itemize}
Thus, evaluating the residual parameter and monitoring its sign changes for different values of $\sigma^{(1)}$, lying in a certain range, we can determine, whether there are any eigenvalues in this range. The smaller the residual is, the more accurately the remaining boundary condition \eqref{bc interface} is satisfied.

%%%%%%%%%%%%%%%%%%%%%%%%%%%%%%%%%%%%%%%%%%%%%%%%%%%%%%%%%%%%%%%%%%%%%%%%%%%%%%%%%%%%%%%%%%%%%%%%%%%%%%%%%%%%%%%%%%%%%%%%%%%%%
%%%%%%%%%%%%%%%%%%%%%%%%%%%%%%%%%%%%%%%%%%%%%%%%%%%%%%%%%%%%%%%%%%%%%%%%%%%%%%%%%%%%%%%%%%%%%%%%%%%%%%%%%%%%%%%%%%%%%%%%%%%%%
%%%%%%%%%%%%%%%%%%%%%%%%%%%%%%%%%%%%%%%%%%%%%%%%%%%%%%%%%%%%%%%%%%%%%%%%%%%%%%%%%%%%%%%%%%%%%%%%%%%%%%%%%%%%%%%%%%%%%%%%%%%%%
\section{Explicit formulas for the formal coefficients in the $r$-mode equations for $l=m$}\label{coefficients}
Below we provide the exact form of the $\Omega$- and $\epsilon$-independent coefficients that appear in the formal system \eqref{formal sys 0} in the discussion of the slow rotation limit:
\begin{gather}
C_1(a)=-\frac{2ak^+_{l+1}}{(l+1)^2}, \qquad C_2(a)=\frac{k^{+}_{l+1}}{(l+1)^2}\biggl[2ag(l+1)-F-2l-1\biggr], \\
C_3(a)=\frac{a e^{-2\nu_0}}{c^2 gl^2(l+1)^2}\biggl[a g\gamma_2+4 l\gamma_1(2a\lambda_0'-F+5)\biggr], \qquad C_4=\frac{2}{l+1}, \\
G_1(a)=A+g(l-1)-\frac{l}{a}, \qquad G_2(a)=\frac{Ac^2g(l+1)^2 e^{2\nu_0}}{4alk^-_l}.
\end{gather}
%

%%%%%%%%%%%%%%%%%%%%%%%%%%%%%%%%%%%%%%%%%%%%%%%%%%%%%%%%%%%%%%%%%%%%%%%%%%%%%%%%%%%%%%%%%%%%%%%%%%%%%%%%%%%%%%%%%%%%%%%%%%%%%
%%%%%%%%%%%%%%%%%%%%%%%%%%%%%%%%%%%%%%%%%%%%%%%%%%%%%%%%%%%%%%%%%%%%%%%%%%%%%%%%%%%%%%%%%%%%%%%%%%%%%%%%%%%%%%%%%%%%%%%%%%%%%
%%%%%%%%%%%%%%%%%%%%%%%%%%%%%%%%%%%%%%%%%%%%%%%%%%%%%%%%%%%%%%%%%%%%%%%%%%%%%%%%%%%%%%%%%%%%%%%%%%%%%%%%%%%%%%%%%%%%%%%%%%%%%
\section{The explicit form of the $r$-mode equations in the $\Omega\to 0$ limit}\label{general eqs}

Using the relativistic $r$-mode ordering, one can derive the $r$-mode equations in the $\Omega\to 0$ limit from the system \eqref{r-modes GR}, retaining only the leading order terms. From the second and third equations of this system we find that $\xi^{(1)}_{l-1}$ is proportional to $\xi^{(1)}_{l+1}$, which allows us to introduce a new useful function
\begin{gather}
\label{tildeXi}
\tilde{\xi}(a)=\frac{\xi^{(1)}_{l-1}(a)}{(l+1)^2k^+_{l+1}}=\frac{\xi^{(1)}_{l+1}(a)}{l^2 k^-_{l}}.
\end{gather}
Then, ignoring small terms in the system \eqref{r-modes GR}, we obtain
\begin{gather}
\label{simple_system}
\left\{
\begin{gathered}
\epsilon[l (l+1) \sigma^{(10)}+2 m \tilde{\omega}(a)]T^{(0)}_l-2 a \gamma _1 l (l+1)\tilde{\xi}'=0 \\
4 a m \epsilon\kappa^2 T^{(0)}_l{}'+A c^2 g l^2 (l+1)^2 e^{2 \nu _0} \tilde{\xi}=0.
\end{gathered}
\right.
\end{gather}
This system is the exact generalization of the system \eqref{formal sys2} to the case of arbitrary $l$ and $m$: if we  consider Eqs.~\eqref{simple_system} with $l=m$, we immediately reproduce the exact form of the coefficients $C_1(a)$, $C_4$ and $G_2(a)$. Once the solution of this system is known, we can use Eqs.~\eqref{Qexcl1}, \eqref{Qexcl2}, \eqref{Texcl1} and  \eqref{Texcl2} to find the functions $Q^{(1)}_{l\pm 1}$ and $T^{(1)}_{l\pm 2}$ in the $\Omega\to 0$ limit:
\begin{gather}
Q^{(1)}_{l-1}=\frac{(l+1)^2  k^{+}_l }{l(l-1)}a\tilde{\xi}', \qquad Q^{(1)}_{l+1}=\frac{l^2  k^-_l }{(l+1)(l+2)}a\tilde{\xi}', \\
T^{(1)}_{l-2}=\frac{i (l-2) l (l+1)^3   k^+_{l-1} k^+_l }{2m(l-1) (2 l-1)}a\tilde{\xi}', \qquad T^{(1)}_{l+2}=-\frac{i l^3 (l+1) (l+3)   k^-_l k^-_{l+1} }{2m(l+2) (2 l+3)}a\tilde{\xi}'.
\end{gather}
It is easy to see that these formulas define the contributions $Q^{(10)}_{l\pm 1}$ and $T^{(10)}_{l\pm 2}$ from the decompositions \eqref{Q1ordering} and \eqref{T1ordering}. Other terms are small in the $\Omega\to 0$ limit and can be ignored.

\section{The relativistic r-mode ordering in the core}\label{ordering appx}

Let us imagine that we are not aware of the relativistic $r$-mode non-analyticity. As we have already mentioned, the small parameter in $r$-mode oscillation equations \eqref{r-modes GR} [or \eqref{r-modes l=m} for $l=m$ case], associated with slow stellar rotation, is $\Omega^2$, not $\Omega$. Then it seems natural to look for the solution to the oscillation equations in the form of a series in this parameter
\begin{gather}
f(a,\Omega)=\sum_n f_{[n]}(a)\Omega^{2n}.
\end{gather}
For example, if we ignore the frame-dragging effect in these equations, we immediately obtain that $\sigma^{(1)}\sim\xi^{(1)}_{l+1}\sim\Omega^2$, which is the traditional $r$-mode ordering, known from the Newtonian theory. In General Relativity, however, we deal with non-analytic functions, for which the leading contribution to the radial displacement $\xi^{(1)}_{l+1}\sim \sqrt{\epsilon}\Omega$ turns out to be of linear order in $\Omega$, and, moreover, the operator $d/da$ should be considered as a ``quantity'' of order $\sqrt{\epsilon}/\Omega$ (recall the discussion in Sec.\ \ref{ordering sec} concerning the validity of terms ``order''and ``series'' when one deals with non-analytic functions). As a result, the terms of linear order in $\Omega$ arise and violate the expected picture: the actual small rotation-associated parameter to be used in $\Omega$-series should be $\Omega$, not $\Omega^2$. This implies that the quantities $T^{(0)}_l(a)$ and $\sigma^{(1)}$ are, generally, allowed to contain linear in $\Omega$ contributions.

It may feel like we have arrived at some internal inconsistency of the theory: in the decomposition $T=T^{(0)}+T^{(1)}$ the function $T^{(0)}=T^{(0)}_l(a)P_{l}^m(\cos\theta)$ that, by definition, should describe the leading order contribution to the toroidal function $T$ in the slow-rotation limit, is allowed to contain a small linear in $\Omega$ term that seemingly should be attributed to $T^{(1)}$. This impression, however, is wrong, since the rotation-associated small parameter, used in the derivation of the general $r$-mode equations, is $\Omega^2$. For slow rotation rates linear in $\Omega$ terms are much larger than $\Omega^2$ and, therefore, can be a part of $T^{(0)}$. Note, however, that it does not necessarily mean that linear in $\Omega$ contributions cannot appear in $T^{(1)}$: they can, but with additional small factors, associated with $\epsilon$.

With all these remarks in mind, let us proceed further with the derivation of the relativistic $r$-mode ordering. As previously, for simplicity we focus on the $l=m$ case, since the analysis of the $l\neq m$ case can be performed in the completely analogous manner. In Sec.\ \ref{ordering sec} we have shown that in the limit of extremely slow rotation the general system of equations
\begin{gather}
\label{formal sys appx}
\left\{
\begin{gathered}
\biggl[C_{1}(a)\frac{d}{d a}+C_{2}(a)\biggr]\xi^{(1)}_{l+1}+\biggl[\Omega^2 C_{3}(a)+\sigma^{(1)}+C_4\epsilon\tilde{\omega}(a)\biggr]T^{(0)}_l=0, \\
\biggl[\frac{d}{da}+G_1(a)\biggr]T^{(0)}_l+\frac{G_2(a)}{\Omega^2}\xi^{(1)}_{l+1}=0,
\end{gathered}
\right.
\end{gather}
reduces to
\begin{gather}
\label{formal sys2 appx}
\left\{
\begin{gathered}
C_{1}(a)\xi^{(1)}_{l+1}{}'+[\sigma^{(1)}+C_4\epsilon\tilde{\omega}(a)]T^{(0)}_l=0 \\
\Omega T^{(0)}_l{}'+\frac{G_2(a)}{\Omega}\xi^{(1)}_{l+1}=0.
\end{gathered}
\right.
\end{gather}
In the latter system $\sigma^{(1)}=\epsilon\sigma^{(10)}$, $T^{(0)}_l=T^{(00)}_l$, and $\xi^{(1)}_{l+1}=\sqrt{\epsilon}\Omega\xi^{(11)}_{l+1}\sim\sqrt{\epsilon}\Omega$ are the leading contributions to, respectively, the $r$-mode eigenfrequency, toroidal function, and radial displacement in this limit, that by definition do not include the generally allowed ``linear'' $\Omega$-order corrections. We have also shown that, because of non-analyticity of the eigenfunctions, the derivative $d/da$ in these equations should be considered as a quantity of order $\sqrt{\epsilon}/\Omega$. For further analysis we introduce a new convenient parameter
\begin{gather}
\kappa=\frac{\Omega}{\sqrt{\epsilon}}
\end{gather}
and look for the solution to the general Eqs.\ \eqref{formal sys appx} in the form
\begin{gather}
\label{decomp appx}
\sigma^{(1)}=\epsilon[\sigma^{(10)}+\delta\sigma], \quad  T^{(0)}_l=T^{(00)}_l+\delta T_l, \quad \xi^{(1)}_{l+1}=\epsilon\kappa[\xi^{(11)}_{l+1}+\delta\xi_{l+1}], \quad \frac{dT^{(00)}_l}{da}\sim \frac{d\xi^{(11)}_{l+1}}{da}\sim \frac{1}{\kappa}.
\end{gather}
Here the first terms correspond to the solution of Eqs.\ \eqref{formal sys2 appx}, while the terms of the form $\delta f$ describe small corrections to this solution of yet unknown order. Note that at this point we do not know, whether the operator $d/da$ changes the order of $\delta T_l$ and $\delta\xi_{l+1}$, like it does with $T^{(00)}_l$ and $\xi^{(11)}_{l+1}$. 

It turns out that when we employ decompositions \eqref{decomp appx} in the general system \eqref{formal sys appx}, the only small parameter left in all equations is $\kappa$. Therefore, the corrections $\delta T_l$, $\delta \xi_{l+1}$ and $\delta\sigma$ should be considered as quantities of non-zero positive $\kappa$-orders (this statement, however, generally does not have to hold for their derivatives, if we allow them to be non-analytic functions of $\kappa$). Having used these decompositions, we can discard small terms in this system, following two selection rules. The first rule is that, in any equation, containing simultaneously the terms $f$ and $\delta f$ (with factors of the same order), the latter can be ignored. The second rule is, that, in each equation, we ignore all the terms smaller than the largest inhomogeneous term in that chosen equation [i.e., the largest term, containing $T^{(00)}_l$ or $\xi^{(11)}$]. For example, in the first equation, we ignore the term $\kappa^2 T^{(00)}_l$ compared to $\kappa \xi^{(11)}$. Eventually, we obtain the following simplified system of equations
\begin{gather}
\left\{
\begin{gathered}
C_1(a)\frac{d}{da}\kappa\delta\xi_{l+1}+[\sigma^{(10)}+C_4\tilde{\omega}(a)]\delta T_l+C_2(a)\kappa\xi^{(11)}_{l+1}+\delta\sigma\cdot T^{(00)}_l=0 \\
\frac{d}{da}\delta T_l+\frac{G_2(a)}{\kappa}\delta\xi_{l+1}+G_1(a)T^{(00)}_l=0.
\end{gathered}
\right.
\end{gather}
This system should be solved with three boundary conditions (two of them imposed at the crust-core interface and one at the stellar center), therefore, its general solution should contain three unknown constants: two integration constants and the correction $\delta\sigma$. This is possible, if and only if the terms with the derivatives of $\delta T_l$ and $\delta\xi_{l+1}$ are not small compared to other terms in the obtained system: otherwise differential equations reduce to algebraic ones, and we are left with only one unknown constant $\delta\sigma$. For the same reason, the term with $\delta\sigma$ in the first equation should be of the same order as the term with the derivative $d\delta\xi_{l+1}/da$. Finally, the order of the correction $\delta T_l$ cannot be larger than $\kappa$, otherwise, $\delta T_l$ should have been attributed to $T^{(1)}$. If we look for the ordering in the form
\begin{gather}
\delta\sigma\sim \kappa^{s}, \quad \delta T_{l}\sim\kappa^{t}, \quad \delta \xi_{l+1}\sim \kappa^x, \quad \frac{d}{da}\sim \kappa^{d},
\end{gather}
the discussed conditions can be formally written as
\begin{gather}
\left\{
\begin{gathered}
1+d+x=s \\
s\leq t
\end{gathered}
\right.
\end{gather}
for the first equation, and
\begin{gather}
\text{\underline{option 1:}}\quad\left\{
\begin{gathered}
d+t=0 \\
x\geq 1, 
\end{gathered}
\right.
\qquad\text{or}\qquad
\text{\underline{option 2:}}\quad
\left\{
\begin{gathered}
d+t=x-1 \\
x\leq 1, 
\end{gathered}
\right.
\end{gather}
for the second equation. Let us consider, for example, the first option. In this case, using the conditions above, we easily obtain the inequallity $2t\geq x+1$. Since $t\leq 1$ and $x\geq 1$, this inequality is satisfied only for $x=t=1$, and we immediately obtain $s=1$ and $d=-1$. The analysis of the second option can be performed in a similar way and leads to the same ordering. 

Thus, we have shown that $\delta\sigma\sim\delta T_l\sim\delta \xi_{l+1}\sim\kappa$, and that corrections $\delta T_l$ and $\delta\xi_{l+1}$ are non-analytic functions of $\kappa$, for which $d/da\sim 1/\kappa$. Now, we can write down the next order terms in the relativistic $r$-mode ordering. First of all, we have
\begin{gather}
\delta\sigma=\kappa \sigma^{(11)}, \qquad \sigma^{(1)}=\epsilon \sigma^{(10)}+\epsilon\kappa\sigma^{(11)} \\
\delta T_l=\kappa T^{(01)}_l, \qquad T^{(0)}_l=T^{(00)}_l+\kappa T^{(01)}_l, \\
\delta\xi_{l+1}=\kappa \xi^{(12)}_{l+1}, \qquad \xi^{(1)}_{l+1}=\epsilon\kappa\xi^{(11)}_{l+1}+\epsilon\kappa^2\xi^{(12)}_{l+1},
\end{gather}
where $\sigma^{(11)}\sim T^{(01)}_l\sim\xi^{(12)}_{l+1}\sim 1$. Secondly, from Eq.\ \eqref{Qexcl2} it follows that, up to the terms, linear in $\kappa$, we have
\begin{gather}
\label{Q1ordering}
Q^{(1)}_{l+1}=\epsilon Q^{(10)}_{l+1}+\epsilon\kappa Q^{(11)}_{l+1}, \\
\quad Q^{(10)}_{l+1}=\frac{a}{(l+1)(l+2)}\kappa\xi^{(11)}_{l+1}{}', \quad Q^{(11)}_{l+1}=\frac{1}{2(l+1)(l+2)}\biggl[(F-1)\xi^{(11)}_{l+1}+2a\kappa\xi^{(12)}_{l+1}{}'\biggr].
\label{E16}
\end{gather}
Note that the ``order'' of the quantities $\kappa\xi^{(11)}_{l+1}{}'$ and $\kappa\xi^{(12)}_{l+1}{}'$ in Eq.\ (\ref{E16}) is 1, since $d/da\sim 1/\kappa$. Finally, from Eq.\eqref{Texcl2} we find that, up to the terms, ``linear'' in $\kappa$, we have
\begin{gather}
\label{T1ordering}
T^{(1)}_{l+2}=\epsilon T^{(10)}_{l+2}+\epsilon\kappa T^{(11)}_{l+2} , \\
\quad T^{(10)}_{l+2}=-\frac{ik_{l+1}^-(l+1)^2(l+3)}{2(2l+3)}Q^{(10)}_{l+1}, \quad
T^{(11)}_{l+2}=-\frac{ik_{l+1}^-(l+1)(l+3)}{2(2l+3)}\biggl[(ag-1)\xi^{(11)}_{l+1}+(l+1)Q^{(11)}_{l+1}\biggr].
\end{gather}

Although we discuss here only the ordering of the coefficients before associated Legendre polynomials, it is easy to see that the same ordering should hold for the whole $\theta$-dependent functions. Indeed, for the $l=m$ case, we have
\begin{gather}
\xi^{(1)}(a,\theta)=\xi^{(1)}_{l+1}(a)P_{l+1}^l(\cos\theta), \qquad Q^{(1)}(a,\theta)=Q^{(1)}_{l+1}(a)P_{l+1}^l(\cos\theta), \\
T^{(0)}(a,\theta)=T^{(0)}_l(a) P_l^l(\cos\theta), \qquad T^{(1)}(a,\theta)=T^{(1)}_l(a) P_l^l(\cos\theta)+T^{(1)}_{l+2}(a)P_{l+2}^l(\cos\theta).
\end{gather}
We know the ordering of all the coefficients in these decompositions, except for $T^{(1)}_l(a)$ that disappears from the equations.  Nevertheless, we still can conclude that the ordering of the toroidal function up to
the terms ``linear'' in $\kappa$ 
%terms 
is defined by the expressions above. The toroidal function correction $T^{(1)}(a,\theta)$, definitely, contains the terms of the order $\epsilon$ and $\epsilon \kappa$, since the terms of the same order appear in $T^{(1)}_{l+2}(a)$. At the same time, it cannot contain the terms of the order $1$ or $\kappa$, since these terms, by definition, should be attributed to the leading order contribution $T^{(0)}(a,\theta)$. One, however, cannot exclude the possibility that $T^{(1)}_l(a)$ is of the higher order than $T^{(1)}_{l+2}(a)$.

Summarizing, we can write the first terms of the relativistic $r$-mode ordering as
\begin{gather}
\sigma/\Omega=\sigma^{(0)}+\epsilon \sigma^{(10)}+\epsilon\kappa\sigma^{(11)}+\dots, \\
T=T^{(00)}+\kappa T^{(01)}+\epsilon T^{(10)}+\epsilon\kappa T^{(11)}+\dots, \\
Q=\epsilon Q^{(10)}+\epsilon\kappa Q^{(11)}+\dots, \\
\xi^a=\epsilon\kappa\xi^{(11)}+\dots, \qquad \frac{\p}{\p a}\sim\frac{1}{\kappa}.
\end{gather}
These formulas do not correspond to any sort of $\Omega$- or $\epsilon$-series. The terms $f^{(ik)}$ in these decompositions, including those for eigenfrequency corrections, generally, are non-analytic functions of $\Omega$ and $\epsilon$ that define the non-analytic contributions of the ``order'' $\epsilon^i \kappa^k$ (but the dependency of these contributions on $\Omega$ and $\epsilon$  does not reduce to the factor $\epsilon^i\kappa^k$). For example, the function $T^{(11)}$ defines the non-analytic contribution to the toroidal function, which is ``linear'' in $\epsilon$ and ``linear'' in $\kappa$. All the terms of the form $f^{(00)}$ and $f^{(01)}$ correspond to the terms in \eqref{new_approach} that were designated as $f^{(0)}$, and all the terms of the form $f^{(10)}$ and $f^{(11)}$ correspond to $f^{(1)}$.  For instance, $T^{(01)}$ is a part of $T^{(0)}$, whereas $T^{(10)}$ is a part of $T^{(1)}$. As we anticipated, ``linear'' in $\kappa$ terms enter the corrections $f^{(1)}$ with small $\epsilon$-associated factors. 

%%%%%%%%%%%%%%%%%%%%%%%%%%%%%%%%%%%%%%%%%%%%%%%%%%%%%%%%%%%%%%%%%%%%%%%%%%%%%%%%%%%%%%%%%%%%%%%%%%%%%%%%%%%%%%%%%%%%%%%%%%%%%
%%%%%%%%%%%%%%%%%%%%%%%%%%%%%%%%%%%%%%%%%%%%%%%%%%%%%%%%%%%%%%%%%%%%%%%%%%%%%%%%%%%%%%%%%%%%%%%%%%%%%%%%%%%%%%%%%%%%%%%%%%%%%
%%%%%%%%%%%%%%%%%%%%%%%%%%%%%%%%%%%%%%%%%%%%%%%%%%%%%%%%%%%%%%%%%%%%%%%%%%%%%%%%%%%%%%%%%%%%%%%%%%%%%%%%%%%%%%%%%%%%%%%%%%%%%
\section{The effect of crust thickness}\label{small crust appx}

Let us take a look at the obtained analytic solution in the crust. According to Eq.\ \eqref{orderingCrust} we have $\xi^{(1)}_{l+1}(a)\approx\epsilon\xi^{(10)}_{l+1}(a)$, while from the definition \eqref{xi1011} it follows that $\xi^{(10)}_{l+1}(a)\sim T^{(0)}_l(a)$. As a result, in the crust, we expect the condition $\xi^{(1)}_{l+1}(a)\sim \epsilon T^{(0)}_l(a)$ to take place. Although the function $\xi^{(10)}_{l+1}(a)$ for the main harmonic does not depend and for the node-possessing nodes depends only weakly on the small parameters $\Omega$ and $\epsilon$, and this condition seems correct, it in fact turns out to be wrong for the considered values $10^{-3}\leq\Omega\leq 10^{-2}$. To see this we use the formulas \eqref{sigma10 main} and \eqref{xi1011} to rewrite the analytic solution for the main harmonic in the $\Omega\to 0$ limit as
\begin{gather}
\xi^{(1)}_{l+1}(a)\approx\epsilon\xi^{(10)}_{l+1}(a)=-\frac{2\epsilon}{l+1}I_2(a_{cc})\Delta(a),
\end{gather}
where
\begin{gather}
\Delta(a)=\frac{I_1(a)}{I_1(a_{cc})}-\frac{I_2(a)}{I_2(a_{cc})}, \quad I_1(a)=\frac{1}{\eta(a)}\int\limits_a^1 g_{21}(a)\eta(a)T^{(0)}_l(a)da, \quad I_2(a)=\frac{1}{\eta(a)}\int\limits_a^1 g_{21}(a)\eta(a)\tilde{\omega}(a) T^{(0)}_l(a)da.
\end{gather}
Our numerical calculations show that the ratios $I_1(a)/I_1(a_{cc})$ and $I_2(a)/I_2(a_{cc})$ both take positive values in the interval $[0;1]$ and only slightly differ from each other. As a result, the function $\Delta(a)$ is, typically, of the order $4\times 10^{-2}$. We also obtain that $I_2(a_{cc})\approx 3\times 10^{-2}$, which, combined with the previous estimate and the fact that in the crust $T^{(0)}_l(a)\sim 1$, allows one to conclude that $\xi^{(1)}_{l+1}(a)\sim 10^{-3}\times\epsilon T^{(0)}_l(a)$. Even for the smallest considered value, $\Omega=10^{-3}$, this means that, effectively, we have $\xi^{(1)}_{l+1}(a)\sim \epsilon\kappa T^{(0)}_l(a)$, which makes the developed analytic theory for the main harmonic less applicable for the considered range of rotation rates. A similar situation emerges when one considers node-possessing harmonics, for which the radial displacement in the $\Omega\to 0$ limit approximately equals
\begin{gather}
\xi^{(1)}_{l+1}(a)\approx \frac{\epsilon}{\eta(a)}\int\limits_a^1 g_{21}(a)\biggl[\sigma^{(10)}+\frac{2\tilde{\omega}(a)}{l+1}\biggr]\eta(a)T^{(0)}_l(a)da \approx -\frac{2\epsilon}{l+1}\biggl[\tilde{\omega}(a_{cc})I_1(a)-I_2(a)\biggr] \sim 10^{-3} \times\epsilon T^{(0)}_l(a).
\end{gather}
Thus, an application of the theory to the node-possessing $r$-modes also turns out to be marginally justified for the considered rotation rates. We expect, however, that the theory provides more accurate results when smaller values of $\Omega\lesssim 10^{-4}$ are considered, but we cannot perform the corresponding calculations due to numerical problems arising at slower rotation rates.

What we can do to test the theory is to consider a slightly different stellar model with a larger crust. Instead of working with a ``genuine'' star with $a_{cc}\approx 0.92$, we can calculate the $r$-modes for an artificial star with a different (smaller) value of $a_{cc}$, by solving the dynamical equations in the core at $a<a_{cc}$ and in the crust at $a>a_{cc}$ (but employing exactly the same equilibrium stellar model as before). Our numerical calculations show that typical values of the function $\Delta(a)$ only slightly depend on the position of the crust-core interface, whereas the quantities $|I_2(a_{cc})|$ and $\tilde{\omega}(a_{cc})$ get larger and larger with the decrease of $a_{cc}$. The functions $|I_1(a)|$ and $|I_2(a)|$ decrease with $a$, so for stellar models with a larger crust they are also allowed to be larger. We, therefore, expect that the $r$-mode spectrum for stellar models with a larger crust should be better described by the analytic formulas derived in the limit $\Omega\to 0$. Moreover, the function $A(a)$ (in our model) exhibits two extremum points in the region $0.87\leq a\leq 0.9$. Therefore, if $a_t<0.9$, the Taylor expansion $q_\sigma(a)=\alpha^2(a_t-a)$ cannot be applied in the vicinity of the crust-core interface and the oscillation spectrum sholuld be calculated according to the quantization rule \eqref{quantization rule} instead of Eq.\eqref{sigma10 nodes1}. In the stellar models with a large crust ($a_{cc}<0.87$), however, extremum points do not arise and the Taylor expansion of $q_\sigma(a)$ remains accurate up to the crust-core interface.

We consider two such stellar models: one with $a_{cc}=0.7$ and the other with $a_{cc}=0.4$. In the model with $a_{cc}=0.7$ according to our estimates (performed in the same manner as previously for the ``genuine'' star), we have  $\xi^{(1)}_{l+1}(a)\sim 10^{-2}\times\epsilon T^{(0)}_l(a)$ for the main harmonic and $\xi^{(1)}_{l+1}(a)\sim 10^{-1}\times\epsilon T^{(0)}_l(a)$ for the node-possessing harmonics. The theory for this model should work better at rotation rates $\Omega\sim 10^{-3}$, but we still cannot expect exact predictions for sufficiently higher rotation rates $\Omega\sim 10^{-2}$, especially for the main harmonic. In the model with $a_{cc}=0.4$ we have $\xi^{(1)}_{l+1}(a)\sim 4\times 10^{-2}\epsilon T^{(0)}_l(a)$ for the main harmonic and $\xi^{(1)}_{l+1}(a)\sim \epsilon T^{(0)}_l(a)$ for the node-possessing $r$-modes. In this case the theory should with high accuracy describe the eigenfrequencies of the node-possessing $r$-modes even at $\Omega\sim 10^{-2}$. Predictions concerning the main harmonic eigenfrequency should also be significantly more precise, than for the other discussed stellar models. All these conclusions are confirmed by Fig.\ \ref{frequencies}, discussed in Sec.\ \ref{verification sec}.


\begin{thebibliography}{67}%
\makeatletter
\providecommand \@ifxundefined [1]{%
 \@ifx{#1\undefined}
}%
\providecommand \@ifnum [1]{%
 \ifnum #1\expandafter \@firstoftwo
 \else \expandafter \@secondoftwo
 \fi
}%
\providecommand \@ifx [1]{%
 \ifx #1\expandafter \@firstoftwo
 \else \expandafter \@secondoftwo
 \fi
}%
\providecommand \natexlab [1]{#1}%
\providecommand \enquote  [1]{``#1''}%
\providecommand \bibnamefont  [1]{#1}%
\providecommand \bibfnamefont [1]{#1}%
\providecommand \citenamefont [1]{#1}%
\providecommand \href@noop [0]{\@secondoftwo}%
\providecommand \href [0]{\begingroup \@sanitize@url \@href}%
\providecommand \@href[1]{\@@startlink{#1}\@@href}%
\providecommand \@@href[1]{\endgroup#1\@@endlink}%
\providecommand \@sanitize@url [0]{\catcode `\\12\catcode `\$12\catcode
  `\&12\catcode `\#12\catcode `\^12\catcode `\_12\catcode `\%12\relax}%
\providecommand \@@startlink[1]{}%
\providecommand \@@endlink[0]{}%
\providecommand \url  [0]{\begingroup\@sanitize@url \@url }%
\providecommand \@url [1]{\endgroup\@href {#1}{\urlprefix }}%
\providecommand \urlprefix  [0]{URL }%
\providecommand \Eprint [0]{\href }%
\providecommand \doibase [0]{https://doi.org/}%
\providecommand \selectlanguage [0]{\@gobble}%
\providecommand \bibinfo  [0]{\@secondoftwo}%
\providecommand \bibfield  [0]{\@secondoftwo}%
\providecommand \translation [1]{[#1]}%
\providecommand \BibitemOpen [0]{}%
\providecommand \bibitemStop [0]{}%
\providecommand \bibitemNoStop [0]{.\EOS\space}%
\providecommand \EOS [0]{\spacefactor3000\relax}%
\providecommand \BibitemShut  [1]{\csname bibitem#1\endcsname}%
\let\auto@bib@innerbib\@empty
%</preamble>
\bibitem [{\citenamefont {{Chandrasekhar}}(1970)}]{chandra1970}%
  \BibitemOpen
  \bibfield  {author} {\bibinfo {author} {\bibfnamefont {S.}~\bibnamefont
  {{Chandrasekhar}}},\ }\bibfield  {title} {\bibinfo {title} {{Solutions of Two
  Problems in the Theory of Gravitational Radiation}},\ }\href
  {https://doi.org/10.1103/PhysRevLett.24.611} {\bibfield  {journal} {\bibinfo
  {journal} {\prl}\ }\textbf {\bibinfo {volume} {24}},\ \bibinfo {pages} {611}
  (\bibinfo {year} {1970})}\BibitemShut {NoStop}%
\bibitem [{\citenamefont {{Friedman}}\ and\ \citenamefont
  {{Schutz}}(1978{\natexlab{a}})}]{fs1978_1}%
  \BibitemOpen
  \bibfield  {author} {\bibinfo {author} {\bibfnamefont {J.~L.}\ \bibnamefont
  {{Friedman}}}\ and\ \bibinfo {author} {\bibfnamefont {B.~F.}\ \bibnamefont
  {{Schutz}}},\ }\bibfield  {title} {\bibinfo {title} {{Lagrangian perturbation
  theory of nonrelativistic fluids.}},\ }\href {https://doi.org/10.1086/156098}
  {\bibfield  {journal} {\bibinfo  {journal} {\apj}\ }\textbf {\bibinfo
  {volume} {221}},\ \bibinfo {pages} {937} (\bibinfo {year}
  {1978}{\natexlab{a}})}\BibitemShut {NoStop}%
\bibitem [{\citenamefont {{Friedman}}\ and\ \citenamefont
  {{Schutz}}(1978{\natexlab{b}})}]{fs1978_2}%
  \BibitemOpen
  \bibfield  {author} {\bibinfo {author} {\bibfnamefont {J.~L.}\ \bibnamefont
  {{Friedman}}}\ and\ \bibinfo {author} {\bibfnamefont {B.~F.}\ \bibnamefont
  {{Schutz}}},\ }\bibfield  {title} {\bibinfo {title} {{Secular instability of
  rotating Newtonian stars.}},\ }\href {https://doi.org/10.1086/156143}
  {\bibfield  {journal} {\bibinfo  {journal} {\apj}\ }\textbf {\bibinfo
  {volume} {222}},\ \bibinfo {pages} {281} (\bibinfo {year}
  {1978}{\natexlab{b}})}\BibitemShut {NoStop}%
\bibitem [{\citenamefont {{Friedman}}(1978)}]{friedman1978}%
  \BibitemOpen
  \bibfield  {author} {\bibinfo {author} {\bibfnamefont {J.~L.}\ \bibnamefont
  {{Friedman}}},\ }\bibfield  {title} {\bibinfo {title} {{Generic instability
  of rotating relativistic stars}},\ }\href
  {https://doi.org/10.1007/BF01202527} {\bibfield  {journal} {\bibinfo
  {journal} {Communications in Mathematical Physics}\ }\textbf {\bibinfo
  {volume} {62}},\ \bibinfo {pages} {247} (\bibinfo {year} {1978})}\BibitemShut
  {NoStop}%
\bibitem [{\citenamefont {{Andersson}}(1998)}]{andersson1998}%
  \BibitemOpen
  \bibfield  {author} {\bibinfo {author} {\bibfnamefont {N.}~\bibnamefont
  {{Andersson}}},\ }\bibfield  {title} {\bibinfo {title} {{A New Class of
  Unstable Modes of Rotating Relativistic Stars}},\ }\href
  {https://doi.org/10.1086/305919} {\bibfield  {journal} {\bibinfo  {journal}
  {\apj}\ }\textbf {\bibinfo {volume} {502}},\ \bibinfo {pages} {708} (\bibinfo
  {year} {1998})},\ \Eprint {https://arxiv.org/abs/gr-qc/9706075}
  {arXiv:gr-qc/9706075 [gr-qc]} \BibitemShut {NoStop}%
\bibitem [{\citenamefont {{Friedman}}\ and\ \citenamefont
  {{Morsink}}(1998)}]{fm1998}%
  \BibitemOpen
  \bibfield  {author} {\bibinfo {author} {\bibfnamefont {J.~L.}\ \bibnamefont
  {{Friedman}}}\ and\ \bibinfo {author} {\bibfnamefont {S.~M.}\ \bibnamefont
  {{Morsink}}},\ }\bibfield  {title} {\bibinfo {title} {{Axial Instability of
  Rotating Relativistic Stars}},\ }\href {https://doi.org/10.1086/305920}
  {\bibfield  {journal} {\bibinfo  {journal} {\apj}\ }\textbf {\bibinfo
  {volume} {502}},\ \bibinfo {pages} {714} (\bibinfo {year} {1998})},\ \Eprint
  {https://arxiv.org/abs/gr-qc/9706073} {arXiv:gr-qc/9706073 [gr-qc]}
  \BibitemShut {NoStop}%
\bibitem [{\citenamefont {{Cowling}}(1941)}]{cowling1941}%
  \BibitemOpen
  \bibfield  {author} {\bibinfo {author} {\bibfnamefont {T.~G.}\ \bibnamefont
  {{Cowling}}},\ }\bibfield  {title} {\bibinfo {title} {{The non-radial
  oscillations of polytropic stars}},\ }\href
  {https://doi.org/10.1093/mnras/101.8.367} {\bibfield  {journal} {\bibinfo
  {journal} {MNRAS}\ }\textbf {\bibinfo {volume} {101}},\ \bibinfo {pages}
  {367} (\bibinfo {year} {1941})}\BibitemShut {NoStop}%
\bibitem [{\citenamefont {{The LIGO Scientific Collaboration, The Virgo
  Collaboration, and the KAGRA Collaboration}}(2021)}]{LIGO2021}%
  \BibitemOpen
  \bibfield  {author} {\bibinfo {author} {\bibnamefont {{The LIGO Scientific
  Collaboration, The Virgo Collaboration, and the KAGRA Collaboration}}},\
  }\bibfield  {title} {\bibinfo {title} {{Searches for continuous gravitational
  waves from young supernova remnants in the early third observing run of
  Advanced LIGO and Virgo}},\ }\href@noop {} {\bibfield  {journal} {\bibinfo
  {journal} {arXiv e-prints}\ } (\bibinfo {year} {2021})},\ \Eprint
  {https://arxiv.org/abs/2105.11641} {arXiv:2105.11641 [astro-ph.HE]}
  \BibitemShut {NoStop}%
\bibitem [{\citenamefont {{The LIGO Scientific Collaboration}}\ \emph
  {et~al.}(2021)\citenamefont {{The LIGO Scientific Collaboration}},
  \citenamefont {{the Virgo Collaboration}},\ and\ \citenamefont {{the KAGRA
  Collaboration}}}]{LIGO21_2}%
  \BibitemOpen
  \bibfield  {author} {\bibinfo {author} {\bibnamefont {{The LIGO Scientific
  Collaboration}}}, \bibinfo {author} {\bibnamefont {{the Virgo
  Collaboration}}},\ and\ \bibinfo {author} {\bibnamefont {{the KAGRA
  Collaboration}}},\ }\bibfield  {title} {\bibinfo {title} {{Search for
  continuous gravitational waves from 20 accreting millisecond X-ray pulsars in
  O3 LIGO data}},\ }\href@noop {} {\bibfield  {journal} {\bibinfo  {journal}
  {arXiv e-prints}\ ,\ \bibinfo {eid} {arXiv:2109.09255}} (\bibinfo {year}
  {2021})},\ \Eprint {https://arxiv.org/abs/2109.09255} {arXiv:2109.09255
  [astro-ph.HE]} \BibitemShut {NoStop}%
\bibitem [{\citenamefont {{Strohmayer}}\ and\ \citenamefont
  {{Mahmoodifar}}(2014{\natexlab{a}})}]{sm2014_1}%
  \BibitemOpen
  \bibfield  {author} {\bibinfo {author} {\bibfnamefont {T.}~\bibnamefont
  {{Strohmayer}}}\ and\ \bibinfo {author} {\bibfnamefont {S.}~\bibnamefont
  {{Mahmoodifar}}},\ }\bibfield  {title} {\bibinfo {title} {{Discovery of a
  Neutron Star Oscillation Mode During a Superburst}},\ }\href
  {https://doi.org/10.1088/2041-8205/793/2/L38} {\bibfield  {journal} {\bibinfo
   {journal} {ApjL}\ }\textbf {\bibinfo {volume} {793}},\ \bibinfo {eid} {L38}
  (\bibinfo {year} {2014}{\natexlab{a}})},\ \Eprint
  {https://arxiv.org/abs/1409.2847} {arXiv:1409.2847 [astro-ph.HE]}
  \BibitemShut {NoStop}%
\bibitem [{\citenamefont {{Strohmayer}}\ and\ \citenamefont
  {{Mahmoodifar}}(2014{\natexlab{b}})}]{sm2014_2}%
  \BibitemOpen
  \bibfield  {author} {\bibinfo {author} {\bibfnamefont {T.}~\bibnamefont
  {{Strohmayer}}}\ and\ \bibinfo {author} {\bibfnamefont {S.}~\bibnamefont
  {{Mahmoodifar}}},\ }\bibfield  {title} {\bibinfo {title} {{A Non-radial
  Oscillation Mode in an Accreting Millisecond Pulsar?}},\ }\href
  {https://doi.org/10.1088/0004-637X/784/1/72} {\bibfield  {journal} {\bibinfo
  {journal} {\apj}\ }\textbf {\bibinfo {volume} {784}},\ \bibinfo {eid} {72}
  (\bibinfo {year} {2014}{\natexlab{b}})},\ \Eprint
  {https://arxiv.org/abs/1310.5147} {arXiv:1310.5147 [astro-ph.HE]}
  \BibitemShut {NoStop}%
\bibitem [{\citenamefont {{Hall}}\ and\ \citenamefont {{Evans}}(2019)}]{he19}%
  \BibitemOpen
  \bibfield  {author} {\bibinfo {author} {\bibfnamefont {E.~D.}\ \bibnamefont
  {{Hall}}}\ and\ \bibinfo {author} {\bibfnamefont {M.}~\bibnamefont
  {{Evans}}},\ }\bibfield  {title} {\bibinfo {title} {{Metrics for
  next-generation gravitational-wave detectors}},\ }\href
  {https://doi.org/10.1088/1361-6382/ab41d6} {\bibfield  {journal} {\bibinfo
  {journal} {Classical and Quantum Gravity}\ }\textbf {\bibinfo {volume}
  {36}},\ \bibinfo {eid} {225002} (\bibinfo {year} {2019})},\ \Eprint
  {https://arxiv.org/abs/1902.09485} {arXiv:1902.09485 [astro-ph.IM]}
  \BibitemShut {NoStop}%
\bibitem [{\citenamefont {{Maggiore}}\ \emph {et~al.}(2020)\citenamefont
  {{Maggiore}}, \citenamefont {{Van Den Broeck}}, \citenamefont {{Bartolo}},
  \citenamefont {{Belgacem}}, \citenamefont {{Bertacca}}, \citenamefont
  {{Bizouard}}, \citenamefont {{Branchesi}}, \citenamefont {{Clesse}},
  \citenamefont {{Foffa}}, \citenamefont {{Garc{\'\i}a-Bellido}}, \citenamefont
  {{Grimm}}, \citenamefont {{Harms}}, \citenamefont {{Hinderer}}, \citenamefont
  {{Matarrese}}, \citenamefont {{Palomba}}, \citenamefont {{Peloso}},
  \citenamefont {{Ricciardone}},\ and\ \citenamefont {{Sakellariadou}}}]{et20}%
  \BibitemOpen
  \bibfield  {author} {\bibinfo {author} {\bibfnamefont {M.}~\bibnamefont
  {{Maggiore}}}, \bibinfo {author} {\bibfnamefont {C.}~\bibnamefont {{Van Den
  Broeck}}}, \bibinfo {author} {\bibfnamefont {N.}~\bibnamefont {{Bartolo}}},
  \bibinfo {author} {\bibfnamefont {E.}~\bibnamefont {{Belgacem}}}, \bibinfo
  {author} {\bibfnamefont {D.}~\bibnamefont {{Bertacca}}}, \bibinfo {author}
  {\bibfnamefont {M.~A.}\ \bibnamefont {{Bizouard}}}, \bibinfo {author}
  {\bibfnamefont {M.}~\bibnamefont {{Branchesi}}}, \bibinfo {author}
  {\bibfnamefont {S.}~\bibnamefont {{Clesse}}}, \bibinfo {author}
  {\bibfnamefont {S.}~\bibnamefont {{Foffa}}}, \bibinfo {author} {\bibfnamefont
  {J.}~\bibnamefont {{Garc{\'\i}a-Bellido}}}, \bibinfo {author} {\bibfnamefont
  {S.}~\bibnamefont {{Grimm}}}, \bibinfo {author} {\bibfnamefont
  {J.}~\bibnamefont {{Harms}}}, \bibinfo {author} {\bibfnamefont
  {T.}~\bibnamefont {{Hinderer}}}, \bibinfo {author} {\bibfnamefont
  {S.}~\bibnamefont {{Matarrese}}}, \bibinfo {author} {\bibfnamefont
  {C.}~\bibnamefont {{Palomba}}}, \bibinfo {author} {\bibfnamefont
  {M.}~\bibnamefont {{Peloso}}}, \bibinfo {author} {\bibfnamefont
  {A.}~\bibnamefont {{Ricciardone}}},\ and\ \bibinfo {author} {\bibfnamefont
  {M.}~\bibnamefont {{Sakellariadou}}},\ }\bibfield  {title} {\bibinfo {title}
  {{Science case for the Einstein telescope}},\ }\href
  {https://doi.org/10.1088/1475-7516/2020/03/050} {\bibfield  {journal}
  {\bibinfo  {journal} {JCAP}\ }\textbf {\bibinfo {volume} {2020}}\bibfield
  {number} {\bibinfo  {number} { (3)},\ \bibinfo {eid} {050}},\ }\Eprint
  {https://arxiv.org/abs/1912.02622} {arXiv:1912.02622 [astro-ph.CO]}
  \BibitemShut {NoStop}%
\bibitem [{\citenamefont {{Andersson}}\ and\ \citenamefont
  {{Kokkotas}}(2001)}]{ak2001}%
  \BibitemOpen
  \bibfield  {author} {\bibinfo {author} {\bibfnamefont {N.}~\bibnamefont
  {{Andersson}}}\ and\ \bibinfo {author} {\bibfnamefont {K.~D.}\ \bibnamefont
  {{Kokkotas}}},\ }\bibfield  {title} {\bibinfo {title} {{The R-Mode
  Instability in Rotating Neutron Stars}},\ }\href
  {https://doi.org/10.1142/S0218271801001062} {\bibfield  {journal} {\bibinfo
  {journal} {International Journal of Modern Physics D}\ }\textbf {\bibinfo
  {volume} {10}},\ \bibinfo {pages} {381} (\bibinfo {year} {2001})},\ \Eprint
  {https://arxiv.org/abs/gr-qc/0010102} {arXiv:gr-qc/0010102 [gr-qc]}
  \BibitemShut {NoStop}%
\bibitem [{\citenamefont {{Haskell}}(2015)}]{haskell14}%
  \BibitemOpen
  \bibfield  {author} {\bibinfo {author} {\bibfnamefont {B.}~\bibnamefont
  {{Haskell}}},\ }\bibfield  {title} {\bibinfo {title} {{R-modes in neutron
  stars: Theory and observations}},\ }\href
  {https://doi.org/10.1142/S0218301315410074} {\bibfield  {journal} {\bibinfo
  {journal} {International Journal of Modern Physics E}\ }\textbf {\bibinfo
  {volume} {24}},\ \bibinfo {eid} {1541007} (\bibinfo {year} {2015})},\ \Eprint
  {https://arxiv.org/abs/1509.04370} {arXiv:1509.04370 [astro-ph.HE]}
  \BibitemShut {NoStop}%
\bibitem [{\citenamefont {{Kojima}}(1997)}]{kojima1997}%
  \BibitemOpen
  \bibfield  {author} {\bibinfo {author} {\bibfnamefont {Y.}~\bibnamefont
  {{Kojima}}},\ }\bibfield  {title} {\bibinfo {title} {{Chapter 4. The
  Rotational Effects of General Relativity on the Stellar Pulsations}},\ }\href
  {https://doi.org/10.1143/PTPS.128.251} {\bibfield  {journal} {\bibinfo
  {journal} {Progress of Theoretical Physics Supplement}\ }\textbf {\bibinfo
  {volume} {128}},\ \bibinfo {pages} {251} (\bibinfo {year}
  {1997})}\BibitemShut {NoStop}%
\bibitem [{\citenamefont {{Kojima}}(1998)}]{kojima1998}%
  \BibitemOpen
  \bibfield  {author} {\bibinfo {author} {\bibfnamefont {Y.}~\bibnamefont
  {{Kojima}}},\ }\bibfield  {title} {\bibinfo {title} {{Quasi-toroidal
  oscillations in rotating relativistic stars}},\ }\href
  {https://doi.org/10.1046/j.1365-8711.1998.01119.x} {\bibfield  {journal}
  {\bibinfo  {journal} {MNRAS}\ }\textbf {\bibinfo {volume} {293}},\ \bibinfo
  {pages} {49} (\bibinfo {year} {1998})},\ \Eprint
  {https://arxiv.org/abs/gr-qc/9709003} {arXiv:gr-qc/9709003 [gr-qc]}
  \BibitemShut {NoStop}%
\bibitem [{\citenamefont {{Beyer}}\ and\ \citenamefont
  {{Kokkotas}}(1999)}]{bk1999}%
  \BibitemOpen
  \bibfield  {author} {\bibinfo {author} {\bibfnamefont {H.~R.}\ \bibnamefont
  {{Beyer}}}\ and\ \bibinfo {author} {\bibfnamefont {K.~D.}\ \bibnamefont
  {{Kokkotas}}},\ }\bibfield  {title} {\bibinfo {title} {{On the r-mode
  spectrum of relativistic stars}},\ }\href
  {https://doi.org/10.1046/j.1365-8711.1999.02739.x} {\bibfield  {journal}
  {\bibinfo  {journal} {MNRAS}\ }\textbf {\bibinfo {volume} {308}},\ \bibinfo
  {pages} {745} (\bibinfo {year} {1999})},\ \Eprint
  {https://arxiv.org/abs/gr-qc/9903019} {arXiv:gr-qc/9903019 [gr-qc]}
  \BibitemShut {NoStop}%
\bibitem [{\citenamefont {{Kojima}}\ and\ \citenamefont
  {{Hosonuma}}(1999)}]{kh1999}%
  \BibitemOpen
  \bibfield  {author} {\bibinfo {author} {\bibfnamefont {Y.}~\bibnamefont
  {{Kojima}}}\ and\ \bibinfo {author} {\bibfnamefont {M.}~\bibnamefont
  {{Hosonuma}}},\ }\bibfield  {title} {\bibinfo {title} {{The r-Mode
  Oscillations in Relativistic Rotating Stars}},\ }\href
  {https://doi.org/10.1086/307481} {\bibfield  {journal} {\bibinfo  {journal}
  {\apj}\ }\textbf {\bibinfo {volume} {520}},\ \bibinfo {pages} {788} (\bibinfo
  {year} {1999})},\ \Eprint {https://arxiv.org/abs/astro-ph/9903055}
  {arXiv:astro-ph/9903055 [astro-ph]} \BibitemShut {NoStop}%
\bibitem [{\citenamefont {{Kojima}}\ and\ \citenamefont
  {{Hosonuma}}(2000)}]{kh2000}%
  \BibitemOpen
  \bibfield  {author} {\bibinfo {author} {\bibfnamefont {Y.}~\bibnamefont
  {{Kojima}}}\ and\ \bibinfo {author} {\bibfnamefont {M.}~\bibnamefont
  {{Hosonuma}}},\ }\bibfield  {title} {\bibinfo {title} {{Approximate equation
  relevant to axial oscillations on slowly rotating relativistic stars}},\
  }\href {https://doi.org/10.1103/PhysRevD.62.044006} {\bibfield  {journal}
  {\bibinfo  {journal} {\prd}\ }\textbf {\bibinfo {volume} {62}},\ \bibinfo
  {eid} {044006} (\bibinfo {year} {2000})},\ \Eprint
  {https://arxiv.org/abs/gr-qc/0004058} {arXiv:gr-qc/0004058 [gr-qc]}
  \BibitemShut {NoStop}%
\bibitem [{\citenamefont {{Lockitch}}\ \emph {et~al.}(2001)\citenamefont
  {{Lockitch}}, \citenamefont {{Andersson}},\ and\ \citenamefont
  {{Friedman}}}]{laf2001}%
  \BibitemOpen
  \bibfield  {author} {\bibinfo {author} {\bibfnamefont {K.~H.}\ \bibnamefont
  {{Lockitch}}}, \bibinfo {author} {\bibfnamefont {N.}~\bibnamefont
  {{Andersson}}},\ and\ \bibinfo {author} {\bibfnamefont {J.~L.}\ \bibnamefont
  {{Friedman}}},\ }\bibfield  {title} {\bibinfo {title} {{Rotational modes of
  relativistic stars: Analytic results}},\ }\href
  {https://doi.org/10.1103/PhysRevD.63.024019} {\bibfield  {journal} {\bibinfo
  {journal} {\prd}\ }\textbf {\bibinfo {volume} {63}},\ \bibinfo {eid} {024019}
  (\bibinfo {year} {2001})},\ \Eprint {https://arxiv.org/abs/gr-qc/0008019}
  {arXiv:gr-qc/0008019 [gr-qc]} \BibitemShut {NoStop}%
\bibitem [{\citenamefont {{Ruoff}}\ and\ \citenamefont
  {{Kokkotas}}(2001)}]{rk2001}%
  \BibitemOpen
  \bibfield  {author} {\bibinfo {author} {\bibfnamefont {J.}~\bibnamefont
  {{Ruoff}}}\ and\ \bibinfo {author} {\bibfnamefont {K.~D.}\ \bibnamefont
  {{Kokkotas}}},\ }\bibfield  {title} {\bibinfo {title} {{On the r-mode
  spectrum of relativistic stars in the low-frequency approximation}},\ }\href
  {https://doi.org/10.1046/j.1365-8711.2001.04909.x} {\bibfield  {journal}
  {\bibinfo  {journal} {MNRAS}\ }\textbf {\bibinfo {volume} {328}},\ \bibinfo
  {pages} {678} (\bibinfo {year} {2001})},\ \Eprint
  {https://arxiv.org/abs/gr-qc/0101105} {arXiv:gr-qc/0101105 [gr-qc]}
  \BibitemShut {NoStop}%
\bibitem [{\citenamefont {{Yoshida}}\ and\ \citenamefont
  {{Lee}}(2002)}]{yl2002}%
  \BibitemOpen
  \bibfield  {author} {\bibinfo {author} {\bibfnamefont {S.}~\bibnamefont
  {{Yoshida}}}\ and\ \bibinfo {author} {\bibfnamefont {U.}~\bibnamefont
  {{Lee}}},\ }\bibfield  {title} {\bibinfo {title} {{Relativistic r-Modes in
  Slowly Rotating Neutron Stars: Numerical Analysis in the Cowling
  Approximation}},\ }\href {https://doi.org/10.1086/338663} {\bibfield
  {journal} {\bibinfo  {journal} {\apj}\ }\textbf {\bibinfo {volume} {567}},\
  \bibinfo {pages} {1112} (\bibinfo {year} {2002})},\ \Eprint
  {https://arxiv.org/abs/gr-qc/0110038} {arXiv:gr-qc/0110038 [gr-qc]}
  \BibitemShut {NoStop}%
\bibitem [{\citenamefont {{Lockitch}}\ \emph {et~al.}(2003)\citenamefont
  {{Lockitch}}, \citenamefont {{Friedman}},\ and\ \citenamefont
  {{Andersson}}}]{lfa2003}%
  \BibitemOpen
  \bibfield  {author} {\bibinfo {author} {\bibfnamefont {K.~H.}\ \bibnamefont
  {{Lockitch}}}, \bibinfo {author} {\bibfnamefont {J.~L.}\ \bibnamefont
  {{Friedman}}},\ and\ \bibinfo {author} {\bibfnamefont {N.}~\bibnamefont
  {{Andersson}}},\ }\bibfield  {title} {\bibinfo {title} {{Rotational modes of
  relativistic stars: Numerical results}},\ }\href
  {https://doi.org/10.1103/PhysRevD.68.124010} {\bibfield  {journal} {\bibinfo
  {journal} {\prd}\ }\textbf {\bibinfo {volume} {68}},\ \bibinfo {eid} {124010}
  (\bibinfo {year} {2003})},\ \Eprint {https://arxiv.org/abs/gr-qc/0210102}
  {arXiv:gr-qc/0210102 [gr-qc]} \BibitemShut {NoStop}%
\bibitem [{\citenamefont {{Yoshida}}\ \emph {et~al.}(2005)\citenamefont
  {{Yoshida}}, \citenamefont {{Yoshida}},\ and\ \citenamefont
  {{Eriguchi}}}]{yoshidaetal2005}%
  \BibitemOpen
  \bibfield  {author} {\bibinfo {author} {\bibfnamefont {S.}~\bibnamefont
  {{Yoshida}}}, \bibinfo {author} {\bibfnamefont {S.}~\bibnamefont
  {{Yoshida}}},\ and\ \bibinfo {author} {\bibfnamefont {Y.}~\bibnamefont
  {{Eriguchi}}},\ }\bibfield  {title} {\bibinfo {title} {{R-mode oscillations
  of rapidly rotating barotropic stars in general relativity: analysis by the
  relativistic Cowling approximation}},\ }\href
  {https://doi.org/10.1111/j.1365-2966.2004.08436.x} {\bibfield  {journal}
  {\bibinfo  {journal} {MNRAS}\ }\textbf {\bibinfo {volume} {356}},\ \bibinfo
  {pages} {217} (\bibinfo {year} {2005})},\ \Eprint
  {https://arxiv.org/abs/astro-ph/0406283} {arXiv:astro-ph/0406283 [astro-ph]}
  \BibitemShut {NoStop}%
\bibitem [{\citenamefont {{Gaertig}}\ and\ \citenamefont
  {{Kokkotas}}(2008)}]{gk2008}%
  \BibitemOpen
  \bibfield  {author} {\bibinfo {author} {\bibfnamefont {E.}~\bibnamefont
  {{Gaertig}}}\ and\ \bibinfo {author} {\bibfnamefont {K.~D.}\ \bibnamefont
  {{Kokkotas}}},\ }\bibfield  {title} {\bibinfo {title} {{Oscillations of
  rapidly rotating relativistic stars}},\ }\href
  {https://doi.org/10.1103/PhysRevD.78.064063} {\bibfield  {journal} {\bibinfo
  {journal} {\prd}\ }\textbf {\bibinfo {volume} {78}},\ \bibinfo {eid} {064063}
  (\bibinfo {year} {2008})},\ \Eprint {https://arxiv.org/abs/0809.0629}
  {arXiv:0809.0629 [gr-qc]} \BibitemShut {NoStop}%
\bibitem [{\citenamefont {{Idrisy}}\ \emph {et~al.}(2015)\citenamefont
  {{Idrisy}}, \citenamefont {{Owen}},\ and\ \citenamefont
  {{Jones}}}]{idrisyetal2015}%
  \BibitemOpen
  \bibfield  {author} {\bibinfo {author} {\bibfnamefont {A.}~\bibnamefont
  {{Idrisy}}}, \bibinfo {author} {\bibfnamefont {B.~J.}\ \bibnamefont
  {{Owen}}},\ and\ \bibinfo {author} {\bibfnamefont {D.~I.}\ \bibnamefont
  {{Jones}}},\ }\bibfield  {title} {\bibinfo {title} {{R -mode frequencies of
  slowly rotating relativistic neutron stars with realistic equations of
  state}},\ }\href {https://doi.org/10.1103/PhysRevD.91.024001} {\bibfield
  {journal} {\bibinfo  {journal} {\prd}\ }\textbf {\bibinfo {volume} {91}},\
  \bibinfo {eid} {024001} (\bibinfo {year} {2015})},\ \Eprint
  {https://arxiv.org/abs/1410.7360} {arXiv:1410.7360 [gr-qc]} \BibitemShut
  {NoStop}%
\bibitem [{\citenamefont {{Yoshida}}\ and\ \citenamefont
  {{Futamase}}(2001)}]{yf2001}%
  \BibitemOpen
  \bibfield  {author} {\bibinfo {author} {\bibfnamefont {S.}~\bibnamefont
  {{Yoshida}}}\ and\ \bibinfo {author} {\bibfnamefont {T.}~\bibnamefont
  {{Futamase}}},\ }\bibfield  {title} {\bibinfo {title} {{R-mode instability of
  slowly rotating nonisentropic relativistic stars}},\ }\href
  {https://doi.org/10.1103/PhysRevD.64.123001} {\bibfield  {journal} {\bibinfo
  {journal} {\prd}\ }\textbf {\bibinfo {volume} {64}},\ \bibinfo {pages}
  {123001} (\bibinfo {year} {2001})},\ \Eprint
  {https://arxiv.org/abs/gr-qc/0106076} {arXiv:gr-qc/0106076 [gr-qc]}
  \BibitemShut {NoStop}%
\bibitem [{\citenamefont {{Ruoff}}\ and\ \citenamefont
  {{Kokkotas}}(2002)}]{rk2002}%
  \BibitemOpen
  \bibfield  {author} {\bibinfo {author} {\bibfnamefont {J.}~\bibnamefont
  {{Ruoff}}}\ and\ \bibinfo {author} {\bibfnamefont {K.~D.}\ \bibnamefont
  {{Kokkotas}}},\ }\bibfield  {title} {\bibinfo {title} {{On the r-mode
  spectrum of relativistic stars: the inclusion of the radiation reaction}},\
  }\href {https://doi.org/10.1046/j.1365-8711.2002.05169.x} {\bibfield
  {journal} {\bibinfo  {journal} {MNRAS}\ }\textbf {\bibinfo {volume} {330}},\
  \bibinfo {pages} {1027} (\bibinfo {year} {2002})},\ \Eprint
  {https://arxiv.org/abs/gr-qc/0106073} {arXiv:gr-qc/0106073 [gr-qc]}
  \BibitemShut {NoStop}%
\bibitem [{\citenamefont {{Lockitch}}\ \emph {et~al.}(2004)\citenamefont
  {{Lockitch}}, \citenamefont {{Andersson}},\ and\ \citenamefont
  {{Watts}}}]{law2004}%
  \BibitemOpen
  \bibfield  {author} {\bibinfo {author} {\bibfnamefont {K.~H.}\ \bibnamefont
  {{Lockitch}}}, \bibinfo {author} {\bibfnamefont {N.}~\bibnamefont
  {{Andersson}}},\ and\ \bibinfo {author} {\bibfnamefont {A.~L.}\ \bibnamefont
  {{Watts}}},\ }\bibfield  {title} {\bibinfo {title} {{Regularizing the r-mode
  problem for nonbarotropic relativistic stars}},\ }\href
  {https://doi.org/10.1088/0264-9381/21/19/012} {\bibfield  {journal} {\bibinfo
   {journal} {Classical and Quantum Gravity}\ }\textbf {\bibinfo {volume}
  {21}},\ \bibinfo {pages} {4661} (\bibinfo {year} {2004})},\ \Eprint
  {https://arxiv.org/abs/gr-qc/0106088} {arXiv:gr-qc/0106088 [gr-qc]}
  \BibitemShut {NoStop}%
\bibitem [{\citenamefont {{Pons}}\ \emph {et~al.}(2005)\citenamefont {{Pons}},
  \citenamefont {{Gualtieri}}, \citenamefont {{Miralles}},\ and\ \citenamefont
  {{Ferrari}}}]{ponsetal2005}%
  \BibitemOpen
  \bibfield  {author} {\bibinfo {author} {\bibfnamefont {J.~A.}\ \bibnamefont
  {{Pons}}}, \bibinfo {author} {\bibfnamefont {L.}~\bibnamefont {{Gualtieri}}},
  \bibinfo {author} {\bibfnamefont {J.~A.}\ \bibnamefont {{Miralles}}},\ and\
  \bibinfo {author} {\bibfnamefont {V.}~\bibnamefont {{Ferrari}}},\ }\bibfield
  {title} {\bibinfo {title} {{Relativistic r modes and shear viscosity:
  regularizing the continuous spectrum}},\ }\href
  {https://doi.org/10.1111/j.1365-2966.2005.09429.x} {\bibfield  {journal}
  {\bibinfo  {journal} {MNRAS}\ }\textbf {\bibinfo {volume} {363}},\ \bibinfo
  {pages} {121} (\bibinfo {year} {2005})},\ \Eprint
  {https://arxiv.org/abs/astro-ph/0504062} {arXiv:astro-ph/0504062 [astro-ph]}
  \BibitemShut {NoStop}%
\bibitem [{\citenamefont {{Gualtieri}}\ \emph {et~al.}(2006)\citenamefont
  {{Gualtieri}}, \citenamefont {{Pons}}, \citenamefont {{Miralles}},\ and\
  \citenamefont {{Ferrari}}}]{gualtierietal2006}%
  \BibitemOpen
  \bibfield  {author} {\bibinfo {author} {\bibfnamefont {L.}~\bibnamefont
  {{Gualtieri}}}, \bibinfo {author} {\bibfnamefont {J.~A.}\ \bibnamefont
  {{Pons}}}, \bibinfo {author} {\bibfnamefont {J.~A.}\ \bibnamefont
  {{Miralles}}},\ and\ \bibinfo {author} {\bibfnamefont {V.}~\bibnamefont
  {{Ferrari}}},\ }\bibfield  {title} {\bibinfo {title} {{Relativistic r-modes
  and shear viscosity}},\ }in\ \href {https://doi.org/10.1063/1.2399636} {\emph
  {\bibinfo {booktitle} {Albert Einstein Century International Conference}}},\
  \bibinfo {series} {American Institute of Physics Conference Series}, Vol.\
  \bibinfo {volume} {861},\ \bibinfo {editor} {edited by\ \bibinfo {editor}
  {\bibfnamefont {J.-M.}\ \bibnamefont {{Alimi}}}\ and\ \bibinfo {editor}
  {\bibfnamefont {A.}~\bibnamefont {{F{\"u}zfa}}}}\ (\bibinfo {year} {2006})\
  pp.\ \bibinfo {pages} {638--645},\ \Eprint
  {https://arxiv.org/abs/gr-qc/0702040} {arXiv:gr-qc/0702040 [gr-qc]}
  \BibitemShut {NoStop}%
\bibitem [{\citenamefont {{Sotani}}\ \emph {et~al.}(2016)\citenamefont
  {{Sotani}}, \citenamefont {{Iida}},\ and\ \citenamefont
  {{Oyamatsu}}}]{sio16}%
  \BibitemOpen
  \bibfield  {author} {\bibinfo {author} {\bibfnamefont {H.}~\bibnamefont
  {{Sotani}}}, \bibinfo {author} {\bibfnamefont {K.}~\bibnamefont {{Iida}}},\
  and\ \bibinfo {author} {\bibfnamefont {K.}~\bibnamefont {{Oyamatsu}}},\
  }\bibfield  {title} {\bibinfo {title} {{Possible identifications of newly
  observed magnetar quasi-periodic oscillations as crustal shear modes}},\
  }\href {https://doi.org/10.1016/j.newast.2015.08.003} {\bibfield  {journal}
  {\bibinfo  {journal} {New Astron.}\ }\textbf {\bibinfo {volume} {43}},\
  \bibinfo {pages} {80} (\bibinfo {year} {2016})},\ \Eprint
  {https://arxiv.org/abs/1508.01728} {arXiv:1508.01728 [astro-ph.HE]}
  \BibitemShut {NoStop}%
\bibitem [{\citenamefont {{Tews}}(2017)}]{tews17}%
  \BibitemOpen
  \bibfield  {author} {\bibinfo {author} {\bibfnamefont {I.}~\bibnamefont
  {{Tews}}},\ }\bibfield  {title} {\bibinfo {title} {{Spectrum of shear modes
  in the neutron-star crust: Estimating the nuclear-physics uncertainties}},\
  }\href {https://doi.org/10.1103/PhysRevC.95.015803} {\bibfield  {journal}
  {\bibinfo  {journal} {\prc}\ }\textbf {\bibinfo {volume} {95}},\ \bibinfo
  {eid} {015803} (\bibinfo {year} {2017})},\ \Eprint
  {https://arxiv.org/abs/1607.06998} {arXiv:1607.06998 [nucl-th]} \BibitemShut
  {NoStop}%
\bibitem [{\citenamefont {{Dommes}}\ \emph {et~al.}(2020)\citenamefont
  {{Dommes}}, \citenamefont {{Gusakov}},\ and\ \citenamefont
  {{Shternin}}}]{dommes2020}%
  \BibitemOpen
  \bibfield  {author} {\bibinfo {author} {\bibfnamefont {V.~A.}\ \bibnamefont
  {{Dommes}}}, \bibinfo {author} {\bibfnamefont {M.~E.}\ \bibnamefont
  {{Gusakov}}},\ and\ \bibinfo {author} {\bibfnamefont {P.~S.}\ \bibnamefont
  {{Shternin}}},\ }\bibfield  {title} {\bibinfo {title} {{Dissipative
  relativistic magnetohydrodynamics of a multicomponent mixture and its
  application to neutron stars}},\ }\href
  {https://doi.org/10.1103/PhysRevD.101.103020} {\bibfield  {journal} {\bibinfo
   {journal} {\prd}\ }\textbf {\bibinfo {volume} {101}},\ \bibinfo {eid}
  {103020} (\bibinfo {year} {2020})},\ \Eprint
  {https://arxiv.org/abs/2006.09840} {arXiv:2006.09840 [astro-ph.HE]}
  \BibitemShut {NoStop}%
\bibitem [{\citenamefont {Ding}\ \emph {et~al.}(2016)\citenamefont {Ding},
  \citenamefont {Rios}, \citenamefont {Dussan}, \citenamefont {Dickhoff},
  \citenamefont {Witte}, \citenamefont {Carbone},\ and\ \citenamefont
  {Polls}}]{drddwcp16}%
  \BibitemOpen
  \bibfield  {author} {\bibinfo {author} {\bibfnamefont {D.}~\bibnamefont
  {Ding}}, \bibinfo {author} {\bibfnamefont {A.}~\bibnamefont {Rios}}, \bibinfo
  {author} {\bibfnamefont {H.}~\bibnamefont {Dussan}}, \bibinfo {author}
  {\bibfnamefont {W.~H.}\ \bibnamefont {Dickhoff}}, \bibinfo {author}
  {\bibfnamefont {S.~J.}\ \bibnamefont {Witte}}, \bibinfo {author}
  {\bibfnamefont {A.}~\bibnamefont {Carbone}},\ and\ \bibinfo {author}
  {\bibfnamefont {A.}~\bibnamefont {Polls}},\ }\bibfield  {title} {\bibinfo
  {title} {Pairing in high-density neutron matter including short- and
  long-range correlations},\ }\href
  {https://doi.org/10.1103/PhysRevC.94.025802} {\bibfield  {journal} {\bibinfo
  {journal} {Phys. Rev. C}\ }\textbf {\bibinfo {volume} {94}},\ \bibinfo
  {pages} {025802} (\bibinfo {year} {2016})}\BibitemShut {NoStop}%
\bibitem [{\citenamefont {{Kantor}}\ and\ \citenamefont
  {{Gusakov}}(2014)}]{kg14}%
  \BibitemOpen
  \bibfield  {author} {\bibinfo {author} {\bibfnamefont {E.~M.}\ \bibnamefont
  {{Kantor}}}\ and\ \bibinfo {author} {\bibfnamefont {M.~E.}\ \bibnamefont
  {{Gusakov}}},\ }\bibfield  {title} {\bibinfo {title} {{Composition
  temperature-dependent g modes in superfluid neutron stars.}},\ }\href
  {https://doi.org/10.1093/mnrasl/slu061} {\bibfield  {journal} {\bibinfo
  {journal} {MNRAS}\ }\textbf {\bibinfo {volume} {442}},\ \bibinfo {pages}
  {L90} (\bibinfo {year} {2014})},\ \Eprint {https://arxiv.org/abs/1404.6768}
  {arXiv:1404.6768 [astro-ph.SR]} \BibitemShut {NoStop}%
\bibitem [{\citenamefont {{Finn}}(1987)}]{finn87}%
  \BibitemOpen
  \bibfield  {author} {\bibinfo {author} {\bibfnamefont {L.~S.}\ \bibnamefont
  {{Finn}}},\ }\bibfield  {title} {\bibinfo {title} {{G-modes in
  zero-temperature neutron stars}},\ }\href
  {https://doi.org/10.1093/mnras/227.2.265} {\bibfield  {journal} {\bibinfo
  {journal} {MNRAS}\ }\textbf {\bibinfo {volume} {227}},\ \bibinfo {pages}
  {265} (\bibinfo {year} {1987})}\BibitemShut {NoStop}%
\bibitem [{\citenamefont {{Hartle}}(1967)}]{hartle1967}%
  \BibitemOpen
  \bibfield  {author} {\bibinfo {author} {\bibfnamefont {J.~B.}\ \bibnamefont
  {{Hartle}}},\ }\bibfield  {title} {\bibinfo {title} {{Slowly Rotating
  Relativistic Stars. I. Equations of Structure}},\ }\href
  {https://doi.org/10.1086/149400} {\bibfield  {journal} {\bibinfo  {journal}
  {\apj}\ }\textbf {\bibinfo {volume} {150}},\ \bibinfo {pages} {1005}
  (\bibinfo {year} {1967})}\BibitemShut {NoStop}%
\bibitem [{\citenamefont {{Tassoul}}(1978)}]{tassoul1978}%
  \BibitemOpen
  \bibfield  {author} {\bibinfo {author} {\bibfnamefont {J.-L.}\ \bibnamefont
  {{Tassoul}}},\ }\href@noop {} {\emph {\bibinfo {title} {{Theory of rotating
  stars}}}}\ (\bibinfo {year} {1978})\BibitemShut {NoStop}%
\bibitem [{\citenamefont {{Hartle}}\ and\ \citenamefont
  {{Thorne}}(1968)}]{hartle1968}%
  \BibitemOpen
  \bibfield  {author} {\bibinfo {author} {\bibfnamefont {J.~B.}\ \bibnamefont
  {{Hartle}}}\ and\ \bibinfo {author} {\bibfnamefont {K.~S.}\ \bibnamefont
  {{Thorne}}},\ }\bibfield  {title} {\bibinfo {title} {{Slowly Rotating
  Relativistic Stars. II. Models for Neutron Stars and Supermassive Stars}},\
  }\href {https://doi.org/10.1086/149707} {\bibfield  {journal} {\bibinfo
  {journal} {\apj}\ }\textbf {\bibinfo {volume} {153}},\ \bibinfo {pages} {807}
  (\bibinfo {year} {1968})}\BibitemShut {NoStop}%
\bibitem [{\citenamefont {{Chandrasekhar}}\ and\ \citenamefont
  {{Roberts}}(1963)}]{cr1963}%
  \BibitemOpen
  \bibfield  {author} {\bibinfo {author} {\bibfnamefont {S.}~\bibnamefont
  {{Chandrasekhar}}}\ and\ \bibinfo {author} {\bibfnamefont {P.~H.}\
  \bibnamefont {{Roberts}}},\ }\bibfield  {title} {\bibinfo {title} {{The
  Ellipticity of a Slowly Rotating Configuration.}},\ }\href
  {https://doi.org/10.1086/147686} {\bibfield  {journal} {\bibinfo  {journal}
  {\apj}\ }\textbf {\bibinfo {volume} {138}},\ \bibinfo {pages} {801} (\bibinfo
  {year} {1963})}\BibitemShut {NoStop}%
\bibitem [{\citenamefont {{Friedman}}\ and\ \citenamefont
  {{Schutz}}(1978{\natexlab{c}})}]{fs1978}%
  \BibitemOpen
  \bibfield  {author} {\bibinfo {author} {\bibfnamefont {J.~L.}\ \bibnamefont
  {{Friedman}}}\ and\ \bibinfo {author} {\bibfnamefont {B.~F.}\ \bibnamefont
  {{Schutz}}},\ }\bibfield  {title} {\bibinfo {title} {{Lagrangian perturbation
  theory of nonrelativistic fluids.}},\ }\href {https://doi.org/10.1086/156098}
  {\bibfield  {journal} {\bibinfo  {journal} {\apj}\ }\textbf {\bibinfo
  {volume} {221}},\ \bibinfo {pages} {937} (\bibinfo {year}
  {1978}{\natexlab{c}})}\BibitemShut {NoStop}%
\bibitem [{\citenamefont {{Unno}}\ \emph {et~al.}(1979)\citenamefont {{Unno}},
  \citenamefont {{Osaki}}, \citenamefont {{Ando}},\ and\ \citenamefont
  {{Shibahashi}}}]{unno1979}%
  \BibitemOpen
  \bibfield  {author} {\bibinfo {author} {\bibfnamefont {W.}~\bibnamefont
  {{Unno}}}, \bibinfo {author} {\bibfnamefont {Y.}~\bibnamefont {{Osaki}}},
  \bibinfo {author} {\bibfnamefont {H.}~\bibnamefont {{Ando}}},\ and\ \bibinfo
  {author} {\bibfnamefont {H.}~\bibnamefont {{Shibahashi}}},\ }\href@noop {}
  {\emph {\bibinfo {title} {{Nonradial oscillations of stars}}}}\ (\bibinfo
  {year} {1979})\BibitemShut {NoStop}%
\bibitem [{\citenamefont {{Cox}}(1980)}]{cox1980}%
  \BibitemOpen
  \bibfield  {author} {\bibinfo {author} {\bibfnamefont {J.~P.}\ \bibnamefont
  {{Cox}}},\ }\href@noop {} {\emph {\bibinfo {title} {{Theory of stellar
  pulsation}}}}\ (\bibinfo {year} {1980})\BibitemShut {NoStop}%
\bibitem [{\citenamefont {{Provost}}\ \emph {et~al.}(1981)\citenamefont
  {{Provost}}, \citenamefont {{Berthomieu}},\ and\ \citenamefont
  {{Rocca}}}]{provost1981}%
  \BibitemOpen
  \bibfield  {author} {\bibinfo {author} {\bibfnamefont {J.}~\bibnamefont
  {{Provost}}}, \bibinfo {author} {\bibfnamefont {G.}~\bibnamefont
  {{Berthomieu}}},\ and\ \bibinfo {author} {\bibfnamefont {A.}~\bibnamefont
  {{Rocca}}},\ }\bibfield  {title} {\bibinfo {title} {{Low Frequency
  Oscillations of a Slowly Rotating Star - Quasi Toroidal Modes}},\ }\href@noop
  {} {\bibfield  {journal} {\bibinfo  {journal} {A\& A}\ }\textbf {\bibinfo
  {volume} {94}},\ \bibinfo {pages} {126} (\bibinfo {year} {1981})}\BibitemShut
  {NoStop}%
\bibitem [{\citenamefont {{Saio}}(1982)}]{saio1982}%
  \BibitemOpen
  \bibfield  {author} {\bibinfo {author} {\bibfnamefont {H.}~\bibnamefont
  {{Saio}}},\ }\bibfield  {title} {\bibinfo {title} {{R-mode oscillations in
  uniformly rotating stars}},\ }\href {https://doi.org/10.1086/159945}
  {\bibfield  {journal} {\bibinfo  {journal} {\apj}\ }\textbf {\bibinfo
  {volume} {256}},\ \bibinfo {pages} {717} (\bibinfo {year}
  {1982})}\BibitemShut {NoStop}%
\bibitem [{\citenamefont {Barrera}\ \emph {et~al.}(1985)\citenamefont
  {Barrera}, \citenamefont {Estevez},\ and\ \citenamefont
  {Giraldo}}]{barrera1985}%
  \BibitemOpen
  \bibfield  {author} {\bibinfo {author} {\bibfnamefont {R.~G.}\ \bibnamefont
  {Barrera}}, \bibinfo {author} {\bibfnamefont {G.~A.}\ \bibnamefont
  {Estevez}},\ and\ \bibinfo {author} {\bibfnamefont {J.}~\bibnamefont
  {Giraldo}},\ }\bibfield  {title} {\bibinfo {title} {Vector spherical
  harmonics and their application to magnetostatics},\ }\href
  {https://doi.org/10.1088/0143-0807/6/4/014} {\bibfield  {journal} {\bibinfo
  {journal} {European Journal of Physics}\ }\textbf {\bibinfo {volume} {6}},\
  \bibinfo {pages} {287} (\bibinfo {year} {1985})}\BibitemShut {NoStop}%
\bibitem [{\citenamefont {Taub}(1969)}]{taub1969}%
  \BibitemOpen
  \bibfield  {author} {\bibinfo {author} {\bibfnamefont {A.~H.}\ \bibnamefont
  {Taub}},\ }\bibfield  {title} {\bibinfo {title} {{Stability of general
  relativistic gaseous masses and variational principles}},\ }\href
  {https://doi.org/cmp/1103841946} {\bibfield  {journal} {\bibinfo  {journal}
  {Communications in Mathematical Physics}\ }\textbf {\bibinfo {volume} {15}},\
  \bibinfo {pages} {235 } (\bibinfo {year} {1969})}\BibitemShut {NoStop}%
\bibitem [{\citenamefont {{Carter}}(1973)}]{carter1973}%
  \BibitemOpen
  \bibfield  {author} {\bibinfo {author} {\bibfnamefont {B.}~\bibnamefont
  {{Carter}}},\ }\bibfield  {title} {\bibinfo {title} {{Elastic perturbation
  theory in General Relativity and a variation principle for a rotating solid
  star}},\ }\href {https://doi.org/10.1007/BF01645505} {\bibfield  {journal}
  {\bibinfo  {journal} {Communications in Mathematical Physics}\ }\textbf
  {\bibinfo {volume} {30}},\ \bibinfo {pages} {261} (\bibinfo {year}
  {1973})}\BibitemShut {NoStop}%
\bibitem [{\citenamefont {{Friedman}}\ and\ \citenamefont
  {{Schutz}}(1975)}]{fs1975}%
  \BibitemOpen
  \bibfield  {author} {\bibinfo {author} {\bibfnamefont {J.~L.}\ \bibnamefont
  {{Friedman}}}\ and\ \bibinfo {author} {\bibfnamefont {B.~F.}\ \bibnamefont
  {{Schutz}}},\ }\bibfield  {title} {\bibinfo {title} {{On the stability of
  relativistic systems.}},\ }\href {https://doi.org/10.1086/153778} {\bibfield
  {journal} {\bibinfo  {journal} {\apj}\ }\textbf {\bibinfo {volume} {200}},\
  \bibinfo {pages} {204} (\bibinfo {year} {1975})}\BibitemShut {NoStop}%
\bibitem [{\citenamefont {{Lindblom}}\ and\ \citenamefont
  {{Splinter}}(1990)}]{lsr1990}%
  \BibitemOpen
  \bibfield  {author} {\bibinfo {author} {\bibfnamefont {L.}~\bibnamefont
  {{Lindblom}}}\ and\ \bibinfo {author} {\bibfnamefont {R.~J.}\ \bibnamefont
  {{Splinter}}},\ }\bibfield  {title} {\bibinfo {title} {{The Accuracy of the
  Relativistic Cowling Approximation}},\ }\href
  {https://doi.org/10.1086/168227} {\bibfield  {journal} {\bibinfo  {journal}
  {\apj}\ }\textbf {\bibinfo {volume} {348}},\ \bibinfo {pages} {198} (\bibinfo
  {year} {1990})}\BibitemShut {NoStop}%
\bibitem [{\citenamefont {{Yoshida}}\ and\ \citenamefont
  {{Kojima}}(1997)}]{yk1997}%
  \BibitemOpen
  \bibfield  {author} {\bibinfo {author} {\bibfnamefont {S.}~\bibnamefont
  {{Yoshida}}}\ and\ \bibinfo {author} {\bibfnamefont {Y.}~\bibnamefont
  {{Kojima}}},\ }\bibfield  {title} {\bibinfo {title} {{Accuracy of the
  relativistic Cowling approximation in slowly rotating stars}},\ }\href
  {https://doi.org/10.1093/mnras/289.1.117} {\bibfield  {journal} {\bibinfo
  {journal} {MNRAS}\ }\textbf {\bibinfo {volume} {289}},\ \bibinfo {pages}
  {117} (\bibinfo {year} {1997})},\ \Eprint
  {https://arxiv.org/abs/gr-qc/9705081} {arXiv:gr-qc/9705081 [gr-qc]}
  \BibitemShut {NoStop}%
\bibitem [{\citenamefont {Jasiulek}\ and\ \citenamefont
  {Chirenti}(2017)}]{jc2017}%
  \BibitemOpen
  \bibfield  {author} {\bibinfo {author} {\bibfnamefont {M.}~\bibnamefont
  {Jasiulek}}\ and\ \bibinfo {author} {\bibfnamefont {C.}~\bibnamefont
  {Chirenti}},\ }\bibfield  {title} {\bibinfo {title} {$r$-mode frequencies of
  rapidly and differentially rotating relativistic neutron stars},\ }\href
  {https://doi.org/10.1103/PhysRevD.95.064060} {\bibfield  {journal} {\bibinfo
  {journal} {Phys. Rev. D}\ }\textbf {\bibinfo {volume} {95}},\ \bibinfo
  {pages} {064060} (\bibinfo {year} {2017})}\BibitemShut {NoStop}%
\bibitem [{\citenamefont {{Regge}}\ and\ \citenamefont
  {{Wheeler}}(1957)}]{rw1957}%
  \BibitemOpen
  \bibfield  {author} {\bibinfo {author} {\bibfnamefont {T.}~\bibnamefont
  {{Regge}}}\ and\ \bibinfo {author} {\bibfnamefont {J.~A.}\ \bibnamefont
  {{Wheeler}}},\ }\bibfield  {title} {\bibinfo {title} {{Stability of a
  Schwarzschild Singularity}},\ }\href
  {https://doi.org/10.1103/PhysRev.108.1063} {\bibfield  {journal} {\bibinfo
  {journal} {Physical Review}\ }\textbf {\bibinfo {volume} {108}},\ \bibinfo
  {pages} {1063} (\bibinfo {year} {1957})}\BibitemShut {NoStop}%
\bibitem [{\citenamefont {{Thorne}}\ and\ \citenamefont
  {{Campolattaro}}(1967)}]{tc1967}%
  \BibitemOpen
  \bibfield  {author} {\bibinfo {author} {\bibfnamefont {K.~S.}\ \bibnamefont
  {{Thorne}}}\ and\ \bibinfo {author} {\bibfnamefont {A.}~\bibnamefont
  {{Campolattaro}}},\ }\bibfield  {title} {\bibinfo {title} {{Non-Radial
  Pulsation of General-Relativistic Stellar Models. I. Analytic Analysis for L
  >= 2}},\ }\href {https://doi.org/10.1086/149288} {\bibfield  {journal}
  {\bibinfo  {journal} {\apj}\ }\textbf {\bibinfo {volume} {149}},\ \bibinfo
  {pages} {591} (\bibinfo {year} {1967})}\BibitemShut {NoStop}%
\bibitem [{\citenamefont {{Yoshida}}(2001)}]{y2001}%
  \BibitemOpen
  \bibfield  {author} {\bibinfo {author} {\bibfnamefont {S.}~\bibnamefont
  {{Yoshida}}},\ }\bibfield  {title} {\bibinfo {title} {{r-Modes of Slowly
  Rotating Nonisentropic Relativistic Stars}},\ }\href
  {https://doi.org/10.1086/322275} {\bibfield  {journal} {\bibinfo  {journal}
  {\apj}\ }\textbf {\bibinfo {volume} {558}},\ \bibinfo {pages} {263} (\bibinfo
  {year} {2001})},\ \Eprint {https://arxiv.org/abs/gr-qc/0101115}
  {arXiv:gr-qc/0101115 [gr-qc]} \BibitemShut {NoStop}%
\bibitem [{\citenamefont {Doneva}\ \emph {et~al.}(2013)\citenamefont {Doneva},
  \citenamefont {Gaertig}, \citenamefont {Kokkotas},\ and\ \citenamefont
  {Kr\"uger}}]{donevaetal2013}%
  \BibitemOpen
  \bibfield  {author} {\bibinfo {author} {\bibfnamefont {D.~D.}\ \bibnamefont
  {Doneva}}, \bibinfo {author} {\bibfnamefont {E.}~\bibnamefont {Gaertig}},
  \bibinfo {author} {\bibfnamefont {K.~D.}\ \bibnamefont {Kokkotas}},\ and\
  \bibinfo {author} {\bibfnamefont {C.}~\bibnamefont {Kr\"uger}},\ }\bibfield
  {title} {\bibinfo {title} {Gravitational wave asteroseismology of fast
  rotating neutron stars with realistic equations of state},\ }\href
  {https://doi.org/10.1103/PhysRevD.88.044052} {\bibfield  {journal} {\bibinfo
  {journal} {Phys. Rev. D}\ }\textbf {\bibinfo {volume} {88}},\ \bibinfo
  {pages} {044052} (\bibinfo {year} {2013})}\BibitemShut {NoStop}%
\bibitem [{\citenamefont {{Kantor}}\ \emph {et~al.}(2021)\citenamefont
  {{Kantor}}, \citenamefont {{Gusakov}},\ and\ \citenamefont
  {{Dommes}}}]{kgd2021}%
  \BibitemOpen
  \bibfield  {author} {\bibinfo {author} {\bibfnamefont {E.~M.}\ \bibnamefont
  {{Kantor}}}, \bibinfo {author} {\bibfnamefont {M.~E.}\ \bibnamefont
  {{Gusakov}}},\ and\ \bibinfo {author} {\bibfnamefont {V.~A.}\ \bibnamefont
  {{Dommes}}},\ }\bibfield  {title} {\bibinfo {title} {{Resonance suppression
  of the r -mode instability in superfluid neutron stars: Accounting for muons
  and entrainment}},\ }\href {https://doi.org/10.1103/PhysRevD.103.023013}
  {\bibfield  {journal} {\bibinfo  {journal} {\prd}\ }\textbf {\bibinfo
  {volume} {103}},\ \bibinfo {eid} {023013} (\bibinfo {year} {2021})},\ \Eprint
  {https://arxiv.org/abs/2102.02716} {arXiv:2102.02716 [astro-ph.HE]}
  \BibitemShut {NoStop}%
\bibitem [{\citenamefont {{Andreev}}\ and\ \citenamefont
  {{Bashkin}}(1975)}]{ab1975}%
  \BibitemOpen
  \bibfield  {author} {\bibinfo {author} {\bibfnamefont {A.~F.}\ \bibnamefont
  {{Andreev}}}\ and\ \bibinfo {author} {\bibfnamefont {E.~P.}\ \bibnamefont
  {{Bashkin}}},\ }\bibfield  {title} {\bibinfo {title} {{Three-velocity
  hydrodynamics of superfluid solutions}},\ }\href@noop {} {\bibfield
  {journal} {\bibinfo  {journal} {Soviet Journal of Experimental and
  Theoretical Physics}\ }\textbf {\bibinfo {volume} {42}},\ \bibinfo {pages}
  {164} (\bibinfo {year} {1975})}\BibitemShut {NoStop}%
\bibitem [{\citenamefont {{Borumand}}\ \emph {et~al.}(1996)\citenamefont
  {{Borumand}}, \citenamefont {{Joynt}},\ and\ \citenamefont
  {{Klu{\'z}niak}}}]{bjk1996}%
  \BibitemOpen
  \bibfield  {author} {\bibinfo {author} {\bibfnamefont {M.}~\bibnamefont
  {{Borumand}}}, \bibinfo {author} {\bibfnamefont {R.}~\bibnamefont
  {{Joynt}}},\ and\ \bibinfo {author} {\bibfnamefont {W.}~\bibnamefont
  {{Klu{\'z}niak}}},\ }\bibfield  {title} {\bibinfo {title} {{Superfluid
  densities in neutron-star matter}},\ }\href
  {https://doi.org/10.1103/PhysRevC.54.2745} {\bibfield  {journal} {\bibinfo
  {journal} {\prc}\ }\textbf {\bibinfo {volume} {54}},\ \bibinfo {pages} {2745}
  (\bibinfo {year} {1996})}\BibitemShut {NoStop}%
\bibitem [{\citenamefont {{Gusakov}}\ and\ \citenamefont
  {{Haensel}}(2005)}]{gh2005}%
  \BibitemOpen
  \bibfield  {author} {\bibinfo {author} {\bibfnamefont {M.~E.}\ \bibnamefont
  {{Gusakov}}}\ and\ \bibinfo {author} {\bibfnamefont {P.}~\bibnamefont
  {{Haensel}}},\ }\bibfield  {title} {\bibinfo {title} {{The entrainment matrix
  of a superfluid neutron proton mixture at a finite temperature}},\ }\href
  {https://doi.org/10.1016/j.nuclphysa.2005.07.005} {\bibfield  {journal}
  {\bibinfo  {journal} {Nucl. Phys. A}\ }\textbf {\bibinfo {volume} {761}},\
  \bibinfo {pages} {333} (\bibinfo {year} {2005})},\ \Eprint
  {https://arxiv.org/abs/astro-ph/0508104} {arXiv:astro-ph/0508104 [astro-ph]}
  \BibitemShut {NoStop}%
\bibitem [{Note1()}]{Note1}%
  \BibitemOpen
  \bibinfo {note} {Consider, for example, the equation $\protect \mathcal
  {A}_l(a)P_l^m(x)+\protect \mathcal {A}_{l+2}(a)P_{l+2}^m(x)+\DOTSB \sum@
  \slimits@ \limits _{L}\protect \mathcal {B}_L(a)P_L^m(x)=0$. In this case the
  1st type equations are $\protect \mathcal {A}_L(a)+\protect \mathcal
  {B}_L(a)=0 \protect \text { for } L\in \{l,l+2\}$, whereas the 2nd type
  equations are $\protect \mathcal {B}_L(a)=0 \protect \text { for } L\not \in
  \{l,l+2\}$}\BibitemShut {NoStop}%
\bibitem [{\citenamefont {{Goriely}}\ \emph {et~al.}(2013)\citenamefont
  {{Goriely}}, \citenamefont {{Chamel}},\ and\ \citenamefont
  {{Pearson}}}]{gorielyetal2013}%
  \BibitemOpen
  \bibfield  {author} {\bibinfo {author} {\bibfnamefont {S.}~\bibnamefont
  {{Goriely}}}, \bibinfo {author} {\bibfnamefont {N.}~\bibnamefont
  {{Chamel}}},\ and\ \bibinfo {author} {\bibfnamefont {J.~M.}\ \bibnamefont
  {{Pearson}}},\ }\bibfield  {title} {\bibinfo {title} {{Further explorations
  of Skyrme-Hartree-Fock-Bogoliubov mass formulas. XIII. The 2012 atomic mass
  evaluation and the symmetry coefficient}},\ }\href
  {https://doi.org/10.1103/PhysRevC.88.024308} {\bibfield  {journal} {\bibinfo
  {journal} {\prc}\ }\textbf {\bibinfo {volume} {88}},\ \bibinfo {eid} {024308}
  (\bibinfo {year} {2013})}\BibitemShut {NoStop}%
\bibitem [{Note2()}]{Note2}%
  \BibitemOpen
  \bibinfo {note} {Note that the opposite is not true, and there are
  non-analytic functions of $\Omega $ with $d_1=0$. Consider, for example, any
  function of the form $f(a,\Omega )=f_1(\Omega )f_2(a)$, where $f_1(\Omega )$
  is some non-analytic function of $\Omega $ that does not depend on $a$, and
  $f_2(a)$ is some $\Omega $-independent function.}\BibitemShut {Stop}%
\bibitem [{\citenamefont {{Landau}}\ and\ \citenamefont
  {{Lifshitz}}(1965)}]{landau1965}%
  \BibitemOpen
  \bibfield  {author} {\bibinfo {author} {\bibfnamefont {L.~D.}\ \bibnamefont
  {{Landau}}}\ and\ \bibinfo {author} {\bibfnamefont {E.~M.}\ \bibnamefont
  {{Lifshitz}}},\ }\href@noop {} {\emph {\bibinfo {title} {{Quantum
  mechanics}}}}\ (\bibinfo {year} {1965})\BibitemShut {NoStop}%
\bibitem [{Note3()}]{Note3}%
  \BibitemOpen
  \bibinfo {note} {Since the function $q_\sigma (a)$ is positive for $a<a_{t}$,
  changes sign only at $a=a_t$, and $a_t\to a_{cc}$ in the $\Omega \to 0$
  limit, its minimal value in the core in this limit is reached at the
  crust-core interface. Therefore, the condition $q_\sigma (a_{cc})\to 0$ for
  obtaining the leading contribution to the correction $\sigma ^{(1)}$ can be
  equivalently presented as \begin {gather*} \label {minQ} \protect \qopname
  \relax m{min}\limits _{a_c<a<a_{cc}}\{q_\sigma (a)\}\to 0 \hskip 2em\relax
  \protect \text {for} \hskip 1em\relax a_t<a_{cc}. \end {gather*} An analogous
  condition determines the spectrum of the superfluid $r$-modes, studied in
  Kantor et al.\ \cite {kgd2021}.}\BibitemShut {Stop}%
\end{thebibliography}
\end{document}